\documentclass[%
 reprint,superscriptaddress,
 amsmath,amssymb,
 aps,
prx,
]{revtex4-1}
\usepackage{amssymb}
\usepackage{amsmath}
\usepackage{bm}
\usepackage{hyperref}
\usepackage{graphicx}
\usepackage{subfigure}
\usepackage{xcolor}
\newcommand{\ket}[1]{|#1\rangle}
\newcommand{\bra}[1]{\langle #1|}

\def\<{\langle}  %% overiding the original command \<
\def\>{\rangle}  %% overiding the original command \>

\begin{document}
% \preprint{APS/123-QED}
%\preprint{}

\title{Incompatibility measures in multi-parameter quantum estimation under hierarchical quantum measurements}
%\title{The uncertainty relation and Quantum Cram\'er-Rao bound}
\author{Hongzhen Chen}
%\email{hzchen@mae.cuhk.edu.hk}
\affiliation{Department of Mechanical and Automation Engineering, The Chinese University of Hong Kong, Shatin, Hong Kong SAR, P.R.China}

\author{Yu Chen}
%\email{anschen@mae.cuhk.edu.hk}
\affiliation{Department of Mechanical and Automation Engineering, The Chinese University of Hong Kong, Shatin, Hong Kong SAR, P.R.China}
%\author{Zhibo Hou}
%\affiliation{Key Laboratory of Quantum Information,University of Science and Technology of China, CAS, Hefei 230026, P. R. China}
%\affiliation{CAS Center For Excellence in Quantum Information and Quantum Physics}
%\author{Guo-Yong Xiang}
%\email{gyxiang@ustc.edu.cn}
%\affiliation{Key Laboratory of Quantum Information,University of Science and Technology of China, CAS, Hefei 230026, P. R. China}
%\affiliation{CAS Center For Excellence in Quantum Information and Quantum Physics}
%\author{Chuan-Feng Li}
%\affiliation{Key Laboratory of Quantum Information,University of Science and Technology of China, CAS, Hefei 230026, P. R. China}
%\affiliation{CAS Center For Excellence in Quantum Information and Quantum Physics}
%\author{Guang-Can Guo}
%\affiliation{Key Laboratory of Quantum Information,University of Science and Technology of China, CAS, Hefei 230026, P. R. China}
%\affiliation{CAS Center For Excellence in Quantum Information and Quantum Physics}
\author{Haidong Yuan}
\email{hdyuan@mae.cuhk.edu.hk}
\affiliation{Department of Mechanical and Automation Engineering, The Chinese University of Hong Kong, Shatin, Hong Kong SAR, P.R.China}
% auth+or. Explanatory text should go in the []'s, actual e-mail

%\author{Hongzhen Chen}
%\email{hzchen@mae.cuhk.edu.hk}
%\author{Haidong Yuan}
%\thanks{}
%\affiliation{Department of Mechanical and Automation Engineering,\\The Chinese University of Hong Kong, Shatin, Hong Kong}


\date{\today}

\begin{abstract}
%Although much progress has been made on the understanding of the Holevo bound, which captures the precision limit in the multi-parameter quantum estimation
%When collective measurements on infinite number of copies of identical quantum states can be performed, the precision limit of multi-parameter quantum estimation is quantified by the Holevo bound.
%In practise, however, the collective measurements are always restricted to a finite number of quantum states, under which the precision limit is still largely unexplored. Here
The incompatibility of the measurements constraints the achievable precisions in multi-parameter quantum estimation. Understanding the tradeoff induced by such incompatibility is a central topic in quantum metrology. Here we provide an approach to study the incompatibility under general $p$-local measurements, which are the measurements that can be performed collectively on at most $p$ copies of quantum states. We demonstrate the power of the approach by presenting a hierarchy of analytical bounds on the tradeoff among the precision limits of different parameters. These bounds lead to a necessary condition for the saturation of the quantum Cram\'er-Rao bound under $p$-local measurements, which recovers the partial commutative condition at p=1 and the weak commutative condition at $p=\infty$. As a further demonstration of the power of the framework, we present another set of tradeoff relations with the right logarithmic operators(RLD). %This paper is an extended version of the companion paper \cite{Companion}
\end{abstract}

%\keywords{}

\maketitle
\section{Introduction}
By utilizing quantum mechanical effects, such as superposition and entanglement, quantum metrology can achieve better precision limit than classical metrology. There is now a good understanding on the local precision limit for single-parameter quantum estimation, where the precision limit can be quantified by the single-parameter quantum Cram\'er-Rao bound\cite{Hels76book,Hole82book,BrauCM96,GiovLM04,GiovLM06,Escher2011,Naga07beating,HiggBBW07,XianHBW11,Slus17unconditional,daryanoosh2018experimental,yuan2017quantum,yuan2017M,Liu2022}. Practical applications, however, typically involve multiple parameters, for which the precision limits are much less understood\cite{Matsumoto_2002,PhysRevLett.111.070403,PhysRevLett.119.130504,Haya05book,Hou20minimal,Federico2021,Houeabd2986,HouSuper2021,Yang2019,PhysRevLett.116.180402,Zhu2018universally,Magdalena2016,PhysRevLett.123.200503,GillM00,ALBARELLI2020126311,Carollo_2019,Lu2021,Ragy2016,Chen_2017,Liu_2019,ChenHZ2019,Francesco2019,Rafal2020,Kok2020,vidrighin2014,crowley2014,Yue2014,Zhang2014,Liu2017}. Due to the incompatibility of the optimal measurements for different parameters, the multi-parameter quantum Cram\'er-Rao bound is in general not achievable. Tradeoffs among the precisions of different parameters have to be made. Quantifying such tradeoff is now one of the main subjects in quantum metrology\cite{GillM00,ALBARELLI2020126311,Carollo_2019,Lu2021,Ragy2016,Chen_2017,Liu_2019,ChenHZ2019,Francesco2019,Rafal2020,Kok2020,vidrighin2014,crowley2014,Yue2014,Zhang2014,Liu2017,Roccia_2017,e22111197,Candeloro2021,Koichi2013,Kahn2009,Yuxiang2019,Suzuki2016,Sidhu2021}.

The incompatibility of the measurements is rooted in the prohibition of simultaneous measurement of non-commutative observables, which is one of the defining features of quantum mechanics. Previous studies on the incompatibility mostly focus on the extreme cases: either the measurement is separable or can be performed collectively on infinite copies of quantum states. 
When the measurements can be performed on infinite number of identical copies of quantum states, the Holevo bound quantifies the achievable precision \cite{Hole82book,Koichi2013,Kahn2009,Yuxiang2019}. Except for few special cases\cite{Suzuki2016,Sidhu2021}, the Holevo bound in general can only be evaluated numerically\cite{PhysRevLett.123.200503}.  %Except a few special cases\cite{GillM00,HayaM05,Zhu2018universally,Lu2021} the tradeoffs under the collective measurements on a finite number of quantum states are little understood .
In practise the measurements typically can only be performed collectively on a finite number of quantum states, under which the Holevo bound is also not achievable in general. In the case of two parameters, Nagaoka provided a bound under the separable measurements which is tighter than the Holevo bound\cite{Nagaoka1,Nagaoka2}. Conlon et al. recently generalized the Nagaoka bound to more than two parameters, which in general requires numerical optimization\cite{Conlon2021}. Gill-Massar bound provided an analytical measure on the tradeoff induced by the incompatibility of the separable measurements\cite{GillM00}. For collective measurements on at most $2$ copies of quantum states, Zhu and Hayashi have obtained a tradeoff relation for completely unknown states\cite{Zhu2018universally}. However, the incompatibilities under general $p$-local measurements, which are the measurements that can be performed collectively on at most $p$ copies of quantum states, are little understood.

Here we provides a framework to study the precision under general $p$-local measurements. %which is based on the Roberston's uncertainty relation.
%. Although the uncertainty relations are connected to the precision limit of the estimation in its origin and have been used to derive the standard quantum limit previously, its connection to the currently widely used tools in quantum metrology, such as the quantum Cram\'er-Rao bound, remains unclear. Our study
This approach leads to new multi-parameter precision bounds which include the Holevo bound and the Nagaoka bound as special cases. We also provide a systematic way to generate hierarchical analytical tradeoff relations under general $p$-local measurements. The obtained tradeoff relations provide a necessary condition  for the saturation of the multi-parameter quantum Cram\'er-Rao bound under $p$-local measurements, which recovers the partial commutative condition\cite{Yang2019} at $p=1$ and the weak commutative condition at $p=\infty$. Our study thus not only provides a framework that can generate new analytical bounds on the tradeoff under general $p$-local measurements, but also provides a coherent picture for the existing results on the extreme cases. %The study thus not only opens the avenue for the studies of the incompatibility in multi-parameter quantum estimation under general $p$-local measurement, but also provides a coherent picture for the existing results on the extreme cases.   % and the obtained results significantly increased our understandings of the incompatibility in multi-parameter quantum estimation.%Specific examples are t to demonstrate the tradeoff relations under $p$-local measurements.

%The approach provides a physical picture for the precision limit in quantum metrology and is expected to cross-fertilize the studies in both the uncertainty relation and quantum metrology.

%The main results obtained are
%\begin{enumerate}
%    \item A systematic way to obtain the precision bounds from the uncertainty relations.\\
%    \item For pure states, we establish a connection between the optimal observables that saturate the uncertainty relations and the optimal measurements that saturate the QCRB. When the QCRB is not saturable, we provide nontrivial bounds to quantify the tradeoff.
 %   \item For mixed states, various tradeoff relations are obtained for the precisions under $p$-local measurements, in a form similar to the Gill-Massar bound.
 %   \item We builds the connection between the partial commutative condition and the weak commutative condition and show that the partial commutative condition reduces to the weak commutative condition when $p\rightarrow \infty$, which solves an open question in previous study.
 %   \item We show how improved matrix bounds can be obtained from the uncertainty relations.
 %   \end{enumerate}
The article is organized as following: in Sec.\ref{sec:review} we introduce the notations and list the main results; in Sec.\ref{sec:purestate} we present analytical tradeoff relations for pure states; in Sec.\ref{sec:mixstate} we provide new multi-parameter precision bounds for mixed states and use it to derive analytical tradeoff relations for mixed states. The tradeoff relations leads to a necessary condition for the saturation of the QCRB and we show how it reduces to the partial commutative condition at $p=1$ and the weak commutative condition at $p\rightarrow \infty$; in Sec. \ref{sec:RLD} we demonstrate the versatility of the approach by presenting another set of tradeoff relations in terms of the right logarithmic derivative; in Sec.\ref{sec:example} some examples are presented and Sec.\ref{sec:summary} concludes. %in multi-parameter quantum estimation.

\section{Precision limit in quantum metrology}\label{sec:review}
We first introduce the notations and terminologies that are used in this article and list the main results.

For the single-parameter quantum estimation, %the local precision limit can be characterized by the quantum Cram\'er-Rao bound\cite{Hels76book,Hole82book}.
given a parametrized state, $\rho_x$, with $x$ as the parameter to be estimated, by performing a positive operator valued measurement(POVM), denoted as $\{M_{\alpha}\}$, on the state, we can get the measurement result, $\alpha$, with a probability $p(\alpha|x)=Tr(\rho_xM_{\alpha})$. The variance of any locally unbiased estimator, $\hat{x}$, is then lower bounded by the Cram\'er-Rao bound\cite{Cram46, Fish22} as $\delta \hat{x}^2\geq \frac{1}{\nu F_C}$, here $\delta \hat{x}^2=E[\hat{x}-x]^2$ is the variance of the unbiased estimator, $F_C=\int_{\alpha} \frac{1}{p(\alpha|x)}(\frac{\partial_x p(\alpha|x)}{\partial x})^2d\alpha$ is the Fisher information\cite{Fish22}, $\nu$ is the number of repetitions of the procedure, which is assumed to be asymptotically large. By optimizing the POVM, we get the quantum Cram\'er-Rao bound(QCRB) \cite{Hels76book,Hole82book},
\begin{equation}\label{eq:QCRB}
\delta \hat{x}^2\geq \frac{1}{\nu F_C}\geq \frac{1}{\nu F_Q},
\end{equation}
here $F_Q$ is the quantum Fisher information (QFI), which is the maximization of the Fisher information over all POVM\cite{Hels76book,Hole82book}. The QFI can be computed directly from the quantum state as $F_Q=Tr(\rho_xL^2)$, here $L$ is the symmetric logarithmic operator(SLD) which is implicitly defined via the equation $\frac{\partial \rho_x}{\partial x}=\frac{1}{2}(L\rho_x+\rho_xL)$. For single-parameter estimation, the QCRB can always be saturated with the POVM performed separately on each copy of the state. One POVM that saturates the single-parameter QCRB is the projective measurement on the eigenvectors of the SLD.
%In general, the optimal measurement is not unique. The necessary and sufficient condition for the POVM, $\{E_y\}$, to saturate the QCRB is \cite{BrauC94}    \begin{equation}\label{eq:optimalPOVM}
%	E_{y}^{1/2}\rho_x^{1/2}=\lambda_{y}E_{y}^{1/2}L\rho_x^{1/2},
%	\lambda_y\in\mathbb{R}\quad \forall y.
%\end{equation}

For multi-parameter quantum estimation, where $x=(x_1,\cdots, x_n)$ with $n\geq 2$, the quantum Fisher information becomes the quantum Fisher information matrix with the $jk$-th entry given by
\begin{equation}
    (F_Q)_{jk}=Tr(\rho_x\frac{L_jL_k+L_kL_j}{2}),
\end{equation}
here $L_q$ is the SLD corresponds to the parameter $x_q$, which satisfies $\partial_{x_q}\rho_x=\frac{1}{2}(\rho_xL_q+L_q\rho_x)$, $\forall q\in\{1,\cdots, n\}$. The multi-parameter quantum Cram\'er-Rao bound is given by
\begin{equation}
    Cov(\hat{x})\geq \frac{1}{\nu}F_Q^{-1},
\end{equation}
where $Cov(\hat{x})$ is the covariance matrix for locally unbiased estimators, $\hat{x}=(\hat{x}_1,\cdots,\hat{x}_n)$, with the $jk$-th entry given by $Cov(\hat{x})_{jk}=E[(\hat{x}_j-x_j)(\hat{x}_k-x_k)]$, $\nu$ is the number of copies of quantum states. In this article, we assume $F_Q$ is non-singular so $F_Q^{-1}$ exists, in which case $Cov(\hat{x})\geq \frac{1}{\nu}F_Q^{-1}>0$ is also non-singular.

Different from the single-parameter quantum estimation, the multi-parameter quantum Cram\'er-Rao bound is in general not saturable. This is due to the incompatibility of the optimal measurements for different parameters. Such incompatibility is rooted in the prohibition of simultaneous measurement of non-commutative observables and its manifested effect in multi-parameter estimation is the tradeoff on the precision limits for the estimation of different parameters.

We can quantify the incompatibility through either $\frac{1}{\nu}Tr[F_Q^{-1}Cov^{-1}(\hat{x})]$\cite{ALBARELLI2020126311} or $\nu Tr[F_Q Cov(\hat{x})]$ \cite{ALBARELLI2020126311,Federico2021,Carollo_2019}, which measures how close $Cov(\hat{x})$ is to $\frac{1}{\nu}F_Q^{-1}$. These two quantities are roughly reciprocal to each another. Compared to the other quantities, such as $\|\nu Cov(\hat{x})-F_Q^{-1}\|$ or $\|\frac{1}{\nu}Cov^{-1}(\hat{x})-F_Q\|$, $\frac{1}{\nu}Tr[F_Q^{-1}Cov^{-1}(\hat{x})]$ and $\nu Tr[F_Q Cov(\hat{x})]$ both have the advantage of being invariant under reparametrization. In this article we will use $\Gamma=\frac{1}{\nu}Tr[F_Q^{-1}Cov^{-1}(\hat{x})]$ as the measure. %Here $\Gamma$ is understood as the largest value it can take with all allowed measurements.
When the QCRB is saturable, $Cov(\hat{x})=\frac{1}{\nu}F_Q^{-1}$, $\Gamma= Tr(I_n)=n$, here $I_n$ denotes the $n\times n$ Identity matrix. This is the maximal value $\Gamma$ can achieve. When the QCRB is not saturable we have $\Gamma<n$. The gap between $n$ and $\Gamma$ quantifies the incompatibility. We will denote the measure under the $p$-local measurement as $\Gamma_p=\frac{1}{\nu}Tr[F_Q^{-1}Cov^{-1}(\hat{x})]$ with $Cov(\hat{x})$ as the covariance matrix achieved under the optimal $p$-local measurement. %that lead to the maximum of $\frac{1}{\nu}Tr[F_Q^{-1}Cov^{-1}(\hat{x})]$.
The gap between $n$ and $\Gamma_p$ decreases with $p$ since the measurements become less restrictive when $p$ increases, we thus have $\Gamma_1\leq \Gamma_2\leq \cdots\leq \Gamma_{\infty}$. For pure states, however, we have $\Gamma_1=\Gamma_2= \cdots= \Gamma_{\infty}$ since for pure states the optimal measurement can be taken as the 1-local measurement \cite{Matsumoto_2002}.

The existing results on the incompatibility are mostly on the extreme cases with either $p=\infty$ or $p=1,2$.

%$\Gamma_{\infty}$, which quantifies the incompatibility of the collective measurements on infinite number of quantum states,
For $p=\infty$, the precision limit can be characterized by the Holevo bound\cite{Hole82book}, which is given by
%\begin{equation}
$\nu Tr[WCov(\hat{x})]\geq \min_{\{X_j\}}\{Tr[W ReZ(X)]+\|\sqrt{W}Im Z(X)\sqrt{W}\|_1\}$, where $W\geq0$ is a weighted matrix, $Z(X)$ is a matrix with its $jk$-th entry given by $Z(X)_{jk}=Tr(\rho_xX_jX_k)$, here $\{X_1,\cdots, X_n\}$ is a set of Hermitian operators that satisfy the local unbiased condition, $Tr(\rho_x X_j)=0$ for any $j\in\{1,\cdots, n\}$ and $Tr(\partial_{x_k}\rho_xX_j)=\delta_k^j$ with $\delta_k^j$ as the Kronecker delta, $\delta_k^j=1$ when $k= j$ and $\delta_k^j=0$ when $k\neq j$, $ReZ(x)=\frac{Z(x)+Z^T(x)}{2}$ is the real part of $Z(x)$, $ImZ(x)=\frac{Z(x)-Z^T(x)}{2i}$ is the imaginary part. The Holevo bound can only be evaluated numerically in general\cite{PhysRevLett.123.200503}. For pure states, the Holevo bound can be saturated by $1$-local measurements\cite{Matsumoto_2002}. For mixed states, the saturation of the Holevo bound in general requires collective measurements on infinite copies of the state.

A necessary and sufficient condition for the Holevo bound to coincide with the QCRB is  $Tr(\rho_x[L_j,L_k])=0$ for any $j, k\in\{1,\cdots, n\}$. This is called the weak commutative condition. When the weak commutative condition holds, there exists collective measurements on infinite copies of quantum states under which the QCRB is saturated and $\Gamma_{\infty}=n$.

As the Holevo bound corresponds to the minimal value upon all choice of $\{X_j\}$, by making a particular choice of $\{X_j\}$ as $X_j=\sum_k(F_Q^{-1})_{jk}L_k$ and $W=F_Q$, we have\cite{Carollo_2019}
\begin{equation}
    \nu Tr[F_QCov(\hat{x})]\leq n+\|F_Q^{-\frac{1}{2}}F_{Im}F_Q^{-\frac{1}{2}}\|_1\leq 2n,
\end{equation}
here $F_{Im}$ is a matrix with the $jk$-th entry given by $(F_{Im})_{jk}=\frac{1}{2i}Tr(\rho_x[L_j,L_k])$. The last inequality is obtained from the fact that $F_Q+iF_{Im}\geq 0$, which leads to $\|F_Q^{-\frac{1}{2}}F_{Im}F_Q^{-\frac{1}{2}}\|_1\leq Tr(F_Q^{-\frac{1}{2}}F_QF_Q^{-\frac{1}{2}})=n$. Through the Cauchy-Schwarz inequality,
\begin{eqnarray}\label{eq:CS}
\aligned
%Tr[F_Q^{\frac{1}{2}}Cov^{\frac{1}{2}}(\hat{x})Cov^{\frac{1}{2}}(\hat{x})F_Q^{\frac{1}{2}}]Tr[F_Q^{-\frac{1}{2}}Cov^{-\frac{1}{2}}(\hat{x})Cov^{-\frac{1}{2}}(\hat{x})F_Q^{-\frac{1}{2}}]
&Tr[F_QCov(\hat{x}]Tr[F_Q^{-1}Cov^{-1}(\hat{x})]\\
&\geq |Tr[F_Q^{\frac{1}{2}}Cov^{\frac{1}{2}}(\hat{x})Cov^{-\frac{1}{2}}(\hat{x})F_Q^{-\frac{1}{2}}]|^2=n^2,
\endaligned
\end{eqnarray}
this leads to a lower bound on $\Gamma_{\infty}$ as\cite{Federico2021}
\begin{equation}\label{eq:lowerbound}
   \Gamma_{\infty}=\frac{1}{\nu} Tr[F_Q^{-1}Cov^{-1}(\hat{x})]\geq \frac{n^2}{n+\|F_Q^{-\frac{1}{2}}F_{Im}F_Q^{-\frac{1}{2}}\|_1}.%\geq \frac{n}{2}.
\end{equation}
We note that the lower bound on $\Gamma_{\infty}$ is not sufficient to decide the incompatibility of the measurements at $p=\infty$ as it can not tell whether $\Gamma_{\infty}$ can reach $n$ and furthermore how close $\Gamma_{\infty}$ is to $n$. The upper bound is more informative in this sense. As if there exists an upper bound which is less than $n$, we can tell for sure that the measurements are incompatible, and furthermore the gap between $n$ and the upper bound provides a measure on the incompatibility. To our knowledge, except the trivial bound, $\Gamma_{\infty}\leq n$, there were no analytical upper bounds on $\Gamma_{\infty}$ even at the extreme case with $p=\infty$.

%A necessary and sufficient condition for the saturation of the QCRB, allowing collective measurements on infinite number of quantum states, is the weak commutative condition, which is  $Tr(\rho_x[L_j,L_k])=0$ for any $j, k\in\{1,\cdots, n\}$. When the measurement is restricted to the $p$-local measurement, it remains unknown when the QCRB can be saturated.

For the other extreme case with $p=1$, Nagaoka provided a bound on the precision limit in the case of two parameters($n=2$)\cite{Nagaoka1, Nagaoka2}, which is given by
\begin{equation}
\aligned
    &\nu Tr[Cov(\hat{x})]\\
    \geq& \min_{\{X_1,X_2\}}Tr(\rho_xX_1^2)+Tr(\rho_xX_2^2)+\|\sqrt{\rho_x}[X_1,X_2]\sqrt{\rho_x}\|_1,
    \endaligned
\end{equation}
where $\{X_1,X_2\}$ are two Hermitian operators satisfying the locally unbiased condition. The Nagaoka bound in general can only be evaluated numerically and is tighter than the Holevo bound. Recently the Nagaoka bound has been generalized to $n$ parameters which also requires numerical evaluation \cite{Conlon2021}.

Gill and Massar provided an analytical upper bound on $\Gamma_1$ as\cite{GillM00}
\begin{equation}
    \Gamma_1%=\frac{1}{\nu}Tr[F_Q^{-1}Cov^{-1}(\hat{x})]
    \leq d-1,
\end{equation}
here $d$ is the dimension of the Hilbert space for a single $\rho_x$. The Gill-Massar bound is nontrivial only when $n> d-1$. Recent studies have also obtained some tradeoff relations with the Ozawa's uncertainty relation for pairs of parameters\cite{Lu2021}.

A necessary condition for the saturation of the QCRB under 1-local measurements is the partial commutative condition\cite{Yang2019}, which requires all SLDs commute on the support of $\rho_x$. Specifically if we write $\rho_x$ in the eigenvalue decomposition as $\rho_x=\sum_1^m\lambda_i|\Psi_i\rangle\langle\Psi_i|$ with $\lambda_i>0$, the partial commutative condition is $\langle \Psi_r|[L_j,L_k]|\Psi_s\rangle=0$ for any $j,k\in\{1,\cdots, n\}$, and any $r,s\in\{1,\cdots, m\}$. The connection between the partial commutative condition and the weak commutative condition remained open\cite{Yang2019}.
%Two questions about the partial commutative condition remain open\cite{Yang2019}: 1) Is the partial commutative condition sufficient for the saturation of the QCRB under the $1$-local measurement? 2) What is the connection between the partial commutative condition and the weak commutative condition?

%The partial commutative condition can be equivalent written as $\sqrt{\rho_x}[L_j,L_k]\sqrt{\rho_x}=0$, or $\|\sqrt{\rho_x}[L_j,L_k]\sqrt{\rho_x}\|_1=0$ for any pair of $j,k$, where $\|\text{ }\|_1$ denotes the trace norm which equals to the sum of the singular values.
%Two questions about the partial commutative condition remain open\cite{Yang2019}: 1) Is the partial commutative condition also sufficient for the saturation of the QCRB under the $1$-local measurement? 2) Does the partial commutative condition for the state $\rho_x^{\otimes p}$ reduces to the weak commutative condition when $p\rightarrow \infty$? We solve the second question in this article and provide an affirmative answer. %which suggests that the answer to the first question may also be affirmative. %partial commutative condition may also be sufficient.

For $p=2$, Zhu and Hayashi provided an upper bound on $\Gamma_2$ as%for completely unknown $d$-dimensional states with $d^2-1$ parameters as\cite{Zhu2018universally}
\begin{equation}
    \Gamma_2=\frac{1}{\nu}Tr[F_Q^{-1}Cov^{-1}(\hat{x})]\leq \frac{3}{2}(d-1),
\end{equation}
which is nontrivial only when $n>\frac{3}{2}(d-1)$.

%on the incompatibility of general $p$-local measurements.

%{\color{red}
%\begin{equation}
%    \Gamma_2=\frac{1}{\nu}Tr[F_Q^{-1}Cov^{-1}(\hat{x})]\leq \frac{3}{2}(d-1).
%\end{equation}
%}

%In the case of pure states, Eq.(\ref{eq:lowerbound}) also puts a constraint on the number of parameters that can have nonsingular $F_Q$, since $\frac{n}{2}\frac{1}{\nu}Tr[F_Q^{-1}Cov^{-1}(\hat{x})]\leq d-1$, which implies $n\leq 2(d-1)$.
% There are, however, little understanding on general $\Gamma_p$\cite{Federico2021}.

%When the weak commutative condition is not satisfied, the ultimate precision limit, allowing collective measurements on infinite number of copies, is provided by the Holevo bound\cite{Hole82book}.
%In general the QCRB, $Cov(\hat{x})\geq \frac{1}{\nu}F_Q^{-1}$, is not saturable due to

%where $\Gamma_{\infty}$ can be characterized by the Holevo bound. Note that for pure states, we have $\Gamma_1=\Gamma_2=\cdots= \Gamma_{\infty}$ since the Holevo bound can be achieved by 1-local measurement for pure states. For mixed states, these quantities are generally not equal.

%The upper bound on $\Gamma_{\infty}$ ($\Gamma_p$ as well), however, are more informative on the incompatibility as it bounds the gap $n-\Gamma$ away from 0.
%$F_{Im}$ then provides a lower bound on the largest $\gamma$, and we will show below that it also provides improved upper bounds on $\gamma$ when it is not 0.

%\end{equation}
For general $p$, the incompatibility is little understood. In this article, we provide a framework to study the incompatibility under general $p$-local measurements. This framework provides new precision bounds that include the Holevo bound and the Nagaoka bound as special cases, and leads to nontrivial analytical upper bounds for general $\Gamma_p$. A necessary condition for the saturation of the QCRB can also be obtained, which recovers the partial commutative condition at $p=1$ and the weak commutative condition at $p\rightarrow \infty$. The new multi-parameter precision bounds are presented in Sec.\ref{subsec:multibounds}, here we first list the analytical upper bounds and the necessary condition for the saturation of QCRB under general p-local measurements.   %which connects the locally unbiased operators, $\{X_j\}$, with the SLDs and RLDs. The connection
\begin{enumerate}
    \item For pure states, we have
    \begin{eqnarray}
    \aligned
    \frac{1}{\nu}Tr[F_Q^{-1}Cov^{-1}(\hat{x})]\leq &n-f(n)\|F_{Q}^{-\frac{1}{2}}F_{Im}F_{Q}^{-\frac{1}{2}}\|_F^2, 
%    \frac{1}{\nu}Tr[F_Q^{-1}Cov^{-1}(\hat{x})]
 %   \leq &n-\frac{1}{4(n-1)}\|F_{Q}^{-\frac{1}{2}}F_{Im}F_{Q}^{-\frac{1}{2}}\|_F^2,\\
  %          \frac{1}{\nu}Tr[F_Q^{-1}Cov^{-1}(\hat{x})]%&\leq n-\frac{n-2}{(n-1)^2}\|\tilde{F}_{Im}\|_F^2\\
%\leq & n-\frac{n-2}{(n-1)^2}\|F_{Q}^{-\frac{1}{2}}F_{Im}F_{Q}^{-\frac{1}{2}}\|_F^2,\\
 %   \frac{1}{\nu}Tr[F_Q^{-1}Cov^{-1}(\hat{x})]
 %   \leq& n-\frac{1}{5}\|F_{Q}^{-\frac{1}{2}}F_{Im}F_{Q}^{-\frac{1}{2}}\|_F^2.
\endaligned
    \end{eqnarray}
    here $\|\text{ }\|_F$ is the Frobenius norm and $n$ is the number of parameters, $f(n)=\max\{\frac{1}{4(n-1)},\frac{n-2}{(n-1)^2},  \frac{1}{5}\}$ which can be equivalently written as 
    $$
f(n)=\left\{\begin{array}{lll}
\frac{1}{4(n-1)} & \text { when } & n=2 \\
\frac{n-2}{(n-1)^{2}} & \text { when } & n=3 \text { or } 4 \\\frac{1}{5} & \text { when } & n \geq 5
\end{array}\right.
$$
    
    %Specifically when $n=2$, the second is tighter than the other two when $n=3$ or $4$, and the third is tighter than the other two for $n\geq 5$.
    %\item For pure states, when $n\geq 3$ the bound can be tightened as
    %\begin{eqnarray}
    %\aligned
    %        &\frac{1}{\nu}Tr[F_Q^{-1}Cov^{-1}(\hat{x})]\\%&\leq n-\frac{n-2}{(n-1)^2}\|\tilde{F}_{Im}\|_F^2\\
%\leq & n-\frac{n-2}{(n-1)^2}\|F_{Q}^{-\frac{1}{2}}F_{Im}F_{Q}^{-\frac{1}{2}}\|_F^2.
%\endaligned
%    \end{eqnarray}
%And we also have
%    \begin{eqnarray}
%    \aligned
%    \frac{1}{\nu}Tr[F_Q^{-1}Cov^{-1}(\hat{x})]
%    \leq n-\frac{1}{5}\|F_{Q}^{-\frac{1}{2}}F_{Im}F_{Q}^{-\frac{1}{2}}\|_F^2,
%\endaligned
%    \end{eqnarray}
%    which is tighter than the above bound for $n\geq 5$.

We note the bounds for pure states do not depend on $p$ since for pure states $\Gamma_1=\Gamma_2=\cdots=\Gamma_{\infty}$.
%   \item For mixed states under the 1-local measurement, we have
%   \begin{equation}
%    \frac{1}{\nu}Tr[F_Q^{-1}Cov^{-1}(\hat{x})]\leq n-\frac{1}{4(n-1)} \|C_{1}\|_F^2,
%\end{equation}
%with the entries of $C_1$ given by
%\begin{eqnarray}
%\aligned
%$(C_1)_{jk}%&=\frac{1}{2}\|\sqrt{\rho_x}[\tilde{L}_j,\tilde{L}_k]\sqrt{\rho_x}\|_1\\
%=\frac{1}{2}\|\sqrt{\rho_x}[\sum_q (F_Q^{-\frac{1}{2}})_{jq}L_q,\sum_q (F_Q^{-\frac{1}{2}})_{kq}L_q]\sqrt{\rho_x}\|_1.$
%\endaligned
%\end{eqnarray}
\item For mixed states under $p$-local measurements, we have
\begin{eqnarray}\label{eq:mainFimps}
\aligned
 \Gamma_p   \leq &n-f(n)\|\frac{F_Q^{-\frac{1}{2}}\mathbf{\bar{F}}_{Imp}F_Q^{-\frac{1}{2}}}{p}\|_F^2,
 %  \Gamma_p   \leq &n-\frac{n-2}{(n-1)^2}\|\frac{F_Q^{-\frac{1}{2}}\mathbf{\bar{F}}_{Imp}F_Q^{-\frac{1}{2}}}{p}\|_F^2, \\%Tr(\tilde{C}_1^T\tilde{C}_1),
   % \endaligned
%\end{eqnarray}
%and
%\begin{eqnarray}
%\aligned
% & \frac{1}{\nu}Tr[F_Q^{-1}Cov^{-1}(\hat{x})]\\
 %\Gamma_p    \leq &n-\frac{1}{5}\|\frac{F_Q^{-\frac{1}{2}}\mathbf{\bar{F}}_{Imp}F_Q^{-\frac{1}{2}}}{p}\|_F^2, %Tr(\tilde{C}_1^T\tilde{C}_1),
    \endaligned
\end{eqnarray}
where $f(n)=\max\{\frac{1}{4(n-1)},\frac{n-2}{(n-1)^2},  \frac{1}{5}\}$, $\mathbf{\bar{F}}_{Imp}$ is the imaginary part of  $\mathbf{\bar{F}}=\sum_q \bar{F}_{u_q}$ with each $\bar{F}_{u_q}$ equal to either $F_{u_q}$ or $F_{u_q}^T$, here $F_{u_q}$ is a $n\times n$ matrix with the $jk$-th entry given by
\begin{equation}
    (F_{u_q})_{jk}=\langle u_q|\sqrt{\rho_x^{\otimes p}}L_{jp}L_{kp}\sqrt{\rho_x^{\otimes p}}|u_q\rangle,
\end{equation}
$L_{jp}$ is the SLD of $\rho_x^{\otimes p}$ corresponding to the parameter $x_j$, and $\{|u_q\rangle\}$ are any set of vectors in $H_d^{\otimes p}$ that satisfies $\sum_q |u_q\rangle\langle u_q|=I_{d^p}$ with $I_{d^p}$ denote the $d^p\times d^p$ Identity matrix.

\item For mixed states under $p$-local measurements, we obtain another bound as
\begin{eqnarray}
\aligned
    %\frac{1}{\nu}Tr[F_Q^{-1}Cov^{-1}(\hat{x})]
    \Gamma_p\leq n-\frac{1}{4(n-1)} \|\frac{C_{p}}{p}\|_F^2, %Tr(\tilde{C}_1^T\tilde{C}_1),
    \endaligned
\end{eqnarray}
here %$C_p$ is a matrix with the entries given by
\begin{equation}%\label{eq:CP}
    (C_p)_{jk}=\frac{1}{2}\|\sqrt{\rho_x^{\otimes p}}[\tilde{L}_{jp}, \tilde{L}_{kp}]\sqrt{\rho_x^{\otimes p}}\|_1,
\end{equation}
%here %$\tilde{L}_{jp}=\sum_{r=1}^p \tilde{L}_j^{(r)}$, here $\tilde{L}_j^{(r)}=I^{\otimes (r-1)}\otimes \tilde{L}_j\otimes I^{\otimes (p-r)}$, $r=1,\cdots, p$, which can also be equivalently written as
$\tilde{L}_{jp}$ is the SLD of $\rho_x^{\otimes p}$ under the reparametrization such that the QFIM of $\rho_x$ equals to the Identity, specifically  $\tilde{L}_{jp}=\sum_q (F_Q^{-\frac{1}{2}})_{jq}L_{qp}$ with $L_{qp}$ as the SLD of $\rho_x^{\otimes p}$ corresponding to the original parameter $x_q$. We note that $\|\frac{C_{p}}{p}\|_F\geq\|\frac{F_Q^{-\frac{1}{2}}\mathbf{\bar{F}}_{Imp}F_Q^{-\frac{1}{2}}}{p}\|_F$, this bound is thus tighter than the bound in Eq.(\ref{eq:mainFimps}) when $f(n)=\frac{1}{4(n-1)}$, however it can be less tighter when $f(n)=\frac{n-2}{(n-1)^2}$ or $\frac{1}{5}$. 
\item From the above bound, we obtain a necessary condition for the saturation of the QCRB under $p$-local measurements, which is $\frac{C_p}{p}=0$. For $p=1$, this reduces to the partial commutative condition. For $p\rightarrow \infty$, we prove that
%\begin{equation}
    %(C_1)_{jk}\geq \frac{(C_2)_{jk}}{2}\geq\cdots
% $   \frac{(C_p)_{jk}}{p}\geq \frac{(C_{p+1})_{jk}}{p+1}$ %\geq \cdots
%\end{equation}
%and
\begin{equation}
\lim_{p\rightarrow \infty}\frac{(C_p)_{jk}}{p}=\frac{1}{2}|Tr(\rho_x[\tilde{L}_j,\tilde{L}_k])|.
\end{equation}
The condition, $\frac{C_p}{p}=0$, thus reduces to the weak commutative condition, $Tr(\rho_x[\tilde{L}_j,\tilde{L}_k])=0$, $\forall j, k$, at $p\rightarrow \infty$. This clarifies the relation between the partial commutative condition and the weak commutative condition, which solves an open question\cite{Yang2019}. %Here $\tilde{L}_j$ is the SLD under the reparameterization which makes the QFIM equals to the Identity, i.e., $\tilde{L}_j=\sum_q (F_Q^{-\frac{1}{2}})_{jq}L_{q}$ with $L_{q}$ as the SLD of $\rho_x$ corresponding to the original parameter $x_q$. This clarifies the relation between the partial commutative condition and the weak commutative condition.
%The weak commutative condition under this parametrization is equivalent to the weak commutative condition in the original parametrization, $Tr(\rho_x[L_j,L_k])=0$, $\forall j, k$. %Here the weak commutative condition is expressed under the parametrization that $\tilde{F}_Q=I$, which is equivalent to the weak commutative condition in the original parametrization, $Tr(\rho_x[L_j,L_k])=0$, $\forall j, k$ .

\item %A set of alternative tradeoff relations are obtained.
%The above bounds for mixed states involve operators on $p$ copies of quantum states, whose dimension grows exponentially with $p$, which can be hard to compute when $p$ is large.
We provide another simpler bound for mixed states which can be calculated with operators only on a single $\rho_x$.

Given $\rho_x=\sum_{q=1}^m\lambda_q |\Psi_q\rangle\langle\Psi_q|$ with $\lambda_q>0$ in the eigenvalue decomposition, under $p$-local measurements we have
\begin{equation}
   % \frac{1}{\nu}Tr[F_Q^{-1}Cov^{-1}(\hat{x})]
    \Gamma_p\leq n-\frac{1}{4(n-1)} \|\frac{T_p}{p}\|_F^2,
\end{equation}
where $T_p$ is a $n\times n$ matrix with the $jk$-th entry given by
%\begin{eqnarray}
%\aligned
 %   &(T_p)_{jk}%=\frac{1}{2}\|D_p^{(jk)}\|_1\\
 %   =&\frac{1}{2}\sum_{v_1,\cdots, v_p%\in\{1,\cdots, m\}
 %   }(\prod_{r=1}^p\lambda_{v_r}) |\sum_{r=1}^p \langle\Psi_{v_r}|[\tilde{L}_j,\tilde{L}_k]|\Psi_{v_r}\rangle|.
 %   \endaligned
%\end{eqnarray}
%here $v_1,\cdots,v_p\in \{1,\cdots, m\}$, 
\begin{equation}
(T_p)_{jk}=\frac{1}{2}E(|\sum_{r=1}^p \langle\Phi_{r}|[\tilde{L}_j,\tilde{L}_k]|\Phi_{r}\rangle|),    
\end{equation}
%&(T_p)_{jk}%=\frac{1}{2}\|D_p^{(jk)}\|_1\\
    %=&\frac{1}{2}\sum_{v_1,\cdots, v_p%\in\{1,\cdots, m\}
    %}(\prod_{r=1}^p\lambda_{v_r}) |\sum_{r=1}^p \langle\Psi_{v_r}|[\tilde{L}_j,\tilde{L}_k]|\Psi_{v_r}\rangle|,
%    \endaligned
%\end{eqnarray}
here $E(\cdot)$ denotes the expected value, each $|\Phi_r\rangle$ is randomly and independently chosen from the eigenvectors of $\rho_x$ with a probability equal to the corresponding eigenvalue, i.e., each $|\Phi_r\rangle$ takes $|\Psi_q\rangle$ with probability $\lambda_q$
, $q\in \{1,\cdots, m\}$. $\tilde{L}_{j}=\sum_{\mu}(F_Q^{-{\frac{1}{2}}})_{j\mu}L_{\mu}$ and $\tilde{L}_{k}=\sum_{\mu}(F_Q^{-{\frac{1}{2}}})_{k\mu}L_{\mu}$. % $(T_p)_{jk}$ basically equals to the expected value of $\frac{1}{2}|\sum_{r=1}^p \langle\Psi_{v_r}|[\tilde{L}_j,\tilde{L}_k]|\Psi_{v_r}\rangle|$ with each eigenvector $|\Psi_{v_r}\rangle$ selected with probability $\lambda_{v_r}$.

%$T_p$ can be computed with operators on a single $\rho_x$.
For large $p$, this bound is almost as tight as the bound with $\frac{C_p}{p}$, the difference between $\frac{T_p}{p}$ and $\frac{C_p}{p}$ is at most of the order  $O(\frac{1}{\sqrt{p}})$ with
\begin{equation}
    \frac{(T_p)_{jk}}{p}\leq \frac{(C_p)_{jk}}{p}\leq \frac{(T_p)_{jk}}{p}+O(\frac{1}{\sqrt{p}}).
\end{equation}
Asymptotically they converge to the same value,
\begin{equation}
\lim_{p\rightarrow \infty}\frac{(T_p)_{jk}}{p}=\lim_{p\rightarrow \infty}\frac{(C_p)_{jk}}{p}=\frac{1}{2}|Tr(\rho_x[\tilde{L}_j,\tilde{L}_k])|.
\end{equation}

\item To demonstrate the versatility of the framework, we provide another set of bounds with the right logarithm derivative(RLD) operators.
\begin{eqnarray}
\aligned
 %&\frac{1}{\nu}Tr[F_{Q}^{-1}Cov^{-1}(\hat{x})]\\
 \Gamma_p &\leq  Tr[F_{Q}^{-1}F_{Re}^{RLD}]-\frac{1}{4(n-1)}\|\frac{C_p^{RLD}}{p}\|_F^2,
\endaligned
\end{eqnarray}
here $F_{Re}^{RLD}$ is the real part of the RLD quantum Fisher information matrix with the $jk$-th entry given by  $(F^{RLD})_{jk}=Tr(\rho_xL_j^{R}L_k^{R\dagger})$, here $L_j^R$ is the RLD operator corresponding to the parameter $x_j$, which can be obtained from $\partial_{x_j}\rho_x=\rho_xL_j^{R}$, $(C_p^{RLD})_{jk}=\min\{\frac{1}{2}\|\sqrt{\rho_x^{\otimes p}}(\tilde{L}_{jp}^R\tilde{L}_{kp}^{R\dagger}-\tilde{L}_{kp}^R\tilde{L}_{jp}^{R\dagger})\sqrt{\rho_x^{\otimes p}}\|_1,2p\}$ with $\tilde{L}^R_{jp}=\sum_q (F_Q^{-\frac{1}{2}})_{jq}L^R_{qp}$ and $\tilde{L}^R_{kp}=\sum_q (F_Q^{-\frac{1}{2}})_{kq}L^R_{qp}$ with $L^R_{qp}$ as the RLD operator of $\rho_x^{\otimes p}$ corresponding to the parameter $x_q$.
\end{enumerate}

%We note that these bounds are in general not saturable, however they provide analytical tighter bounds than the QCRB and are nontrivial regardless of the number of the parameters and the dimension of the quantum states.
These bounds are in general not saturable, however they are nontrivial regardless of the number of the parameters and the dimension of the quantum states. The upper bounds can also be directly transformed to the lower bounds for various other measures via the Cauchy-Schwarz inequality. For example, 
%We note that when the QCRB is saturable $\nu Tr[F_QCov(\hat{x})]$ achieves the minimal value, $n$, and when the QCRB is not saturable the gap between $\nu Tr[F_QCov(\hat{x})]$ and $n$ provides a measure on the incompatibility. For example,
from the upper bound
\begin{eqnarray}
\aligned
    \frac{1}{\nu}Tr[F_Q^{-1}Cov^{-1}(\hat{x})]\leq n-\frac{1}{4(n-1)} \|\frac{C_{p}}{p}\|_F^2, %Tr(\tilde{C}_1^T\tilde{C}_1),
    \endaligned
\end{eqnarray}
we can obtain a lower bound for $\nu Tr[F_QCov(\hat{x})]$ via the Cauchy-Schwarz inequality as
\begin{eqnarray}
\aligned
\nu Tr[F_QCov(\hat{x})]&\geq \frac{n^2}{\frac{1}{\nu}Tr[F_{Q}^{-1}Cov^{-1}(\hat{x})]}\\
    &\geq \frac{n^2}{n-\frac{1}{4(n-1)} \|\frac{C_{p}}{p}\|_F^2}\\
    &\geq n+\frac{1}{4(n-1)} \|\frac{C_{p}}{p}\|_F^2,
    \endaligned
\end{eqnarray}
which provides a lower bound on $\nu Tr[F_QCov(\hat{x})]$ under p-local measurements. $\nu Tr[F_QCov(\hat{x})]$ achieves the minimal value, $n$, when the QCRB is saturable and the gap between $\nu Tr[F_QCov(\hat{x})]$ and $n$ provides a measure on the incompatibility. We note that the transformation from the upper bound to the lower bound via the Cauchy-Schwarz inequality does not work the other way, i.e., the lower bound on $\nu Tr[F_QCov(\hat{x})]$ can not be directly transformed to the upper bound on $\frac{1}{\nu}Tr[F_Q^{-1}Cov^{-1}(\hat{x})]$ via the Cauchy-Schwarz inequality. This is one advantage of choosing $\frac{1}{\nu}Tr[F_Q^{-1}Cov^{-1}(\hat{x})]$ over $\nu Tr[F_QCov(\hat{x})]$ as the measure of the incompatibility.

Similarly we can obtain lower bounds on the weighted covariance matrix, $\nu Tr[W Cov(\hat{x})]$, via the Cauchy-Schwarz inequality as
\begin{equation}
   \nu Tr[W Cov(\hat{x})]\geq \frac{(Tr \sqrt{F_Q^{-\frac{1}{2}}WF_Q^{-\frac{1}{2}}})^2}{\frac{1}{\nu}Tr[F_Q^{-1}Cov^{-1}(\hat{x})]}.
\end{equation}
For example, from the upper bound
\begin{eqnarray}
\aligned
    \frac{1}{\nu}Tr[F_Q^{-1}Cov^{-1}(\hat{x})]\leq n-\frac{1}{4(n-1)} \|\frac{C_{p}}{p}\|_F^2, %Tr(\tilde{C}_1^T\tilde{C}_1),
    \endaligned
\end{eqnarray}
we can obtain a lower bound,
\begin{equation}
   \nu Tr[W Cov(\hat{x})]\geq \frac{(Tr \sqrt{F_Q^{-\frac{1}{2}}WF_Q^{-\frac{1}{2}}})^2}{n-\frac{1}{4(n-1)} \|\frac{C_{p}}{p}\|_F^2},
\end{equation}
which constraints the precision that can be achieved under p-local measurements.
%As an alternative incompatibility measure $\nu Tr[F_QCov(\hat{x})]$ has some similar properties: it achieves the minimal value, $n$, when the QCRB is saturable and the gap between $\nu Tr[F_QCov(\hat{x})]$ and $n$ provides a measure on the incompatibility.
%However, the Cauchy-Schwarz inequality, $\frac{1}{\nu}Tr[F_Q^{-1}Cov^{-1}(\hat{x})]\geq \frac{n^2}{\nu Tr[F_QCov(\hat{x})]}$, can not directly transform the lower bound on $\nu Tr[F_QCov(\hat{x})]$ to upper bounds on $\frac{1}{\nu}Tr[F_Q^{-1}Cov^{-1}(\hat{x})]$. This is one reason that we choose  $\frac{1}{\nu}Tr[F_Q^{-1}Cov^{-1}(\hat{x})]$ over $\nu Tr[F_QCov(\hat{x})]$ as the incompatibility measure in this article.

%We note that these bounds are in general not saturable, however they are nontrivial regardless of the number of the parameters and the dimension of the quantum states.

Besides these analytical bounds, new multi-parameter precision bounds for mixed states, which requires numerical optimization, are presented in Sec.\ref{subsec:multibounds}. %which provides a useful tool for multi-parameter quantum estimation.
%Generically $\frac{C_p}{p}$ has $n(n-1)$ nontrivial entries (total $n^2$ entries with $n$ zero diagonal entries), so generically $\|\frac{C_p}{p}\|_F^2$ is of the order $O(n^2)$. %thus $\frac{1}{4(n-1)}\|\frac{C_p}{p}\|_F^2$ in the order of $O(n)$.
%use the uncertainty relation, in particular the Robertson's uncertainty relation with multiple observables\cite{PhysRev.46.794,Trifonov:2002aa}, to quantify the incompatibility for general $p$-local measurements.

\section{Analytical bounds for pure states }\label{sec:purestate}
We start the derivation of the bounds for pure states, then generalize it to mixed states in the next section.

Given a probe state $|\Psi_x\rangle$ with $x=(x_1,x_2,\cdots, x_n)$, and $q$ operators $\{Y_1,Y_2,\cdots, Y_q\}$, we have
\begin{eqnarray}\label{eq:S}
\aligned
S&=\left(\begin{array}{ccc}
Y_1|\Psi_x\rangle & \cdots & Y_q|\Psi_x\rangle \end{array}\right)^\dagger \left(\begin{array}{ccc}Y_1|\Psi_x\rangle & \cdots & Y_q|\Psi_x\rangle \end{array}\right)\\
&\geq 0,
\endaligned
\end{eqnarray}
%\begin{eqnarray}
%S\geq 0,
%\end{eqnarray}
here $S$ is a $q\times q$ matrix with its $jk$-th entry given by $(S)_{jk}=\langle\Psi_x|Y_j^\dagger Y_k|\Psi_x\rangle=Tr(\rho_xY_j^\dagger Y_k)$ with $\rho_x=|\Psi_x\rangle\langle\Psi_x|$. %Since the optimal measurement for pure states can be taken as 1-local measurements,
We note that $S\geq 0$ also forms the basis for the generalized Robertson uncertainty relation\cite{PhysRev.46.794,Trifonov:2002aa}. %GIBILISCO20081706}.
%the corresponding generator of $x_i$ is $H_i$ as
%\begin{eqnarray}
%\partial_{x_i}|\Psi(x)\rangle=-iH_i|\Psi(x)\rangle.
%\partial_{x_2}|\Psi(x_1,x_2)\rangle=-iH_2|\Psi(x_1,x_2)\rangle.
%\end{eqnarray}

We first consider a single copy of the state, for $\nu$ copies of the states, we can just replace $|\Psi(x)\rangle$ with $|\Psi(x)\rangle^{\otimes \nu}$. Given a measurement, $\{M_{\alpha}\}$ with $\sum_{\alpha} M_{\alpha}=I$, we can construct $n$ observables as
\begin{eqnarray}
X_j=\sum_{\alpha}[\hat{x}_j(\alpha)-x_j]M_{\alpha},
%\hat{X}_2=\sum_p(\hat{x}_2(p)-x_2)M_p.
\end{eqnarray}
where $\hat{x}_j$ is the estimator for $x_j$.
For locally unbiased estimator, we have
\begin{eqnarray}
Tr(\rho_xX_j)=0,\qquad j=1,\cdots, n%\langle\Psi|\hat{X}_2|\Psi\rangle=0,
\end{eqnarray}
and
\begin{eqnarray}\label{eq:localmain}
\aligned
Tr(\partial_{x_j}\rho_xX_k)=\delta^j_k.
\endaligned
\end{eqnarray}

Let $L_j$ be the SLD for $x_j$ with $j\in\{1,\cdots, n\}$, then by replacing the set of $\{Y_j\}$ in Eq.(\ref{eq:S}) with the $2n$ operators, $\{X_1,\cdots, X_n,L_1,\cdots, L_n\}$, we have
\begin{equation}
    S=\left(\begin{array}{cc}
      A & B  \\
      B^\dagger & F \\
          \end{array}\right)\geq 0,
\end{equation}
where $A,B,F$ are $n\times n$ matrices with the entries given by
\begin{eqnarray}
\aligned
(A)_{kj}&=Tr(\rho_xX_kX_j),\\
(B)_{kj}&=Tr(\rho_xX_kL_j),\\
(F)_{kj}&=Tr(\rho_xL_kL_j).
%A=&\left(\begin{array}{cc}
%      Tr(\rho_x \hat{X}_1^2) & Tr(\rho_x\hat{X}_1\hat{X}_2) % \\
%      Tr(\rho_x\hat{X}_2\hat{X}_1) & Tr(\rho_x \hat{X}_2^2) %\\
%          \end{array}\right),\\
%          B=&\left(\begin{array}{cc}
%      Tr(\rho_x \hat{X}_1L_1) & Tr(\rho_x\hat{X}_1L_2)  \\
%      Tr(\rho_x\hat{X}_2L_1) & Tr(\rho_x \hat{X}_2L_2) \\
%          \end{array}\right)=B_Q+iB_{im}, \\
%          F=&\left(\begin{array}{cc}
%      Tr(\rho_x L_1^2) & Tr(\rho_x L_1L_2)  \\
%      Tr(\rho_xL_2L_1) & Tr(\rho_x L_2^2) \\
%          \end{array}\right)=F_Q+iF_{im},
\endaligned
\end{eqnarray}  %$L_i=2i[H_i,\rho_x]$
We can write these matrices in terms of the real and imaginary part as $A=A_{Re}+iA_{Im}$, $B=B_{Re}+iB_{Im}$, $F=F_Q+iF_{Im}$, where
\begin{eqnarray}
\aligned
&(A_{Re})_{kj}=\frac{1}{2}Tr(\rho_x\{X_k,X_j\}),\\ &(B_{Re})_{kj}=\frac{1}{2}Tr(\rho_x\{X_k,L_j\}),\\ &(F_Q)_{kj}=\frac{1}{2}Tr(\rho_x\{L_k,L_j\}),
\endaligned
\end{eqnarray}
here $\{X,Y\}=XY+YX$ is the anti-commutator, and
\begin{eqnarray}
\aligned
& (A_{Im})_{kj}=\frac{1}{2i}Tr(\rho_x[X_k,X_j]),\\ &(B_{Im})_{kj}=\frac{1}{2i}Tr(\rho_x[X_k,L_j]),\\  &(F_{Im})_{kj}=\frac{1}{2i}Tr(\rho_x[L_k,L_j]),
\endaligned
\end{eqnarray}
here $[X,Y]=XY-YX$ is the commutator. It is easy to see that $F_Q$ is exactly the quantum Fisher information matrix, and the local unbiased condition in
Eq.(\ref{eq:localmain}) can be equivalently written as
\begin{eqnarray}
\label{eq:lu}
\aligned
Tr(\rho_x\frac{1}{2}\{L_j,X_k\})=\delta^j_k,
\endaligned
\end{eqnarray}
which means $B_{Re}=I$. $A$ is the same as $Z(X)$ in the Holevo bound, however we use a different notation here as in the case of mixed states it can be different from $Z(X)$.

As $Cov(\hat{x})\geq A$\cite{Hole82book,Haya05book,HayaM05}, we have
\begin{equation}
\left(\begin{array}{cc}
    Cov(\hat{x}) & B  \\
    B^\dagger & F \\
        \end{array}\right)=\left(\begin{array}{cc}
    Cov(\hat{x})-A & 0  \\
    0 & 0 \\
        \end{array}\right)+\left(\begin{array}{cc}
    A & B  \\
    B^\dagger & F \\
        \end{array}\right)\geq 0.
\end{equation}
%In this article, we assume $F_Q$ is nonsingular. This implies that $Cov(\hat{x})\geq F_Q^{-1}$ is also nonsingular, then using
Using the Schur's complement\cite{Boyd} we have
\begin{equation}
    F-B^\dagger Cov^{-1}(\hat{x})B\geq 0,%=Cov(\hat{x})^{-1}+B_{Im}^TCov(\hat{x})^{-1}B_{Im}+i(Cov(\hat{x})^{-1}B_{Im}-B_{Im}^TCov(\hat{x})^{-1})
\end{equation}
%As $F=F_Q+iF_{Im}$, $B=I+iB_{Im}$, this can be written as
this can be equivalently written as \begin{eqnarray}\label{eq:maininequality}
\aligned
&F_Q-Cov^{-1}(\hat{x})-B_{Im}^TCov^{-1}(\hat{x})B_{Im}\\
&+i[F_{Im}+B_{Im}^TCov^{-1}(\hat{x})-Cov^{-1}(\hat{x})B_{Im}]\geq 0.
\endaligned
\end{eqnarray}
Since for a positive semidefinite matrix, $M\geq 0$, the real part is also positive semidefinite, i.e., $M_{Re}=\frac{M+M^T}{2}\geq 0$. We thus have
$F_Q-Cov^{-1}(\hat{x})-B_{Im}^TCov^{-1}(\hat{x})B_{Im}\geq 0$, which can be equivalently written as
\begin{equation}\label{eq:QCRBImproved}
F_Q-Cov^{-1}(\hat{x})\geq B_{Im}^TCov^{-1}(\hat{x})B_{Im}.
\end{equation}
Note that $B_{Im}^TCov^{-1}(\hat{x})B_{Im}\geq 0$, thus the real part of Eq.(\ref{eq:maininequality}) already gives a tighter bound than the QCRB.

By multiplying $F_Q^{-\frac{1}{2}}$ from both the left and the right of Eq.(\ref{eq:maininequality}), we get
\begin{eqnarray}\label{eq:maintradeoff}
\aligned
&I-\tilde{Cov}^{-1}(\hat{x})-\tilde{B}_{Im}^T\tilde{Cov}^{-1}(\hat{x})\tilde{B}_{Im}\\
&+i[\tilde{F}_{Im}+\tilde{B}_{Im}^T\tilde{Cov}^{-1}(\hat{x})-\tilde{Cov}^{-1}(\hat{x})\tilde{B}_{Im}]\geq 0,
\endaligned
\end{eqnarray}
here $\tilde{Cov}^{-1}(\hat{x})=F_Q^{-\frac{1}{2}}Cov^{-1}(\hat{x})F_Q^{-\frac{1}{2}}$, $\tilde{B}_{Im}=F_Q^{\frac{1}{2}}B_{Im}F_Q^{-\frac{1}{2}}$, $\tilde{F}_{Im}=F_Q^{-\frac{1}{2}}F_{Im}F_Q^{-\frac{1}{2}}$. This is equivalent to the reparametrization which changes the QFIM to the Identity, and $\tilde{Cov}(\hat{x})$ can be regarded as the covariance matrix under the reparametrization. %and its inverse can be associated with the classical Fisher information matrix. %$I-\tilde{Cov}(\hat{x})^{-1}$ then is the difference between QFIM and  .
Various tradeoff relations can be obtained from Eq.(\ref{eq:maintradeoff}). In the appendix, we show that Eq.(\ref{eq:maintradeoff}) implies
\begin{equation}
\label{eq:mainjk}
    1-[\tilde{Cov}^{-1}(\hat{x})]_{jj}+1-[\tilde{Cov}^{-1}(\hat{x})]_{kk}\geq \frac{1}{2}|(\tilde{F}_{Im})_{jk}|^2.
\end{equation}
This describes a tradeoff between $[\tilde{Cov}^{-1}(\hat{x})]_{jj}$ and $[\tilde{Cov}^{-1}(\hat{x})]_{kk}$ as they can not reach 1 simultaneously when $(\tilde{F}_{Im})_{jk}\neq 0$. 
%When the QCRB is saturable, which requires $(\tilde{F}_{Im})_{jk}=0$, $\tilde{Cov}^{-1}(\hat{x})$ can reach the Identity, i.e., $[\tilde{Cov}^{-1}(\hat{x})]_{jj}$ and $[\tilde{Cov}^{-1}(\hat{x})]_{kk}$ can reach 1 simultaneously.
%When $(\tilde{F}_{Im})_{jk}\neq 0$ this places a tradeoff between $[\tilde{Cov}^{-1}(\hat{x})]_{jj}$ and $[\tilde{Cov}^{-1}(\hat{x})]_{kk}$.

By summing Eq.(\ref{eq:mainjk}) over all different choices of $j,k$, we can obtain 
\begin{eqnarray}\label{eq:tradeoffpure}
\aligned
Tr[F_Q^{-1}Cov^{-1}(\hat{x})]&=Tr[\tilde{Cov}^{-1}(\hat{x})]\\
&\leq n-\frac{1}{4(n-1)}\|F_Q^{-\frac{1}{2}}F_{Im}F_Q^{-\frac{1}{2}}\|_F^2,%\\ &=n-\frac{1}{4(n-1)}Tr(F_Q^{-1}F_{Im}^TF_Q^{-1}F_{Im}),
\endaligned
\end{eqnarray}
here $\|\cdot{}\|_F=\sqrt{\sum_{j,k}|(\cdot)_{jk}|^2}$ is the Frobenius norm.
%    where $\|F_Q^{-\frac{1}{2}}F_{Im}F_Q^{-\frac{1}{2}}\|_F=\sqrt{\sum_{j,k}|(F_Q^{-\frac{1}{2}}F_{Im}F_Q^{-\frac{1}{2}})_{jk}|^2}$ is the Frobenius norm.

When there are $\nu$ copies of the state, we can replace $|\Psi(x)\rangle$ with $|\Psi(x)\rangle^{\otimes \nu}$ and repeat the procedure to get the tradeoff relation as
    \begin{eqnarray}
\aligned
&Tr[F_{Q\nu}^{-1}Cov^{-1}(\hat{x})]\\
&\leq n-\frac{1}{4(n-1)}\|F_{Q\nu}^{-\frac{1}{2}}F_{Im\nu}F_{Q\nu}^{-\frac{1}{2}}\|_F^2,%\\ %&=n-\frac{1}{4(n-1)}Tr(F_{Q\nu}^{-1}F_{Im\nu}^TF_{Q\nu}^{-1}F_{Im\nu}),
\endaligned
\end{eqnarray}
where $F_{\nu}=F_{Q\nu}+iF_{Im\nu}$ is the corresponding operator associate with $|\Psi(x)\rangle^{\otimes \nu}$.
It is easy to verify that $F_{Q\nu}=\nu F_Q$, which is the QFIM for $|\Psi(x)\rangle^{\otimes \nu}$, and $F_{Im\nu}=\nu F_{Im}$. Thus when there are $\nu$ copies of the state, the tradeoff relation is given by
    \begin{eqnarray}\label{eq:tradepffpurenu}
\aligned
&\frac{1}{\nu}Tr[F_Q^{-1}Cov^{-1}(\hat{x})]\\
&\leq n-\frac{1}{4(n-1)}\|F_{Q}^{-\frac{1}{2}}F_{Im}F_{Q}^{-\frac{1}{2}}\|_F^2,%\\ &=n-\frac{1}{4(n-1)}Tr(F_Q^{-1}F_{Im}^TF_Q^{-1}F_{Im}),
\endaligned
\end{eqnarray}
This tradeoff relation holds for arbitrary measurements on $\nu$ copies of the states.

The tradeoff relation for $\nu$ copies of the pure state can also be obtained in an alternative way. Note that for pure states the optimal measurement can be taken as the 1-local measurement\cite{Matsumoto_2002}, if we repeat the 1-local measurement $\nu$ times with $\nu$ copies of the state, $Cov(\hat{x})$ will be reduced by $\nu$ times. Eq.(\ref{eq:tradeoffpure}), which is the tradeoff relation for a single state, then directly becomes Eq.(\ref{eq:tradepffpurenu}) since $Cov(\hat{x})$ is reduced by $\nu$ times. The two ways to get Eq.(\ref{eq:tradepffpurenu}), however, have different meanings. The derivation that uses $|\Psi(x)\rangle^{\otimes \nu}$ allows arbitrary measurement on $|\Psi(x)\rangle^{\otimes \nu}$ while the derivation with the repetition of the 1-local measurement only uses 1-local measurement. The reason that they lead to the same tradeoff relation is that for pure states 1-local measurement is already optimal, allowing collective measurements does not improve the precision. The situation is different for mixed states as we will see in the next section.% in particular for mixed states the corresponding $\bar{F}_{Im}$ no longer increases linearly.

%the same tradeoff relation then holds even the measurement is restricted to be 1-local. This can also be seen from Eq.(\ref{eq:tradeoffpure}), which is the tradeoff relation for a single copy of $|\Psi(x)\rangle$, i.e., with 1-local measurement on $|\Psi(x)\rangle$. If we repeat the 1-local measurement $\nu$ times on $\nu$ copies of the state, $Cov(\hat{x})$ will be reduced by $\nu$ times, which leads to the same tradeoff relation as Eq.(\ref{eq:tradepffpurenu}).
%Intuitively, when 1-local measurement is already optimal, allowing collective measurements does not improve the precision. Since 1-local measurement is a subset of $p$-local measurement, the same tradeoff relations thus hold under any $p$-local measurements for pure states, i.e., for pure states we have $\Gamma_1=\Gamma_2=\cdots= \Gamma_{\infty}$. We will show that the situation is different for mixed states.
%{\color{purple} It is hard to understand context(Since 1-local measurement is a subset of $p$-local measurement, the same tradeoff relations thus hold under any $p$-local measurements for pure states, i.e., for pure states we have $\Gamma_1=\Gamma_2=\cdots= \Gamma_{\infty}$.). In my opinion, if Eq.(\ref{eq:tradepffpurenu}) holds for arbitrary measurements, it must hold for $p$-local measurement, why do we need to explain? Besides, we can not conclude that $\Gamma_1 = \Gamma_2 = \cdots = \Gamma_{\infty}$ from point that Eq.(\ref{eq:tradepffpurenu}) is independent of $p$, although it is a fact from previous paper.}

The bound in Eq.(\ref{eq:tradepffpurenu}) is obtained by summing the tradeoff relations between pairs of parameters in Eq.(\ref{eq:mainjk}), which ignores the correlations with the other parameters. The presence of other parameters, however, can affect the precisions. In the appendix we show that when $n\geq 3$, by including the correlations among the parameters, the bound can be improved as
    \begin{eqnarray}\label{eq:tradeoffn3}
\aligned
&\frac{1}{\nu}Tr[F_Q^{-1}Cov^{-1}(\hat{x})]%&\leq n-\frac{n-2}{(n-1)^2}\|\tilde{F}_{Im}\|_F^2\\
\\&\leq n-\frac{n-2}{(n-1)^2}\|F_{Q}^{-\frac{1}{2}}F_{Im}F_{Q}^{-\frac{1}{2}}\|_F^2.
%Tr(F_Q^{-1}F_{Im}^TF_Q^{-1}F_{Im}).
\endaligned
\end{eqnarray}
Since $\frac{n-2}{n-1}>\frac{1}{4}$ when $n\geq 3$, this is tighter than the bound in Eq.(\ref{eq:tradepffpurenu}). It is also tighter than summing the tightest bound for a pair of parameters in previous study\cite{Lu2021}.

We can obtain even tighter tradeoff relation for large $n$ as(see appendix for detailed derivation) 
%From Eq.(\ref{eq:maintradeoff}) we have
%\begin{eqnarray}
%\aligned
%&I-\tilde{Cov}^{-1}(\hat{x})-\tilde{B}_{Im}^T\tilde{Cov}^{-1}(\hat{x})\tilde{B}_{Im}\\
%&\geq -i[\tilde{F}_{Im}+\tilde{B}_{Im}^T\tilde{Cov}^{-1}(\hat{x})-\tilde{Cov}^{-1}(\hat{x})\tilde{B}_{Im}],
%\endaligned
%\end{eqnarray}
%which leads to
%\begin{eqnarray}
%\aligned
%&Tr[I-\tilde{Cov}^{-1}(\hat{x})-\tilde{B}_{Im}^T\tilde{Cov}^{-1}(\hat{x})\tilde{B}_{Im}]\\
%&\geq \|\tilde{F}_{Im}+\tilde{B}_{Im}^T\tilde{Cov}^{-1}(\hat{x})-\tilde{Cov}^{-1}(\hat{x})\tilde{B}_{Im}\|_1,
%\endaligned
%\end{eqnarray}
%thus
%\begin{equation}
%\aligned
%Tr[&\tilde{Cov}^{-1}(\hat{x})]\leq n-(Tr[\tilde{B}_{Im}^T\tilde{Cov}^{-1}(\hat{x})\tilde{B}_{Im}]\\
%&+\|\tilde{F}_{Im}+\tilde{B}_{Im}^T\tilde{Cov}^{-1}(\hat{x})-\tilde{Cov}^{-1}(\hat{x})\tilde{B}_{Im}\|_1).
%\endaligned
%\end{equation}
%This leads to(see appendix for derivation) %the summation of the last terms is at least $\|\tilde{F}_{Im}\|_F^2$, thus
%\begin{equation}
%    Tr[\tilde{Cov}^{-1}(\hat{x})]\leq n-\frac{1}{5} \|\tilde{F}_{Im}\|_F^2.
%\end{equation}
%\begin{eqnarray}
%Tr[\tilde{B}_{Im}^T\tilde{Cov}^{-1}(\hat{x})\tilde{B}_{Im}]+\|\tilde{F}_{Im}+\tilde{B}_{Im}^T\tilde{Cov}^{-1}(\hat{x})-\tilde{Cov}^{-1}(\hat{x})\tilde{B}_{Im}\|_1\geq \frac{1}{5} \|\tilde{F}_{Im}\|_F^2,
%\end{eqnarray}
%With $\nu$ copies of quantum states, this can be written as
\begin{eqnarray}\label{eq:tradeoffpuren5}
\aligned
\frac{1}{\nu}Tr[F_Q^{-1}Cov^{-1}(\hat{x})]%&\leq n-\frac{n-2}{(n-1)^2}\|\tilde{F}_{Im}\|_F^2\\
&\leq n-\frac{1}{5}\|F_{Q}^{-\frac{1}{2}}F_{Im}F_{Q}^{-\frac{1}{2}}\|_F^2,
%Tr(F_Q^{-1}F_{Im}^TF_Q^{-1}F_{Im}).
\endaligned
\end{eqnarray}
which is tighter than Eq.(\ref{eq:tradeoffn3}) when $n\geq 5$.

The three bounds in Eq.(\ref{eq:tradepffpurenu}), Eq.(\ref{eq:tradeoffn3}) and Eq.(\ref{eq:tradeoffpuren5}) can be written concisely as
\begin{eqnarray}
    \aligned
    \frac{1}{\nu}Tr[F_Q^{-1}Cov^{-1}(\hat{x})]\leq &n-f(n)\|F_{Q}^{-\frac{1}{2}}F_{Im}F_{Q}^{-\frac{1}{2}}\|_F^2, 
   % \frac{1}{\nu}Tr[F_Q^{-1}Cov^{-1}(\hat{x})]
   %\leq &n-\frac{1}{4(n-1)}\|F_{Q}^{-\frac{1}{2}}F_{Im}F_{Q}^{-\frac{1}{2}}\|_F^2,\\
    %        \frac{1}{\nu}Tr[F_Q^{-1}Cov^{-1}(\hat{x})]
%\leq & n-\frac{n-2}{(n-1)^2}\|F_{Q}^{-\frac{1}{2}}F_{Im}F_{Q}^{-\frac{1}{2}}\|_F^2,\\
 %  \frac{1}{\nu}Tr[F_Q^{-1}Cov^{-1}(\hat{x})]
  % \leq& n-\frac{1}{5}\|F_{Q}^{-\frac{1}{2}}F_{Im}F_{Q}^{-\frac{1}{2}}\|_F^2.
\endaligned
    \end{eqnarray}
where $f(n)=\frac{1}{4(n-1)}, \frac{n-2}{(n-1)^2}$ or $\frac{1}{5}$. These bounds are all valid for any $n$, since larger f(n) leads to tighter bound we can take $f(n)=\max\{\frac{1}{4(n-1)},\frac{n-2}{(n-1)^2},  \frac{1}{5}\}$ to get a tighter upper bound.

%However, for any $n$all the bounds are nontrivial . %which is $\frac{1}{\nu}Tr[F_Q^{-1}Cov(\hat{x})^{-1}]\leq d-1$ with $d$ as the dimension of the Hilbert space. The Gill-Massar bound is only nontrivial when $n\geq d$ while the bound derived here is nontrivial for all $n$.
%\subsection{Optimal observables for pure states}\label{sec:pure}
%For mixed states, however, this is not the case, as we will show in the next section.

\section{Precision bounds for mixed states}\label{sec:mixstate}
%We now consider the mixed states. %where the state, $\rho_x$, is mixed.
For pure states, the ultimate precision under the local measurement can be quantified by the Holevo bound since  for pure states the Holevo bound can be saturated with the 1-local measurement. For mixed states, however, the Holevo bound is in general not saturable under the local measurement. We will first provide a tighter bound for the mixed states under local measurement, then use it to obtain the upper bounds for the incompatibility measures. %This bound includes the Holevo bound

%For mixed states, the 1-local measurement is in general not optimal and the precision can in general be improved with $p$-local measurement when $p$ increases. We now derive bounds with general $p$-local measurement for mixed states.

\subsection{Multi-parameter precision bound for mixed states}\label{subsec:multibounds}
For a mixed state, $\rho_x$, with $x=(x_1,\cdots, x_n)$, given any POVM, $\{M_\alpha\}$, and any $|u\rangle$, we define $Cov_u$ as a $n\times n$ matrix with the $jk$-th entry given by
\begin{equation}
    (Cov_u)_{jk}=\sum_{\alpha}(\hat{x}_j(\alpha)-x_j)(\hat{x}_k(\alpha)-x_k)\langle u|\sqrt{\rho_x}M_{\alpha} \sqrt{\rho_x}|u\rangle,
\end{equation}
and $A_u$ as a $n\times n$ matrix with the $jk$-th entry given by
\begin{eqnarray}\label{eq:Aumain}
\aligned
&(A_u)_{jk}=\langle u|\sqrt{\rho_x}X_j^\dagger X_k\sqrt{\rho_x}|u\rangle\\
&=\frac{1}{2}\langle u|\sqrt{\rho_x}\{X_j,X_k\}\sqrt{\rho_x}|u\rangle+i\frac{1}{2i}\langle u|\sqrt{\rho_x}[X_j,X_k]\sqrt{\rho_x}|u\rangle,
\endaligned
\end{eqnarray}
here $X_j=\sum_{\alpha}[\hat{x}_j(\alpha)-x_j]M_{\alpha}$ is locally unbiased.

We first note that for any set of $\{|u_q\rangle\}$ that satisfies $\sum_q |u_q\rangle\langle u_q|=I$, we have $Cov(\hat{x})=\sum_q Cov_{u_q}$. %in particular this holds when $\{|u_q\rangle\}$ form a complete basis. %, $\{|u_1\rangle, \cdots, |u_d\rangle\}$,
This can be verified by comparing $\sum_q (Cov_{u_q})_{jk}$ and $Cov(\hat{x})_{jk}$ as
\begin{eqnarray}\label{eq:Covu}
\aligned
&\sum_q (Cov_{u_q})_{jk}\\
&=\sum_q\sum_{\alpha}(\hat{x}_j(\alpha)-x_j)(\hat{x}_k(\alpha)-x_k)\langle u_q|\sqrt{\rho_x}M_{\alpha} \sqrt{\rho_x}|u_q\rangle\\
&=\sum_{\alpha}(\hat{x}_j(\alpha)-x_j)(\hat{x}_k(\alpha)-x_k)Tr(\rho_xM_{\alpha})\\
&=Cov(\hat{x})_{jk}.
\endaligned
\end{eqnarray}

And for any $|u\rangle$, we have $Cov_u\geq A_u$(see appendix \ref{sec:cov>A}). %since
%for any vector $b=(b_1, \cdots, b_n)^T$,
%\begin{widetext}
%\begin{eqnarray}
%\aligned
%&b^\dagger Cov_u b-b^\dagger A_u b\\
%=&\langle u|\sum_{j,k} b_j^*b_k\{\sum_{\alpha}(\hat{x}_j(\alpha)-x_j)(\hat{x}_k(\alpha)-x_k)\sqrt{\rho_x}M_{\alpha} \sqrt{\rho_x}- \sum_{\beta}(\hat{x}_j(\beta)-x_j)\sqrt{\rho_x}M_{\beta}\sum_{\gamma}(\hat{x}_k(\gamma)-x_k)M_{\gamma}]\sqrt{\rho_x}\}|u\rangle \\
%=&\langle u|\sum_{j,k} \{\sum_{\alpha}(\hat{x}_j(\alpha)-x_j)b_j^*(\hat{x}_k(\alpha)-x_k)b_k\sqrt{\rho_x}M_{\alpha} \sqrt{\rho_x}\\&- \sum_{\beta}(\hat{x}_j(\beta)-x_j)b_j^*\sqrt{\rho_x}M_{\beta}(\sum_{\alpha}M_{\alpha})\sum_{\gamma}(\hat{x}_k(\gamma)-x_k)b_kM_{\gamma}]\sqrt{\rho_x}\}|u\rangle \\
%=&\langle u|\sum_{\alpha}\{[\sum_j(\hat{x}_j(\alpha)-x_j)b_j^*\sqrt{\rho_x}-\sum_j\sum_{\beta}(\hat{x}_j(\beta)-x_j)b_j^*\sqrt{\rho_x}M_{\beta}]M_{\alpha}[\sum_k (x_k(\alpha)-x_k)b_k\sqrt{\rho_x}\\& -\sum_k \sum_{\gamma}(\hat{x}_k(\gamma)-x_k)b_kM_{\gamma}\sqrt{\rho_x}]\}|u\rangle\\
%=&\langle u|\sum_{\alpha} M^\dagger(b) M_{\alpha} M(b)]|u\rangle \geq 0,
%\endaligned
%\end{eqnarray}
%\end{widetext}
%here $M(b)=\sum_k (x_k(\alpha)-x_k)b_k\sqrt{\rho_x}-\sum_k \sum_{\gamma}(\hat{x}_k(\gamma)-x_k)b_kM_{\gamma}\sqrt{\rho_x}$. %This then implies $Cov_u\geq A_u$.
Since $Cov_{u}$ is symmetric, we also have $Cov_{u}=Cov_{u}^T\geq A_{u}^T$. %i.e., %$Cov_u\geq \mathbf{\bar{A}}_{u_q}$. Since $Cov_{u_q}\geq A_{u_q}$ and $Cov_{u_q}\geq A^T_{u_q}$, we have
%\begin{equation}
%    Cov_{u_q}\geq \mathbf{\bar{A}}_{u_q},
%\end{equation}
%here $\mathbf{\bar{A}}_{u_q}$ equal to either $A_{u_q}$ or $A^T_{u_q}$.
Thus for any set of $\{|u_q\rangle\}$ that satisfies $\sum_q|u_q\rangle\langle u_q|=I$ and any choices of $\mathbf{\bar{A}}_{u_q}\in \{A_{u_q}, A^T_{u_q}\}$, we have
\begin{equation}
Cov(\hat{x})=\sum_q Cov_{u_q} \geq \mathbf{\bar{A}}=\sum_q\mathbf{\bar{A}}_{u_q},
\end{equation}
where $\mathbf{\bar{A}}_{u_q}$ equal to either $A_{u_q}$ or $A^T_{u_q}$.
We can write $\mathbf{\bar{A}}$ in terms of the real and imaginary part as  $\bold{\bar{A}}=\bold{\bar{A}}_{Re}+i\bold{\bar{A}}_{Im}$, then
\begin{equation}\label{eq:bound}
    \nu Tr[WCov(\hat{x})]\geq \min_{\{X_j\}}Tr[W\bold{\bar{A}}_{Re}]+\|\sqrt{W}\bold{\bar{A}}_{Im}\sqrt{W}\|_1.
    \end{equation}
where $W\geq 0$ is the weight matrix and the number of repetition, $\nu$, has been included.
%    which can be equivalent formulated as
%\begin{equation}\label{eq:bound2}
%    Tr[Cov(\hat{x})]\geq \min_{\{X_j, G\}}\{Tr[G]|G\geq \bold{\bar{A}}, G \text{is real symmetric}\}.
%    \end{equation}
%Here we take the weight matrix as the Identity, any non-Identity positive-definite weight matrix can be reduced to the Identity by reparametrization.

This includes the Holevo bound\cite{Hole82book} and the Nagaoka bound\cite{Nagaoka1,Nagaoka2} as special cases. To see the connection with the Holevo bound, we just choose $\mathbf{\bar{A}}_{u_q}=A_{u_q}$ for all $q$, then for any set of $\{|u_q\rangle\}$ that satisfies $\sum_q |u_q\rangle\langle u_q|=I$, we have $\mathbf{\bar{A}}=\sum_q A_{u_q}=Z(X)$ since (note that $X_j$ is Hermitian)
\begin{eqnarray}
\aligned
\mathbf{\bar{A}}_{jk}&=\sum_q (A_{u_q})_{jk}\\
&=\sum_q \langle u_q|\sqrt{\rho_x}X_j^\dagger X_k\sqrt{\rho_x}|u_q\rangle\\
&=Tr(\rho_xX_j^\dagger X_k)\\
&=Z(X)_{jk}.
\endaligned
\end{eqnarray}
    %Z(X)_{jk}=Tr(\rho_xX_jX_k).
%\end{equation}
Eq.(\ref{eq:bound}) then reduces to the Holevo bound. When there are only two parameters, $x_1$ and $x_2$, we can choose the set of $\{|u_q\rangle\}$ as the eigenvectors of $\sqrt{\rho_x}[X_1,X_2]\sqrt{\rho_x}$ and choose $\bold{\bar{A}}_{u_q}$ as
\begin{eqnarray}\label{eq:Suq}
\bold{\bar{A}}_{u_q}:=\{\begin{array}{cc}
A_{u_q}, & \text{for } \frac{1}{2i}\langle u_q|\sqrt{\rho_x}[X_1,X_2]\sqrt{\rho_x}|u_q\rangle\geq 0,\\
A^T_{u_q}, & \text{for } \frac{1}{2i}\langle u_q|\sqrt{\rho_x}[X_1,X_2]\sqrt{\rho_x}|u_q\rangle< 0.
\end{array}
\end{eqnarray}
Intuitively $A_{u_q}$ can be written as the real and imaginary part as $A_{u_q}=A_{u_qRe}+iA_{u_qIm}$, where $A_{u_qIm}$ is a $2\times 2$ skew symmetric matrix $\left(\begin{array}{cc} 0 & a_q\\ -a_q & 0 \end{array}\right)$ with $a_q=\frac{1}{2i}\langle u_q|\sqrt{\rho_x}[X_1,X_2]\sqrt{\rho_x}|u_q\rangle$. $\bold{\bar{A}}_{u_q}$ is then chosen according to the sign of $a_q$, $\bold{\bar{A}}_{u_q}=A_{u_q}$ when $a_q\geq 0$ and $\bold{\bar{A}}_{u_q}=A_{u_q}^T$ when $a_q\le 0$. The imaginary parts of different $\bold{\bar{A}}_{u_q}$ are then aligned and add up to $\frac{1}{2}\|\sqrt{\rho_x}[X_1,X_2]\sqrt{\rho_x}\|_1$. With this choice we then have
\begin{eqnarray}
\aligned
\bold{\bar{A}}&=\sum_q \bold{\bar{A}}_{u_q}\\
&=\begin{pmatrix}
      Tr(\rho_xX_1^2) & \frac{1}{2}Tr[\rho_x\{X_1,X_2\}]\\
      \frac{1}{2}Tr[\rho_x\{X_1,X_2\}] & Tr(\rho_xX_2^2)
    \end{pmatrix}\\
    &+i\begin{pmatrix}
      0 & \frac{1}{2}\|\sqrt{\rho_x}[X_1,X_2]\sqrt{\rho_x}\|_1\\
      -\frac{1}{2}\|\sqrt{\rho_x}[X_1,X_2]\sqrt{\rho_x}\|_1 & 0
    \end{pmatrix}.
\endaligned
\end{eqnarray}
Eq.(\ref{eq:bound}) then becomes(with $W=I$)
\begin{eqnarray}
\aligned
     &\nu Tr[Cov(\hat{x})]\\
     &\geq \min_{\{X_1,X_2\}}Tr[\bold{\bar{A}}_{Re}]+\|\bold{\bar{A}}_{Im}\|_1\\
          &=\min_{\{X_1,X_2\}}Tr(\rho_xX_1^2)+Tr(\rho_xX_2^2)+\|\sqrt{\rho_x}[X_1,X_2]\sqrt{\rho_x}\|_1,
     \endaligned
\end{eqnarray}
which recovers the Nagaoka bound\cite{Nagaoka1,Nagaoka2}. The general bound thus establishes a connection between the Holevo bound and the Nagaoka bound and improves our understanding on these existing bounds. 

The optimal choice of $|u_q\rangle$ and $\mathbf{\bar{A}}_{u_q}$ provides the tightest bound, but any choice lead to a valid bound. %although the optimal choice can be hard to identify,
We now show how non-trivial analytical upper bounds on $\Gamma_p$ can be obtained by making particular choices of $|u_q\rangle$ and $\mathbf{\bar{A}}_{u_q}$. %We start with the 1-local measurement.

%where the real part of $\mathbf{\bar{A}}$ does not change with the choices with its $jk$-th entry always given by $(\mathbf{\bar{A}}_{Re})_{jk}=\frac{1}{2}Tr(\rho_x \{X_j,X_k\})$, the imaginary part depends on the choices that can be optimized to tighten the bound.

\subsection{Incompatibility under 1-local measurements}
Given a mixed state, $\rho_x$, we can make a reparametrization with $\tilde{x}=F_Q^{\frac{1}{2}}x$ under which  $\tilde{F}_Q=I$, and $\tilde{Cov}(\hat{x})=F_Q^{\frac{1}{2}}Cov(\hat{x})F_Q^{\frac{1}{2}}$.
%the SLDs are given by $\tilde{L}_j=\sum_k (F_Q^{-\frac{1}{2}})_{jq}L_q$.  %and $\tilde{F}_{Im}=(F_Q^{-\frac{1}{2}})F_{Im}(F_Q^{-\frac{1}{2}})$.
Thus without loss of generality, we start with the case that the QFIM equals to the Identity.

We first consider the precision under 1-local measurements, i.e., separable measurements. Note that for any vector $|u\rangle$, we have
\begin{widetext}
\begin{eqnarray}
\aligned
S_u&=\left(\begin{array}{cccccc}
X_1\sqrt{\rho_x}|u\rangle & \cdots & X_n\sqrt{\rho_x}|u\rangle & L_1\sqrt{\rho_x}|u\rangle & \cdots & L_n\sqrt{\rho_x}|u\rangle \end{array}\right)^\dagger \left(\begin{array}{cccccc}X_1\sqrt{\rho_x}|u\rangle & \cdots & X_n\sqrt{\rho_x}|u\rangle & L_1\sqrt{\rho_x}|u\rangle & \cdots & L_n\sqrt{\rho_x}|u\rangle \end{array}\right)\\
&=\left(\begin{array}{cc}
A_u & B_u\\
B_u^\dagger & F_u
\end{array}\right)\geq 0,
%\left(\begin{array}{ccc}
%\langle u|\sqrt{\rho}Y_1^\dagger Y_1\sqrt{\rho}|u\rangle & \cdots & \langle u|\sqrt{\rho}Y_1^\dagger Y_n\sqrt{\rho}|u\rangle\\
%\vdots & \vdots & \vdots\\
%\langle u|\sqrt{\rho}Y_1^\dagger Y_1\sqrt{\rho}|u\rangle & \cdots & \sqrt{\rho}Y_j^\dagger Y_n\sqrt{\rho}\\
%\vdots & \vdots & \vdots\\
%\sqrt{\rho}Y_n^\dagger Y_1\sqrt{\rho} & \cdots & \sqrt{\rho}Y_n^\dagger Y_n\sqrt{\rho}\end{array}\right)
\endaligned
\end{eqnarray}
\end{widetext}
here %the entries of $A_u$ are given by Eq.(\ref{eq:Aumain}), 
$A_u$, $B_u$ and $F_u$ are $n\times n$ matrices with the entries given by
%$(A_u)_{jk}=\langle u|\sqrt{\rho_x}X_j^\dagger X_k\sqrt{\rho_x}|u\rangle$, $(B_u)_{jk}=\langle u|\sqrt{\rho_x}X_j^\dagger L_k\sqrt{\rho_x}|u\rangle$, $(F_u)_{jk}=\langle u|\sqrt{\rho_x}L_j^\dagger L_k\sqrt{\rho_x}|\u\rangle$. Note that $X_j$, $L_k$, $j, k\in\{1,\cdots, n\}$, are all Hermitian, we can write the entries
%\begin{eqnarray}
%\aligned
%&(A_u)_{jk}=\langle u|\sqrt{\rho_x}X_j^\dagger X_k\sqrt{\rho_x}|u\rangle\\
%&=\frac{1}{2}\langle u|\sqrt{\rho_x}\{X_j,X_k\}\sqrt{\rho_x}|u\rangle+i\frac{1}{2i}\langle u|\sqrt{\rho_x}[X_j,X_k]\sqrt{\rho_x}|u\rangle
%\endaligned
%\end{eqnarray}
\begin{eqnarray}
\aligned
&(A_u)_{jk}=\langle u|\sqrt{\rho_x}X_j^\dagger X_k\sqrt{\rho_x}|u\rangle\\
&=\frac{1}{2}\langle u|\sqrt{\rho_x}\{X_j,X_k\}\sqrt{\rho_x}|u\rangle+i\frac{1}{2i}\langle u|\sqrt{\rho_x}[X_j,X_k]\sqrt{\rho_x}|u\rangle,\\
&(B_u)_{jk}=\langle u|\sqrt{\rho_x}X_j^\dagger L_k\sqrt{\rho_x}|u\rangle\\&=\frac{1}{2}\langle u|\sqrt{\rho_x}\{X_j,L_k\}\sqrt{\rho_x}|u\rangle+i\frac{1}{2i}\langle u|\sqrt{\rho_x}[X_j,L_k]\sqrt{\rho_x}|u\rangle\\
%\endaligned
%\end{eqnarray}
%\begin{eqnarray}
%\aligned
&(F_u)_{jk}=\langle u|\sqrt{\rho_x}L_j^\dagger L_k\sqrt{\rho_x}|u\rangle\\&=\frac{1}{2}\langle u|\sqrt{\rho_x}\{L_j,L_k\}\sqrt{\rho_x}|u\rangle+i\frac{1}{2i}\langle u|\sqrt{\rho_x}[L_j,L_k]\sqrt{\rho_x}|u\rangle.\\
\endaligned
\end{eqnarray}
For a set of $\{|u_q\rangle\}$ with $\sum_q |u_q\rangle\langle u_q|=I$, we obtain a corresponding set of $\{S_{u_1},\cdots,S_{u_d}\}$.  We then let $\mathbf{\bar{S}}=\sum_q \bold{\bar{S}}_{u_q}$ where $\bold{\bar{S}}_{u_q}\in \{S_{u_q}, S_{u_q}^T\}$. Since $S_{u_q}\geq 0$ and $S_{u_q}^T\geq 0$, it is then easy to see that 
\begin{eqnarray}
\mathbf{\bar{S}}=\sum_q \bold{\bar{S}}_{u_q}=\left(\begin{array}{cc}
\bold{\bar{A}} & \bold{\bar{B}}\\
\bold{\bar{B}}^\dagger & \bold{\bar{F}}
\end{array}\right)\geq 0,
\end{eqnarray}
here $\bold{\bar{F}}=\sum_q\bold{\bar{F}}_{u_q}$ with $\bold{\bar{F}}_{u_q}$ equal to either $F_{u_q}$ or $F^T_{u_q}$, $\bold{\bar{A}}=\sum_q \bold{\bar{A}}_{u_q}$ with $\bold{\bar{A}}_{u_q}$ equals to either $A_{u_q}$ or $A^T_{u_q}$, and $\bold{\bar{B}}=\sum_q \bold{\bar{B}}_{u_q}$. Since $\mathbf{\bar{S}}_{u_q}$ has the same real part as $S_{u_q}$, the real part of $\mathbf{\bar{S}}$ is independent of the choices of $\mathbf{\bar{S}}_{u_q}$. In particular the real part of $\mathbf{\bar{F}}$ always equals to the QFIM as
\begin{eqnarray}
\aligned
(\bold{\bar{F}}_{Re})_{jk}&=\sum_q \frac{1}{2}\langle u_q|\sqrt{\rho_x}\{L_j,L_k\}\sqrt{\rho_x}|u_q\rangle\\
&=Tr[\rho_x\frac{1}{2}\{L_j,L_k\}]\\
&=(F_Q)_{jk}.
\endaligned
\end{eqnarray}
Similarly it is straightforward to see that the real part of $\mathbf{\bar{B}}$ also remains the same as
\begin{eqnarray}
\aligned
(\mathbf{\bar{B}}_{Re})_{jk}&=\sum_q \frac{1}{2}\langle u_q|\sqrt{\rho_x}\{X_j,L_k\}\sqrt{\rho_x}|u_q\rangle\\
&=Tr[\rho_x\frac{1}{2}\{X_j,L_k\}]\\
&=\delta_k^j,
\endaligned
\end{eqnarray}
where the last equality is the locally unbiased condition. We can thus write $\mathbf{\bar{B}}=I+i\mathbf{\bar{B}}_{Im}$.
%We also have $\bar{A}=\sum_q\bar{A}_{u_q}$ with $\bar{A}_{u_q}$ equals to either $A_{u_q}$ or $A^T_{u_q}$ and
%Since $Cov_{u_q}\geq A_{u_q}$ and $Cov_{u_q}\geq A^T_{u_q}$, we have $Cov(\hat{x})=\sum_q Cov_{u_q} \geq \bold{\bar{A}}=\sum_q\bold{\bar{A}}_{u_q}$ with $\bold{\bar{A}}_{u_q}$ equals to either $A_{u_q}$ or $A^T_{u_q}$.  %$\bar{F}=F_Q+i\bar{F}_{Im}$ with %(note that each time we take a fixed pair of $j,k$, other entries of $\bar{F}_{Im}$ may not achieve this value simultaneously).
Since $Cov(\hat{x})\geq \mathbf{\bar{A}}$, we have \begin{eqnarray}
\left(\begin{array}{cc}
Cov(\hat{x}) & \mathbf{\bar{B}}\\
\mathbf{\bar{B}}^\dagger & \mathbf{\bar{F}}
\end{array}\right)\geq 0.
\end{eqnarray}
Then by following the same derivation as in the previous section we have
\begin{eqnarray}\label{eq:mainboundF}
\aligned
&\frac{1}{\nu}Tr[F_Q^{-1}Cov^{-1}(\hat{x})]\\&\leq n-f(n)\|F_Q^{-\frac{1}{2}}\bold{\bar{F}}_{Im}F_Q^{-\frac{1}{2}}\|_F^2, %&=n-\frac{n-2}{(n-1)^2}Tr(F_Q^{-1}F_{Im}^TF_Q^{-1}F_{Im}).
\endaligned
\end{eqnarray}
where $
f(n)=\max\{\frac{1}{4(n-1)},\frac{n-2}{(n-1)^2},\frac{1}{5}\}$, $\bold{\bar{F}}_{Im}$ is the imaginary part of $\bold{\bar{F}}=\sum_q\bold{\bar{F}}_{u_q}$ with each $\bold{\bar{F}}_{u_q}$ equals to either $F_{u_q}$ or $F^T_{u_q}$ which can be optimized to get the maximal $\|F_Q^{-\frac{1}{2}}\bold{\bar{F}}_{Im}F_Q^{-\frac{1}{2}}\|_F$.

We can also obtain additional bounds by combining different choices of $\{|u_q\rangle\}$. In particular, we can choose different set of $\{|u_q\rangle\}$ according to different pair of indexes, say $\alpha\neq \beta \in \{1,2,\cdots, n\}$. Specifically, for a given pair of index, $\alpha$ and $\beta$, we choose a set of  $\{|u_1\rangle, \cdots, |u_d\rangle\}$ as the orthonormal eigenvectors of  $\sqrt{\rho_x}[L_{\alpha},L_{\beta}]\sqrt{\rho_x}$. Note that $\sqrt{\rho_x}[L_{\alpha},L_{\beta}]\sqrt{\rho_x}$ is skew Hermitian whose eigenvalues are pure imaginary, thus for any eigenvector $|u_q\rangle$, $\langle u_q|\sqrt{\rho_x}[L_{\alpha},L_{\beta}]\sqrt{\rho_x}|u_q\rangle=ia_q$ with $a_q$ a real number. The imaginary axis of $(F_{u_q})_{\alpha\beta}=\langle u_q|\sqrt{\rho_x}L_{\alpha}L_{\beta}\sqrt{\rho_x}|u_q\rangle$ is then $\frac{1}{2i}\langle u_q|\sqrt{\rho_x}[L_{\alpha},L_{\beta}]\sqrt{\rho_x}|u_q\rangle=\frac{1}{2}a_q$. We then let
%\begin{eqnarray}
%\mathbf{\bar{S}}=\sum_q \bold{\bar{S}}_{u_q}=\left(\begin{array}{cc}
%\bold{\bar{A}} & \bold{\bar{B}}\\
%\bold{\bar{B}}^\dagger & \bold{\bar{F}}
%\end{array}\right)\geq 0,
%\end{eqnarray}
%here $\bold{\bar{F}}=\sum_q\bold{\bar{F}}_{u_q}$ with $\bold{\bar{F}}_{u_q}$ equal to either $F_{u_q}$ or $F^T_{u_q}$ where the imaginary part of $(\bold{\bar{F}}_{u_q})_{\alpha\beta}$ is always positive due to the choices in Eq.(\ref{eq:Suq}), $\bold{\bar{A}}=\sum_q \bold{\bar{A}}_{u_q}$ with $\bold{\bar{A}}_{u_q}$ equals to either $A_{u_q}$ or $A^T_{u_q}$, and $\bold{\bar{B}}=\sum_q \bold{\bar{B}}_{u_q}$
%$\bar{S}_{u_q}:=S_{u_q}$ when $\frac{1}{2i}\langle u_q|\sqrt{\rho_x}[L_j,L_k]\sqrt{\rho_x}|u_q\rangle\geq 0$, $\bar{S}_{u_q}:=S_{u_q}^T$ when $\frac{1}{2i}\langle u_q|\sqrt{\rho_x}[L_j,L_k]\sqrt{\rho_x}|u_q\rangle<0$. %$\bar{S}_{u_j}$ thus has the same real part as $S_{u_j}$ and
\begin{eqnarray}\label{eq:Suq}
\mathbf{\bar{S}}_{u_q}:=\{\begin{array}{cc}
S_{u_q}, & \text{for } a_q\geq 0,\\ %\frac{1}{2i}\langle u_q|\sqrt{\rho_x}[L_{\alpha},L_{\beta}]\sqrt{\rho_x}|u_q\rangle\geq 0,\\
S^T_{u_q}, & \text{for } a_q<0,%\frac{1}{2i}\langle u_q|\sqrt{\rho_x}[L_{\alpha},L_{\beta}]\sqrt{\rho_x}|u_q\rangle< 0,
\end{array}
\end{eqnarray}
and sum $\bold{\bar{S}}_{u_q}$ to get %$\bar{S}=\sum_j \bar{S}_{u_j}$.
\begin{eqnarray}
\mathbf{\bar{S}}=\sum_q \bold{\bar{S}}_{u_q}=\left(\begin{array}{cc}
\bold{\bar{A}} & \bold{\bar{B}}\\
\bold{\bar{B}}^\dagger & \bold{\bar{F}}
\end{array}\right)\geq 0,
\end{eqnarray}
here $\bold{\bar{F}}=\sum_q\bold{\bar{F}}_{u_q}$ with $\bold{\bar{F}}_{u_q}$ equal to either $F_{u_q}$ or $F^T_{u_q}$ which are determined by the choices in Eq.(\ref{eq:Suq}) so that the imaginary parts of all $(\bold{\bar{F}}_{u_q})_{\alpha\beta}$ are all positive, $\bold{\bar{A}}=\sum_q \bold{\bar{A}}_{u_q}$ with $\bold{\bar{A}}_{u_q}$ equals to either $A_{u_q}$ or $A^T_{u_q}$, and $\bold{\bar{B}}=\sum_q \bold{\bar{B}}_{u_q}$. %with $\bar{B}_{u_q}$ equals to either $B_{u_q}$ or $A^T_{u_q}$. %which equals to one-half of the absolute value of the eigenvalue that corresponds to the eigenvector $|u_q\rangle$ of $\sqrt{\rho_x}[L_j,L_k]\sqrt{\rho_x}$.
It is easy to verify that according to the choices in Eq.(\ref{eq:Suq}), which aligns the imaginary part of the $\alpha\beta$-th entry of each $\mathbf{\bar{F}}_{u_q}$ with the same sign, we have
\begin{equation}\label{eq:mainFjk}
(\bold{\bar{F}}_{Im})_{\alpha\beta}=\sum_q\frac{1}{2}|a_q|= \frac{1}{2}\|\sqrt{\rho_x}[L_{\alpha},L_{\beta}]\sqrt{\rho_x}\|_1,
\end{equation}
where $\|\text{ }\|_1$ is the trace norm which equals to the sum of singular values and for the skew-Hermitian matrix just equals to the sum of the absolute value of the eigenvalues. 
%Since $\mathbf{\bar{S}}_{u_q}$ has the same real part as $S_{u_q}$, the real part of $\mathbf{\bar{F}}$ still equals to the QFIM as
%\begin{eqnarray}
%\aligned
%(\bold{\bar{F}}_Q)_{jk}&=\sum_q \frac{1}{2}\langle u_q|\sqrt{\rho_x}\{L_j,L_k\}\sqrt{\rho_x}|u_q\rangle\\
%&=Tr[\rho_x\frac{1}{2}\{L_j,L_k\}]\\
%&=(F_Q)_{jk}.
%\endaligned
%\end{eqnarray}
%As the real part of $\mathbf{\bar{B}}$ remains unchanged, we can write $\mathbf{\bar{B}}=I+i\mathbf{\bar{B}}_{Im}$.
%We also have $\bar{A}=\sum_q\bar{A}_{u_q}$ with $\bar{A}_{u_q}$ equals to either $A_{u_q}$ or $A^T_{u_q}$ and
%Since $Cov_{u_q}\geq A_{u_q}$ and $Cov_{u_q}\geq A^T_{u_q}$, we have $Cov(\hat{x})=\sum_q Cov_{u_q} \geq \bold{\bar{A}}=\sum_q\bold{\bar{A}}_{u_q}$ with $\bold{\bar{A}}_{u_q}$ equals to either $A_{u_q}$ or $A^T_{u_q}$.  %$\bar{F}=F_Q+i\bar{F}_{Im}$ with %(note that each time we take a fixed pair of $j,k$, other entries of $\bar{F}_{Im}$ may not achieve this value simultaneously).
Again since $Cov(\hat{x})\geq \mathbf{\bar{A}}$, we have \begin{eqnarray}
\left(\begin{array}{cc}
Cov(\hat{x}) & \mathbf{\bar{B}}\\
\mathbf{\bar{B}}^\dagger & \mathbf{\bar{F}}
\end{array}\right)\geq 0.
\end{eqnarray}
Then by following the same derivation as in the previous section, under the parametrization that $F_Q=I$, we can get the same tradeoff relation, similar as Eq.(\ref{eq:mainjk}). In particular for the entries associated with $\alpha$ and $\beta$, we have
\begin{equation}\label{eq:maintradeoffab}
%\label{eq:mainjk}
    \aligned
    &1-[Cov^{-1}(\hat{x})]_{\alpha\alpha}+1-[Cov^{-1}(\hat{x})]_{\beta\beta}
    \geq& \frac{1}{2}|(\bold{\bar{F}}_{Im})_{\alpha\beta}|^2
    %=\frac{1}{2}|\frac{1}{2}\|\sqrt{\rho_x}[L_{\alpha},L_{\beta}]\sqrt{\rho_x}\|_1|^2.
    \endaligned
\end{equation}
with $(\bold{\bar{F}}_{Im})_{\alpha\beta}=\frac{1}{2}\|\sqrt{\rho_x}[L_{\alpha},L_{\beta}]\sqrt{\rho_x}\|_1$. We note that here we make the choices of $\{|u_q\rangle\}$ and $\{\mathbf{\bar{S}}_{u_q}\}$ according to a particular pair of indexes $\alpha$ and $\beta$, thus only the imaginary part of $(\bold{\bar{F}}_{Im})_{\alpha\beta}$ equals to $\frac{1}{2}\|\sqrt{\rho_x}[L_{\alpha},L_{\beta}]\sqrt{\rho_x}\|_1$, for other indexes $(j,k)\neq (\alpha,\beta)$, in general $(\bold{\bar{F}}_{Im})_{jk}\neq \frac{1}{2}\|\sqrt{\rho_x}[L_{j},L_{k}]\sqrt{\rho_x}\|_1$.  However, for different pairs of indexes, we can repeat the procedure, i.e., choose another set of $\{|u_q\rangle\}$ and $\mathbf{\bar{S}}_{u_q}$, to get the same tradeoff relations with different indexes as
\begin{eqnarray}\label{eq:maintradeoffjk}
\aligned
    &1-[Cov^{-1}(\hat{x})]_{jj}+1-[Cov^{-1}(\hat{x})]_{kk}\\
    \geq& \frac{1}{2}(\frac{1}{2}\|\sqrt{\rho_x}[L_{j},L_{k}]\sqrt{\rho_x}\|_1)^2.
\endaligned
\end{eqnarray}
We note that these tradeoff relations are on the same covariance matrix as the choices of $\{|u_q\rangle\}$ and $\mathbf{\bar{S}}_{u_q}$ do not affect the covariance matrix itself, they are only used to obtain the bounds. %Although the tradeoff relations take the same form as Eq.(\ref{eq:maintradeoffab}) but now with different $\alpha$ and $\beta$. 

By summing the tradeoff relations in Eq.(\ref{eq:maintradeoffjk}) over all pairs of indexes we get
\begin{equation}\label{eq:tradeoff1}
    \frac{1}{\nu}Tr[Cov^{-1}(\hat{x})]\leq n-\frac{1}{4(n-1)} \|C_{1}\|_F^2, %Tr(\tilde{C}_1^T\tilde{C}_1),
\end{equation}
here $\nu$ comes from repeating the 1-local measurement on $\nu$ copies of the state, $C_1$ is a matrix with its entries given by
\begin{equation}\label{eq:Cjk}
    (C_1)_{jk}=\frac{1}{2}\|\sqrt{\rho_x}[L_j,L_k]\sqrt{\rho_x}\|_1.
\end{equation}
We note that $C_1$ is different from any particular $\mathbf{\bar{F}}_{Im}$. We get different $\mathbf{\bar{F}}_{Im}$ by choosing different $\{|u_q\rangle\}$ and $\mathbf{\bar{S}}_{u_q}$ for different pairs of indexes. $C_1$ is obtained by combining the tradeoff relations in Eq.(\ref{eq:maintradeoffjk}) which are obtained by choosing different $\{|u_q\rangle\}$ and $\mathbf{\bar{S}}_{u_q}$ for different indexes.

As stated at the beginning of this section, when $F_Q\neq I$ in the original parametrization, we can make a reparametrization, $\tilde{x}=F_Q^{\frac{1}{2}}x$, under which $\tilde{F}_Q=I$, $\tilde{Cov}(\hat{\tilde{x}})=F_Q^{\frac{1}{2}}Cov(\hat{x})F_Q^{\frac{1}{2}}$, $\tilde{L}_j=\sum_q (F_Q^{-\frac{1}{2}})_{jq}L_q$, the tradeoff relation in Eq.(\ref{eq:tradeoff1}) can be written in the original parametrization as
\begin{eqnarray}\label{eq:tradeofforigin}
\aligned
    \frac{1}{\nu}Tr[F_Q^{-1}Cov^{-1}(\hat{x})]&=\frac{1}{\nu}Tr[\tilde{Cov}^{-1}(\hat{x})]\\
    &\leq n-\frac{1}{4(n-1)} \|C_{1}\|_F^2,
    \endaligned
\end{eqnarray}
with the entries of $C_1$ given by
\begin{eqnarray}
\aligned
(C_1)_{jk}&=\frac{1}{2}\|\sqrt{\rho_x}[\tilde{L}_j,\tilde{L}_k]\sqrt{\rho_x}\|_1\\
&=\frac{1}{2}\|\sqrt{\rho_x}[\sum_q (F_Q^{-\frac{1}{2}})_{jq}L_q,\sum_q (F_Q^{-\frac{1}{2}})_{kq}L_q]\sqrt{\rho_x}\|_1.
\endaligned
\end{eqnarray}

The tradeoff relation immediately gives a necessary condition for the saturation of the QCRB under the 1-local measurement. To saturate the QCRB, i.e., for $Cov(\hat{x})=\frac{1}{\nu}F_Q^{-1}$, it requires $\frac{1}{\nu}Tr[F_Q^{-1}Cov^{-1}(\hat{x})]=n$, which is only possible when $C_1=0$, i.e., $\sqrt{\rho_x}[\tilde{L}_j,\tilde{L}_k]\sqrt{\rho_x}=0$ for any $j,k$. This is the partial commutative condition expressed under the parameterization where $\tilde{F}_Q=I$, and is equivalent to the partial commutative condition in the original parametrization as $\sqrt{\rho_x}[L_q,L_s]\sqrt{\rho_x}=0$ for any $q$ and $s$. %It is equivalent to the standard partial commutative condition expressed in the original parameterization as $\sqrt{\rho_x}[L_q,L_m]\sqrt{\rho_x}=0$ for any $q$ and $m$. %First it is obvious that when the partial commutative condition holds, then $C_1=0$. On the other side, if
%Since when $C_1=0$, we have $\sqrt{\rho_x}[\tilde{L}_j,\tilde{L}_k]\sqrt{\rho_x}=0$ for any $j,k$, then by writing
The equivalence can be seen by writing $L_q=\sum_j(F_Q^{\frac{1}{2}})_{qj}\tilde{L}_j$ and  $L_s=\sum_k(F_Q^{\frac{1}{2}})_{sk}\tilde{L}_k$, it is then easy to seen that when $\sqrt{\rho_x}[\tilde{L}_j,\tilde{L}_k]\sqrt{\rho_x}=0$ for any $j,k$ we have $\sqrt{\rho_x}[L_q,L_s]\sqrt{\rho_x}=0$ for any $q$ and $s$, and vice versa.

\subsection{Incompatibility measures under p-local measurements}
For p-local measurements, which are the collective measurements on at most $p$ copies of the state, we can get the tradeoff relation by replacing $\rho_x$ with $\rho_x^{\otimes p}$ in the previous section. %and repeat the measurement by $\nu/p$ times for a total of $\nu$ copies of the state.
Again we first assume $F_Q=I$ for $\rho_x$, then $F_{Qp}=pI$ for $\rho_x^{\otimes p}$. Following the same procedure as the previous section, for a fixed pair of $j,k$, by substituting $\tilde{Cov}^{-1}(\hat{x})=F_{Qp}^{-\frac{1}{2}}Cov^{-1}(\hat{x})F_{Qp}^{-\frac{1}{2}}=\frac{Cov^{-1}(\hat{x})}{p}$ and $\tilde{F}_{Imp}=F_{Qp}^{-\frac{1}{2}}\bold{\bar{F}}_{Imp}F_{Qp}^{-\frac{1}{2}}=\frac{\bold{\bar{F}}_{Imp}}{p}$ in Eq.(\ref{eq:mainjk}) we can get
\begin{equation}
%\label{eq:mainjk}
    1-\frac{Cov^{-1}(\hat{x})_{jj}}{p}+1-\frac{Cov^{-1}(\hat{x})_{kk}}{p}\geq \frac{1}{2}|\frac{(\bold{\bar{F}}_{Imp})_{jk}}{p}|^2
\end{equation}
with $(\bold{\bar{F}}_{Imp})_{jk}=\frac{1}{2}\|\sqrt{\rho_x^{\otimes p}}[L_{jp}, L_{kp}]\sqrt{\rho_x^{\otimes p}}\|_1$, here $L_{jp}$ is the SLD corresponding to the parameter $x_j$ for $\rho_x^{\otimes p}$, which can be written as $L_{jp}=\sum_{r=1}^p L_j^{(r)}$ with $L_j^{(r)}=I^{\otimes (r-1)}\otimes L_j\otimes I^{\otimes (p-r)}$, $r=1,\cdots, p$, $L_j$ is the SLD for a single copy of the state.

Again we can repeat the procedure for different pairs of $j,k$ and sum over all pairs of $j,k$ to get the tradeoff relation. Under the parameterization that $F_Q=I$, we have
\begin{eqnarray}\label{eq:tradeoffpr}
\aligned
    \frac{1}{\nu/p}\frac{Tr[Cov^{-1}(\hat{x})]}{p}&=\frac{1}{\nu}Tr[Cov^{-1}(\hat{x})]\\
    &\leq n-\frac{1}{4(n-1)} \|\frac{C_{p}}{p}\|_F^2, %Tr(\tilde{C}_1^T\tilde{C}_1),
    \endaligned
\end{eqnarray}
where the factor $\frac{1}{\nu/p}$ comes from repeating the $p$-local measurement $\nu/p$ times on a total $\nu$ copies of the state, $C_p$ is a matrix with the entries given by
\begin{equation}
    (C_p)_{jk}=\frac{1}{2}\|\sqrt{\rho_x^{\otimes p}}[L_{jp}, L_{kp}]\sqrt{\rho_x^{\otimes p}}\|_1.
\end{equation}

If $F_Q\neq I$ in the initial parametrization, we can again make a reparametrization $\tilde{x}=F_Q^{\frac{1}{2}}x$ first, under which $\tilde{L}_j=\sum_q (F_Q^{-\frac{1}{2}})_{jq}L_q$. The tradeoff relation can then be written as
\begin{eqnarray}\label{eq:tradeoffp}
\aligned
    \frac{1}{\nu}Tr[F_Q^{-1}Cov^{-1}(\hat{x})]\leq n-\frac{1}{4(n-1)} \|\frac{C_{p}}{p}\|_F^2, %Tr(\tilde{C}_1^T\tilde{C}_1),
    \endaligned
\end{eqnarray}
with %$C_p$ is a matrix with the entries given by
\begin{equation}\label{eq:CP}
    (C_p)_{jk}=\frac{1}{2}\|\sqrt{\rho_x^{\otimes p}}[\tilde{L}_{jp}, \tilde{L}_{kp}]\sqrt{\rho_x^{\otimes p}}\|_1,
\end{equation}
here %$\tilde{L}_{jp}=\sum_{r=1}^p \tilde{L}_j^{(r)}$, here $\tilde{L}_j^{(r)}=I^{\otimes (r-1)}\otimes \tilde{L}_j\otimes I^{\otimes (p-r)}$, $r=1,\cdots, p$, which can also be equivalently written as
$\tilde{L}_{jp}=\sum_q (F_Q^{-\frac{1}{2}})_{jq}L_{qp}$.

$\|\frac{C_{p}}{p}\|_F$ determines the gap between the bound and $n$, which measures the incompatibility of the measurements. Since $p$-local measurement is a subset of $(p+1)$-local measurement, we expect that $\|\frac{C_{p+1}}{p+1}\|_F\leq \|\frac{C_p}{p}\|_F$ since there should be less incompatibility when more measurements are allowed. This can be verified as
\begin{eqnarray}
\aligned
&\frac{\|\sqrt{\rho_x^{\otimes (p+1)}}[\tilde{L}_{j(p+1)}, \tilde{L}_{k(p+1)}]\sqrt{\rho_x^{\otimes (p+1)}}\|_1}{p+1}\\
=&\frac{\|\sum_{r=1}^{p+1}\sqrt{\rho_x^{\otimes (p+1)}}[\tilde{L}^{(r)}_{j}, \tilde{L}^{(r)}_{k}]\sqrt{\rho_x^{\otimes (p+1)}}\|_1}{p+1}\\
=&\frac{\|(1/p)\sum_{q=1}^{p+1} \sum_{r\neq q} \sqrt{\rho_x^{\otimes (p+1)}}[\tilde{L}^{(r)}_{j}, \tilde{L}^{(r)}_{k}]\sqrt{\rho_x^{\otimes (p+1)}}\|_1}{p+1}\\
\leq &\frac{\sum_{q=1}^{p+1} \|\sum_{r\neq q} \sqrt{\rho_x^{\otimes (p+1)}}[\tilde{L}^{(r)}_{j}, \tilde{L}^{(r)}_{k}]\sqrt{\rho_x^{\otimes (p+1)}}\|_1}{p(p+1)}\\
=&\frac{(p+1) \|\sqrt{\rho_x^{\otimes p}}[\tilde{L}_{jp}, \tilde{L}_{kp}]\sqrt{\rho_x^{\otimes p}}\|_1}{p(p+1)}\\
=&\frac{\|\sqrt{\rho_x^{\otimes p}}[\tilde{L}_{jp}, \tilde{L}_{kp}]\sqrt{\rho_x^{\otimes p}}\|_1}{p},
%\leq& \frac{\|\sum_{r=1}^p\sqrt{\rho_x^{\otimes (p+1)}}[\tilde{L}^{(r)}_{j}, \tilde{L}^{(r)}_{k}]\sqrt{\rho_x^{\otimes (p+1)}}\|_1}{p+1}\\
%&+\frac{\sqrt{\rho_x^{\otimes p+1}}[\tilde{L}^{(p+1)}_{j}, \tilde{L}^{(p+1)}_{k}]\sqrt{\rho_x^{\otimes (p+1)}}\|_1}{p+1}\\
%=&\frac{\|\sqrt{\rho_x^{\otimes p}}[\tilde{L}_{jp}, \tilde{L}_{kp}]\sqrt{\rho_x^{\otimes (p)}}\otimes \rho_x\|_1}{p+1}\\
%&+\frac{\|\rho_x^{\otimes p}\otimes \sqrt{\rho_x}[\tilde{L}_{j}, \tilde{L}_{k}]\sqrt{\rho_x}\|_1}{p+1}\\
%=&\frac{\|\sqrt{\rho_x^{\otimes p}}[\tilde{L}_{jp}, \tilde{L}_{kp}]\sqrt{\rho_x^{\otimes (p)}}\|_1}{p+1}\\
%&+\frac{\| \sqrt{\rho_x}[\tilde{L}_{j}, \tilde{L}_{k}]\sqrt{\rho_x}\|_1}{p+1}\\
%\leq&\frac{\|\sqrt{\rho_x^{\otimes p}}[\tilde{L}_{jp}, \tilde{L}_{kp}]\sqrt{\rho_x^{\otimes (p)}}\|_1}{p+1}\\
%&+\frac{1/p\|\sqrt{\rho_x^{\otimes p}}[\tilde{L}_{jp}, \tilde{L}_{kp}]\sqrt{\rho_x^{\otimes (p)}}\|_1}{p+1}\\
%=&\frac{\|\sqrt{\rho_x^{\otimes p}}[\tilde{L}_{jp}, \tilde{L}_{kp}]\sqrt{\rho_x^{\otimes (p)}}\|_1}{p},
\endaligned
\end{eqnarray}
i.e., $\frac{(C_{p+1})_{jk}}{p+1}\leq \frac{(C_{p})_{jk}}{p}$, which implies $\|\frac{C_{p+1}}{p+1}\|_F\leq \|\frac{C_p}{p}\|_F$.
%where the last inequality comes from the fact that
%\begin{eqnarray}
%\aligned
%&\|\sqrt{\rho_x^{\otimes p}}[\tilde{L}_{jp}, %\tilde{L}_{kp}]\sqrt{\rho_x^{\otimes (p)}}\|_1\\
%\leq
%\endaligned
%\end{eqnarray}

%From the tradeoff relations, we can also immediately get a necessary condition for the saturation of the QCRB under the p-local measurements, which is $\frac{C_p}{p}=0$, i.e., $\frac{\|\sqrt{\rho_x^{\otimes p}}[\tilde{L}_{jp}, \tilde{L}_{kp}]\sqrt{\rho_x^{\otimes p}}\|_1}{p}=0$ for any $j,k$. Again this is equivalent to $\frac{\|\sqrt{\rho_x^{\otimes p}}[L_{qp}, L_{mp}]\sqrt{\rho_x^{\otimes p}}\|_1}{p}=0$ for any $q,m$ in the original parameterization. %For $n\geq 3$ the bound can be similarly improved.
%\begin{equation}
%    \frac{1}{\mu/p}Tr[F_{Qp}^{-1}Cov(\hat{x})^{-1}]\leq n-\frac{1}{4(n-1)^2} Tr(C_{p}^TC_{p}),
%\end{equation}
%where $F_{Qp}$ is the quantum Fisher information matrix of $\rho_x^{\otimes p}$ which equals to $pF_Q$, $C_{p}$ is a matrix with the entries given by
%$(C_{p})_{jk}=\frac{1}{2}\|\sqrt{\rho_x^{\otimes p}}[L_{jp},L_{kp}]|\sqrt{\rho_x^{\otimes p}}\|_1$ with  The tradeoff relation thus can be rewritten as
%\begin{equation}\label{eq:tradep}
%    \frac{1}{\nu}Tr[F_Q^{-1}Cov(\hat{x})^{-1}]\leq n-\frac{1}{4(n-1)^2} Tr(F_{Q}^{-1}\frac{C_{p}}{p}^TF_{Q}^{-1}\frac{C_{p}}{p}
%\end{equation}

%\section{Connection between the partial commutative condition and the weak commutative condition}\label{sec:commutative}
A necessary condition for the saturation of the QCRB under the p-local measurement is $\frac{C_p}{p}=0$, which implies $\frac{\|\sqrt{\rho_x^{\otimes p}}[\tilde{L}_{jp}, \tilde{L}_{kp}]\sqrt{\rho_x^{\otimes p}}\|_1}{p}=0$ for any $j,k$. This is equivalent to $\frac{\|\sqrt{\rho_x^{\otimes p}}[L_{jp}, L_{kp}]\sqrt{\rho_x^{\otimes p}}\|_1}{p}=0$ for any $j,k$ in the original parameterization, and can be seen as the partial commutative condition under the p-local measurement.
%As we have shown that a necessary condition for the saturation of the QCRB under the p-local measurements is $\frac{\|\sqrt{\rho_x^{\otimes p}}[L_{jp}, L_{kp}]\sqrt{\rho_x^{\otimes p}}\|_1}{p}=0$ for any $j,k$ and  %$\sqrt{\rho_x^{\otimes p}}[L_{jp}, L_{kp}]\sqrt{\rho_x^{\otimes p}}=0$, which can be equivalent written as
%$\frac{\|\sqrt{\rho_x^{\otimes p}}[L_{jp}, L_{kp}]\sqrt{\rho_x^{\otimes p}}\|_1}{p}=0$ for any $j,k\in\{1,\cdots, n\}$.

At $p=1$, the condition $\frac{C_p}{p}=0$ is equivalent to the partial commutative condition. It is natural to ask whether this condition recovers the weak commutative condition at $p\rightarrow \infty$. %On the other hand, it is well known that the weak commutative condition, $Tr(\rho_x[L_j,L_k])=0$ for any $j,k\in\{1,\cdots, n\}$, is necessary and sufficient for the saturation of the QCRB when $p\rightarrow \infty$\cite{Ragy2016}.
%Note that if this condition is also sufficient for the saturation of the QCRB under the $p$-local measurement, the answer will be obviously affirmative, although even in this case an explicit proof can also add insights. %it should reduce to the weak commutative condition when $p\rightarrow \infty$.
%Since it is still unknown whether this condition is sufficient for the saturation of the QCRB, it is nontrivial the connection between this condition and the weak commutative condition remained unclear\cite{Yang2019}.
In the appendix we explicitly show that this condition indeed reduces to the weak commutative condition when $p\rightarrow \infty$. Specifically we show that (regardless of the parametrization)
\begin{equation}
\lim_{p\rightarrow \infty}\frac{\|\sqrt{\rho_x^{\otimes p}}[\tilde{L}_{jp}, \tilde{L}_{kp}]\sqrt{\rho_x^{\otimes p}}\|_1}{p}=|Tr(\rho_x[\tilde{L}_j,\tilde{L}_k])|. \end{equation}
%which holds under any parametrization.

When $p\rightarrow \infty$ the partial commutative condition, $\frac{C_p}{p}=0$, is then equivalent to the weak commutative condition, $\tilde{F}_{Im}=0$, here $(\tilde{F}_{Im})_{jk}=\frac{1}{2}Tr(\rho_x[\tilde{L}_j,\tilde{L}_k])$. % which is equivalent to  $F_{Im}=0$ in the original parametrization since $\tilde{F}_{Im}=F_Q^{-\frac{1}{2}}F_{Im}F_Q^{-\frac{1}{2}}$. 
This clarifies the connection between the partial commutative condition and the weak commutative condition and solves an open question\cite{Yang2019}. The connection also suggests that the partial commutative condition under p-local measurements, $\frac{C_p}{p}=0$, is likely also sufficient for the saturation of QCRB under $p$-local measurements, although we do not have a proof.
%except for qubit system. For qubit system the partial commutative condition is sufficient  since for  2-dimensional $\rho_x$, it is either a pure state or full-rank mixed state ($\rho_x^{\otimes p}$ is also either pure or full rank). For pure states the partial commutative condition is equivalent to the weak commutative condition, which is known to be sufficient for the saturation of the QCRB under the $1$-local measurement (thus also for the $p$-local measurement), and for full-rank mixed states the partial commutative condition is equivalent to the condition that all SLDs commute with each other, in which case the QCRB is also saturable.

Since $\|\frac{C_{p}}{p}\|_F$ is monotone, we have
\begin{eqnarray}
\|C_1\|_F\geq \|\frac{C_2}{2}\|_F\geq \cdots \geq \lim_{p\rightarrow \infty}\|\frac{C_{p}}{p}\|_F=\|\tilde{F}_{Im}\|_F,
\end{eqnarray}
where $(C_1)_{jk}=\frac{1}{2}\|\sqrt{\rho_x}[\tilde{L}_j,\tilde{L}_k]\sqrt{\rho_x}\|_1$ and $(\tilde{F}_{Im})_{jk}=\frac{1}{2i}Tr(\rho_x[\tilde{L}_j,\tilde{L}_k])$ with $\tilde{L}_j$ and $\tilde{L}_k$ as the SLDs under the reparametrization that $\tilde{F}_Q=I$. All values of $\frac{(C_p)_{jk}}{p}$ are thus between $\frac{1}{2}|Tr(\sqrt{\rho_x}[\tilde{L}_j,\tilde{L}_k]\sqrt{\rho_x})|$ and $\frac{1}{2}\|\sqrt{\rho_x}[\tilde{L}_j,\tilde{L}_k]\sqrt{\rho_x}\|_1$, i.e., between the absolute value of the trace and the trace norm of the same matrix, $\frac{1}{2}\sqrt{\rho_x}[\tilde{L}_j,\tilde{L}_k]\sqrt{\rho_x}$.

When $p\rightarrow \infty$, by substituting $\lim_{p\rightarrow \infty}\|\frac{C_{p}}{p}\|_F=\|\tilde{F}_{Im}\|_F$ into the bound
\begin{eqnarray}
\aligned
    \Gamma_p\leq n-\frac{1}{4(n-1)} \|\frac{C_{p}}{p}\|_F^2, %Tr(\tilde{C}_1^T\tilde{C}_1),
    \endaligned
\end{eqnarray}
we have
\begin{eqnarray}
\aligned
     \Gamma_{\infty}%=\frac{1}{\nu}Tr[F_Q^{-1}Cov^{-1}(\hat{x})]
     \leq n-\frac{1}{4(n-1)} \|\tilde{F}_{Im}\|_F^2. %Tr(\tilde{C}_1^T\tilde{C}_1),
    \endaligned
\end{eqnarray}
Combined with the lower bound in Eq.(\ref{eq:lowerbound})\cite{Carollo_2019}, which is \begin{eqnarray}
\aligned
   \Gamma_{\infty}&%=\frac{1}{\nu} Tr[F_Q^{-1}Cov^{-1}(\hat{x})]\\
   \geq & \frac{n^2}{n+\|\tilde{F}_{Im}\|_1}
   \geq n-\|\tilde{F}_{Im}\|_1,
   \endaligned
\end{eqnarray}
we get
\begin{equation}\label{eq:lowerupper}
    n-\|\tilde{F}_{Im}\|_1\leq \Gamma_{\infty}\leq n-\frac{1}{4(n-1)} \|\tilde{F}_{Im}\|_F^2,
\end{equation}
here $\tilde{F}_{Im}=F_Q^{-\frac{1}{2}}F_{Im}F_Q^{-\frac{1}{2}}$. It can be easily seen that the QCRB is saturable (in which case $\Gamma_{\infty}=n$) if and only if $\tilde{F}_{Im}=0$, which is just the weak commutative condition. This provides an alternative way of showing the weak commutative condition is necessary and sufficient for the saturation of QCRB at $p\rightarrow \infty$.

$\tilde{F}_{Im}$ has been proposed as a measure of quantumness based on the lower bound,  $\Gamma_{\infty} \geq n-\|\tilde{F}_{Im}\|_1$\cite{Carollo_2019}. The upper bound obtained here adds another layer on the interpretation of $\tilde{F}_{Im}$ as the quantumness when $p\rightarrow \infty$. We note that if $\frac{C_p}{p}=0$ is also sufficient for the saturation of the QCRB under $p$-local measurements, $\frac{C_p}{p}$ can be used as a measure of the quantumness under $p$-local measurements.

Similarly we also have
\begin{eqnarray}\label{eq:mainImproved}
\aligned
    \frac{1}{\nu}Tr[F_Q^{-1}Cov^{-1}(\hat{x})]
    \leq &n-f(n)\|\frac{F_Q^{-\frac{1}{2}}\bold{\bar{F}}_{Imp}F_Q^{-\frac{1}{2}}}{p}\|_F^2,
%    \frac{1}{\nu}Tr[F_Q^{-1}Cov^{-1}(\hat{x})]
 %   \leq &n-\frac{n-2}{(n-1)^2}\|\frac{F_Q^{-\frac{1}{2}}\bold{\bar{F}}_{Imp}F_Q^{-\frac{1}{2}}}{p}\|_F^2,\\ %Tr(\tilde{C}_1^T\tilde{C}_1),
  %  \frac{1}{\nu}Tr[F_Q^{-1}Cov^{-1}(\hat{x})]
   % \leq &n-\frac{1}{5}\|\frac{F_Q^{-\frac{1}{2}}\bold{\bar{F}}_{Imp}F_Q^{-\frac{1}{2}}}{p}\|_F^2, %Tr(\tilde{C}_1^T\tilde{C}_1),
    \endaligned
\end{eqnarray}
where $
f(n)=\max\{\frac{1}{4(n-1)},\frac{n-2}{(n-1)^2},\frac{1}{5}\}$, $\bold{\bar{F}}_{Imp}$ is the imaginary part of  $\bold{\bar{F}}=\sum_q \bold{\bar{F}}_{u_q}$ with $\bold{\bar{F}}_{u_q}$ equal to either $F_{u_q}$ or $F_{u_q}^T$, here $F_{u_q}$ is a $n\times n$ matrix with the $jk$-th entry given by
\begin{equation}
    (F_{u_q})_{jk}=\langle u_q|\sqrt{\rho_x^{\otimes p}}L_{jp}L_{kp}\sqrt{\rho_x^{\otimes p}}|u_q\rangle,
\end{equation}
$L_{jp}$ is the SLD of $\rho_x^{\otimes p}$ corresponding to the parameter $x_j$, and $\{|u_q\rangle\}$ are a set of vectors in $H_d^{\otimes p}$ that satisfies $\sum_q |u_q\rangle\langle u_q|=I_{d^p}$ with $I_{d^p}$ denote the $d^p\times d^p$ Identity matrix.
% that for pure states and full rank mixed states, the partial commutative condition is also sufficient.

\subsection{Simpler bounds of the incompatibility measures}
The obtained tradeoff relation under the p-local measurement in Eq.(\ref{eq:tradeofforigin}) needs to compute $\|\sqrt{\rho_x^{\otimes p}}[\tilde{L}_{jp}, \tilde{L}_{kp}]\sqrt{\rho_x^{\otimes p}}\|_1$, which involves operators whose dimension increases exponentially with $p$. Here we provide an alternative tradeoff relation, which only uses operators on a single quantum state thus easier to compute.

If we write $\sqrt{\rho_x^{\otimes p}}[\tilde{L}_{jp}, \tilde{L}_{kp}]\sqrt{\rho_x^{\otimes p}}=D_p^{(jk)}+O_p^{(jk)}$ with $D_p^{(jk)}$ as the diagonal part and $O_p^{(jk)}$ as the off-diagonal part, we have(see appendix)
\begin{eqnarray}
\|D_p^{(jk)}\|_1\leq \|D_p^{(jk)}+O_p^{(jk)}\|_1\leq \|D_p^{(jk)}\|_1+\|O_p^{(jk)}\|_1.
\end{eqnarray}
In the appendix we show that with the eigenvalue decomposition, $\rho_x=\sum_{q=1}^m\lambda_q |\Psi_q\rangle\langle\Psi_q|$ with $\lambda_q>0$,
\begin{eqnarray}
\aligned
   \|D_p^{(jk)}\|_1=\sum_{v_1,\cdots, v_p%\in\{1,\cdots, m\}
    }(\prod_{r=1}^p\lambda_{v_r}) |\sum_{r=1}^p \langle\Psi_{v_r}|[\tilde{L}_j,\tilde{L}_k]|\Psi_{v_r}\rangle|,
    \endaligned
\end{eqnarray}
where $v_1,\cdots,v_p\in \{1,\cdots, m\}$, $\tilde{L}_{j}=\sum_{\mu}(F_Q^{-{\frac{1}{2}}})_{j\mu}L_{\mu}$ and $\tilde{L}_{k}=\sum_{\mu}(F_Q^{-{\frac{1}{2}}})_{k\mu}L_{\mu}$.  As shown in the appendix, $\|O_p^{(jk)}\|_1 \approx O(\sqrt{p})$, the difference between $\frac{\|D_{p}^{(jk)}\|_1}{p}$ and $\frac{\|\sqrt{\rho_x^{\otimes p}}[\tilde{L}_{jp}, \tilde{L}_{kp}]\sqrt{\rho_x^{\otimes p}}\|_1}{p}$ is then within the order of $\frac{1}{\sqrt{p}}$, i.e.,
\begin{equation}
   \frac{\|D_{p}^{(jk)}\|_1}{p}\leq  \frac{\|\sqrt{\rho_x^{\otimes p}}[\tilde{L}_{jp}, \tilde{L}_{kp}]\sqrt{\rho_x^{\otimes p}}\|_1}{p}\leq \frac{\|D_{p}^{(jk)}\|_1}{p}+O(\frac{1}{\sqrt{p}}).
\end{equation}
Here $\|D_p^{(jk)}\|_1$ is quantitatively equivalent to the expected value of $|\sum_{r=1}^p \langle\Psi_{v_r}|[\tilde{L}_j,\tilde{L}_k]|\Psi_{v_r}\rangle|$ with each eigenvector $|\Psi_{v_r}\rangle$ selected independently with probability $\lambda_{v_r}$, i.e.,
\begin{equation}
    \|D_p^{(jk)}\|_1=E(|\sum_{r=1}^p \langle\Psi_{v_r}|[\tilde{L}_j,\tilde{L}_k]|\Psi_{v_r}\rangle|)
\end{equation}
where $E(\cdot)$ denotes the expectation, each $|\Psi_{v_r}\rangle$ is randomly and independently chosen from the eigenvectors of $\rho_x$ with a probability of the  corresponding eigenvalue, $\lambda_{v_r}$.  

By replacing $\|\sqrt{\rho_x^{\otimes p}}[\tilde{L}_{jp}, \tilde{L}_{kp}]\sqrt{\rho_x^{\otimes p}}\|_1$  with $\|D_p^{(jk)}\|_1$, we then obtain the alternative bound
\begin{equation}\label{eq:mainTp}
    \frac{1}{\nu}Tr[F_Q^{-1}Cov^{-1}(\hat{x})]\leq n-\frac{1}{4(n-1)} \|\frac{T_p}{p}\|_F^2,
\end{equation}
with
\begin{eqnarray}
\aligned
    (T_p)_{jk}% &=\frac{1}{2}\|D_p^{(jk)}\|_1\\
    &=\frac{1}{2}E(|\sum_{r=1}^p \langle\Psi_{v_r}|[\tilde{L}_j,\tilde{L}_k]|\Psi_{v_r}\rangle|).
%    \sum_{v_1,\cdots, v_p%\in\{1,\cdots, m\}
 %   }(\prod_{r=1}^p\lambda_{v_r}) |\sum_{r=1}^p \langle\Psi_{v_r}|[\tilde{L}_j,\tilde{L}_k]|\Psi_{v_r}\rangle|.
    \endaligned
\end{eqnarray}

Here $T_p$ is also monotonically decreasing with $p$ as
\begin{eqnarray}
\|T_1\|_F\geq \|\frac{T_2}{2}\|_F\geq \cdots \geq \lim_{p\rightarrow \infty}\|\frac{T_p}{p}\|_F=\|\tilde{F}_{Im}\|_F.
\end{eqnarray}
%where $(T_1)_{jk}=\frac{1}{2}\sum_{q=1}^m \lambda_q |\langle \Psi_q|[\tilde{L}_j,\tilde{L}_k]|\Psi_q\rangle|$, %=\frac{1}{2}\sum_{q=1}^m|\langle \Psi_q|\sqrt{\rho_x}[\tilde{L}_j,\tilde{L}_k]\sqrt{\rho_x}|\Psi_q\rangle|$, 
where $(\tilde{F}_{Im})_{jk}=\frac{1}{2}|Tr(\rho_x[\tilde{L}_j,\tilde{L}_k])|$. %=\frac{1}{2}|\sum_{q=1}^m \langle \Psi_q|\sqrt{\rho_x}[\tilde{L}_j,\tilde{L}_k]\sqrt{\rho_x}|\Psi_q\rangle|$. 
%All values of $\frac{(T_p)_{jk}}{p}$ are thus between $|\frac{1}{2}\sum_{q=1}^m \langle \Psi_q|\sqrt{\rho_x}[\tilde{L}_j,\tilde{L}_k]\sqrt{\rho_x}|\Psi_q\rangle|$ and $\frac{1}{2}\sum_{q=1}^m|\langle \Psi_q|\sqrt{\rho_x}[\tilde{L}_j,\tilde{L}_k]\sqrt{\rho_x}|\Psi_q\rangle|$, i.e., between the absolute value of the summation and the summation of the absolute values of the diagonal entries of $\frac{1}{2}\sqrt{\rho_x}[\tilde{L}_j,\tilde{L}_k]\sqrt{\rho_x}$ in the basis of $\{|\Psi_q\rangle\}$.
%Since the difference between $\frac{T_p}{p}$ and $\frac{C_p}{p}$ is $O(\frac{1}{\sqrt{p}}$, we also have $\lim_{p\rightarrow \infty}\|\frac{T_p}{p}\|_F=\|\tilde{F}_{Im}\|_F$.

%Compared to $C_p$, $T_p$ is expressed only with operators on a single copy of the state, which is easier to compute.
%For large $p$ where the computation of $C_p$ becomes difficult, we can use the bound with $T_p$, which is almost as tight for large $p$.
We note that this bound can be equivalently obtained by choosing the set of $\{|u_q\rangle\}$ in Eq.(\ref{eq:Covu}) %differently. Specifically, it can be obtained by choosing $\{|u_j\rangle\}$
as the eigenvectors of $\rho_x$ instead of the eigenvectors of $\sqrt{\rho_x}[\tilde{L}_j,\tilde{L}_k]\sqrt{\rho_x}$. %and $T_p$ by choosing $\{|u_j\rangle\}$ as the tensor product of the eigenvectors of $\rho_x$).

\section{Incompatibility measures with RLDs}\label{sec:RLD}
The approach can be used to obtain various other incompatibility measures with different operators. Here we demonstrate it with the right logarithmic operators(RLD)\cite{Hels76book,Yuen1973}. %Various tradeoff relations can be obtained by choosing different operators.

%The bound provided by the right logarithm operator(RLD) can be similarly obtained from the uncertainty relation.
The quantum Cramer-Rao bound in terms of the RLD quantum Fisher information is given by
\begin{equation}\label{eq:RLDQCRB}
    Cov(\hat{x})\geq \frac{1}{\nu}(F^{RLD})^{-1}.
\end{equation}
where $(F^{RLD})_{jk}=Tr(\rho_x L_j^R L_k^{R\dagger})$, $L^R_j$($L^R_k$) is the RLD associated with the parameter $x_j$($x_k$), which can be obtained from the equation $\partial_{x_j}\rho_x=\rho_x L^R_j$\cite{Hels76book,Yuen1973,Haya05book}. 
Different from the SLD quantum Fisher information matrix, the RLD quantum Fisher information matrix is in general a complex matrix. If we decompose the inverse of the RLD quantum Fisher information matrix into the real and imaginary part as  $(F^{RLD})^{-1}=(F^{RLD})^{-1}_{Re}+i(F^{RLD})^{-1}_{Im}$, Eq.(\ref{eq:RLDQCRB}) then leads to the standard RLD lower bound on the weighted covariance matrix as
\begin{eqnarray}
\aligned
&\nu Tr[W Cov(\hat{x})]\\
&\geq Tr[W(F^{RLD})^{-1}_{Re}]+\|\sqrt{W}(F^{RLD})^{-1}_{Im}\sqrt{W}\|_1.
\endaligned
\end{eqnarray}

For single-parameter estimation the standard RLD bound is always less tighter than the SLD bound. For multi-parameter quantum estimation, however, the RLD bound can be tighter than the SLD bound\cite{Yuen1973,Rafal2020,Sidhu2021}.  

We can obtain an upper bound on $\Gamma_p$ from the standard RLD bound. As $Cov^{-1}(\hat{x})\leq \nu F^{RLD}$, by writing $F^{RLD}=F^{RLD}_{Re}+iF^{RLD}_{Im}$ as the real and imaginary parts, we have 
\begin{equation}\label{eq:mainRLDstd}
\aligned
    \frac{1}{\nu}Tr[F_Q^{-1}Cov^{-1}(\hat{x})]\leq Tr[F_Q^{-1}F^{RLD}_{Re}]-\|F_Q^{-\frac{1}{2}}F^{RLD}_{Im}F_Q^{-\frac{1}{2}}\|_1.
\endaligned
\end{equation}
This bound is independent of $p$ since the RLD bound holds under any measurements.

We now show how the standard RLD bound can be improved in a similar way. By choosing the operators as $(X_1,\cdots,X_n,L_1^{R\dagger},\cdots, L_n^{R\dagger})$, we have
%\begin{equation}
%    S=\left(\begin{array}{cc}
%      A & B  \\
%      B^\dagger & F^R \\
%          \end{array}\right)\geq 0,
%\end{equation}
\begin{widetext}
\begin{eqnarray}
\aligned
S_u&=\left(\begin{array}{cccccc}
X_1\sqrt{\rho_x}|u\rangle & \cdots & X_n\sqrt{\rho_x}|u\rangle & L_1^{R\dagger}\sqrt{\rho_x}|u\rangle & \cdots & L_n^{R\dagger}\sqrt{\rho_x}|u\rangle \end{array}\right)^\dagger \left(\begin{array}{cccccc}X_1\sqrt{\rho}|u\rangle & \cdots & X_n\sqrt{\rho}|u\rangle & L_1^{R\dagger}\sqrt{\rho}|u\rangle & \cdots & L_n^{R\dagger}\sqrt{\rho}|u\rangle \end{array}\right)\\
&=\left(\begin{array}{cc}
A_u & B_u\\
B_u^\dagger & F_u
\end{array}\right)\geq 0,
\endaligned
\end{eqnarray}
\end{widetext}
with $(A_u)_{jk}=\langle u|\sqrt{\rho_x} X_jX_k\sqrt{\rho_x}|u\rangle$, $(B_u)_{jk}=\langle u|\sqrt{\rho_x} X_jL_k^{R\dagger}\sqrt{\rho_x}|u\rangle$, $(F_u)_{jk}=\langle u|\sqrt{\rho_x} L_j^{R}L_k^{R\dagger}\sqrt{\rho_x}|u\rangle$.

Similarly, if we choose a set of $\{|u_q\rangle\}$ with $\sum_q |u_q\rangle\langle u_q|=I$, we can get $\mathbf{\bar{S}}=\sum_q \mathbf{\bar{S}}_{u_q}$ with $\mathbf{\bar{S}}_{u_q}\in \{S_{u_q},S^T_{u_q}\}$. The standard RLD bound corresponds to choosing $\mathbf{\bar{S}}_{u_q}=S_{u_q}$ for all $q$. In this case  $\mathbf{\bar{S}}=\sum_q S_{u_q}= \left(\begin{array}{cc}
A & B\\
B^\dagger & F^{RLD}
\end{array}\right)\geq 0$,
where $(A)_{jk}=Tr(\rho_xX_jX_k)$, $(B)_{jk}=Tr(\rho_xX_jL_k^{R\dagger})$, $(F^{RLD})_{jk}=Tr(\rho_xL_j^RL_k^{R\dagger})$. From the local unbiased condition,
\begin{eqnarray}
\label{eq:rlu}
\aligned
&Tr(\rho_xL_j^R \hat{X}_k)=\delta^k_j,\qquad
%Tr(\rho_xL_1^R\hat{X}_2)=0,\\
%&Tr(\rho_xL_2^R\hat{X}_1)=0,\qquad
%Tr(\rho_xL_2^R\hat{X}_2)=1.\\
\endaligned
\end{eqnarray} 
we can get
\begin{eqnarray}
\aligned
(B)_{jk}&=Tr(\rho_xX_jL_k^{R\dagger}) \\
&=Tr(\rho_xL_k^{R}X_j)^*\\
&=\delta_k^j,
\endaligned
\end{eqnarray}
thus in this case $B=I$. The standard RLD bound can then be obtained via the Schur's complement as
\begin{equation}
    Cov(\hat{x})\geq A\geq B(F^{RLD})^{-1}B^\dagger=(F^{RLD})^{-1}.
\end{equation}
If it is repeated with $\nu$ times, we then obtain the standard RLD bound, 
\begin{equation}
    Cov(\hat{x})\geq \frac{1}{\nu}(F^{RLD})^{-1},
\end{equation}
which then leads to the upper bound on $\Gamma_p$ as in Eq.(\ref{eq:mainRLDstd}).
%This leads to a tradeoff relation as in 
%\begin{equation}\label{eq:tradeoffRLD1}
%    Tr[F_Q^{-1}Cov^{-1}(\hat{x})]\leq Tr[F_Q^{-1}F^{RLD}_{Re}]-\|F_Q^{-\frac{1}{2}}F^{RLD}_{Im}F_Q^{-\frac{1}{2}}\|_1,
%\end{equation}
%where $F^{RLD}_{Re}=\frac{1}{2}[F^{RLD}+(F^{RLD})^T]$ and $F^{RLD}_{Im}=\frac{1}{2i}[F^{RLD}-(F^{RLD})^T]$ are the real and imaginary part of $F^{RLD}$ respectively. If there are $\nu$ copies of the state, then by repeating the 1-local measurement $\nu$ times, we can get the tradeoff relation as
%\begin{equation}\label{eq:RLDtradeoff}
 %   \frac{1}{\nu}Tr[F_Q^{-1}Cov^{-1}(\hat{x})]\leq Tr[F_Q^{-1}F^{RLD}_{Re}]-\|F_Q^{-\frac{1}{2}}F^{RLD}_{Im}F_Q^{-\frac{1}{2}}\|_1.
%\end{equation}
For any p-local measurements, we can replace $\rho_x$ with $\rho_x^{\otimes p}$ and repeat the measurement $\nu/p$ times, which leads to the same tradeoff relation as in Eq.(\ref{eq:mainRLDstd}). This is consistent with the fact the standard RLD bound holds for any measurements.

%{\color{blue}

The standard RLD bound can be improved by making proper choices on $\{|u_q\rangle\}$ and $\{\mathbf{\bar{S}}_{u_j}\}$. Here we make a particular choice as an illustration. Again we first assume $F_Q=I$ and for a fixed pair of indexes, $j,k$, choose a complete basis, $\{|u_1\rangle, \cdots, |u_d\rangle\}$, as the orthonormal eigenvectors of  $\sqrt{\rho_x}(L_j^RL_k^{R\dagger}-L_k^RL_j^{R\dagger})\sqrt{\rho_x}$. For any $|u_q\rangle$,
the imaginary part of $(F_{u_q})_{jk}$ is $\frac{1}{2i}\langle u_q|\sqrt{\rho_x}(L_j^RL_k^{R\dagger}-L_k^RL_j^{R\dagger})\sqrt{\rho_x}|u_q\rangle$, which we denote as $t^q_{jk}$. We then let
\begin{eqnarray}\label{eq:maint}
\bold{\bar{S}}_{u_q}:=\{\begin{array}{cc}
S_{u_q},  \text{when } t^q_{jk}\geq 0,\\%\frac{1}{2i}\langle u_q|\sqrt{\rho_x}(L_j^RL_k^{R\dagger}-L_k^RL_j^{R\dagger})\sqrt{\rho_x}|u_q\rangle\geq 0\\
S^T_{u_q}, \text{when } t^q_{jk}<0.%\frac{1}{2i}\langle u_q|\sqrt{\rho_x}(L_j^RL_k^{R\dagger}-L_k^RL_j^{R\dagger})\sqrt{\rho_x}|u_q\rangle< 0
\end{array}
\end{eqnarray}
In this case we get
\begin{eqnarray}\label{eq:suqRLD}
\bold{\bar{S}}=\sum_q \bold{\bar{S}}_{u_q}=\left(\begin{array}{cc}
\bold{\bar{A}} & \bold{\bar{B}}\\
\bold{\bar{B}}^\dagger & \bold{\bar{F}}^{RLD}
\end{array}\right),
\end{eqnarray}
here $\bold{\bar{B}}=I+i\bold{\bar{B}}_{Im}$, $\bold{\bar{F}}^{RLD}=\sum_q\bold{\bar{F}}_{u_q}$ with $\bold{\bar{F}}_{u_q}$ equals to either $F_{u_q}$ or $F^T_{u_q}$ according to the choices in Eq.(\ref{eq:maint})(which makes the imaginary part of $(\bold{\bar{F}}_{u_q})_{jk}$ always positive).
The real part of $\bold{\bar{F}}^{RLD}$ remains the same as $F^{RLD}_{Re}$, the imaginary part of the $jk$-th entry of $\bold{\bar{F}}^{RLD}$ is
\begin{equation}\label{eq:Fjk}
(\bold{\bar{F}}^{RLD}_{Im})_{jk}=\frac{1}{2}\|\sqrt{\rho_x}(L_j^RL_k^{R\dagger}-L_k^RL_j^{R\dagger})\sqrt{\rho_x}\|_1.
\end{equation}
%We then get  $Cov(\hat{x})\geq \bold{\bar{A}}$, which further gives $\bold{\bar{F}}^{RLD}-\bold{\bar{B}}^\dagger Cov^{-1}(\hat{x})\bold{\bar{B}}\ge 0$.
By following the same procedure, we can obtain the tradeoff relation under the 1-local measurement(under the parametrization such that $F_{Q}=I$) as
\begin{equation}
    Tr[Cov^{-1}(\hat{x})]\leq Tr[\bold{\bar{F}}^{RLD}_{Re}]-\frac{1}{4(n-1)}\|C^{RLD}_1\|_F^2,
\end{equation}
where $(C^{RLD}_1)_{jk}=\min\{\frac{1}{2}\|\sqrt{\rho_x}(L_j^RL_k^{R\dagger}-L_k^RL_j^{R\dagger})\sqrt{\rho_x}\|_1,2\}$(see appendix).
If we repeat the 1-local measurement on $\nu$ copies of the state, the tradeoff relation under 1-local measurements, with the parametrization such that $F_Q=I$, is then
\begin{equation}\label{eq:tradeoffRLDmain}
    \frac{1}{\nu}Tr[Cov^{-1}(\hat{x})]\leq Tr[F^{RLD}_{Re}]-\frac{1}{4(n-1)}\|C^{RLD}_1\|_F^2.
    %\frac{1}{2(n-1)}\sum_{j,k}|(C^{RLD}_1)_{jk}|.
\end{equation}
When $F_Q\neq I$ initially, we can first make a reparametrization with $\tilde{x}=F_Q^{-\frac{1}{2}}x$.  The tradeoff relation in Eq.(\ref{eq:tradeoffRLDmain}) can then be expressed in the original parametrization as
\begin{eqnarray}\label{eq:tradeoffRLDoriginmain}
\aligned
    &\frac{1}{\nu}Tr[F_Q^{-1}Cov^{-1}(\hat{x})]\\
    \leq &Tr[F_Q^{-1}F^{RLD}_{Re}]-\frac{1}{4(n-1)}\|C^{RLD}_1\|_F^2
    %\frac{1}{2(n-1)} \sum_{j,k}|(C^{RLD}_{1})_{jk}|,
    \endaligned
\end{eqnarray}
with the entries of $C^{RLD}_1$ given by
%\begin{eqnarray}
%\aligned
$(C^{RLD}_1)_{jk}=\min \{\frac{1}{2}\|\sqrt{\rho_x}(\tilde{L}_j^R\tilde{L}_k^{R\dagger}-\tilde{L}_k^R\tilde{L}_j^{R\dagger})\sqrt{\rho_x}\|_1, 2\}$,
%&=\frac{1}{2}\|\sqrt{\rho_x}[\sum_q (F_Q^{-\frac{1}{2}})_{jq}L^{RLD}_q,\sum_q (F_Q^{-\frac{1}{2}})_{kq}L^{RLD}_q]\sqrt{\rho_x}\|_1.
%\endaligned
%\end{eqnarray}
where $\tilde{L}^R_j=\sum_q (F_Q^{-\frac{1}{2}})_{jq}L^R_q$ and $\tilde{L}^R_k=\sum_q (F_Q^{-\frac{1}{2}})_{kq}L^R_q$(see appendix for detail).

%{\color{blue}
For p-local measurements, we can similarly get
\begin{equation}
\aligned
&  \frac{1}{\nu}Tr[F_{Q}^{-1}Cov^{-1}(\hat{x})]\\
&\le Tr[F_{Q}^{-1}F_{Re}^{RLD}]-\frac{1}{4(n-1)}\|\frac{C_p^{RLD}}{p}\|_F^2,
\endaligned
\end{equation}
where $(C_p^{RLD})_{jk}=\min\{\frac{1}{2}\|\sqrt{\rho_x^{\otimes p}}(\tilde{L}_{jp}^R\tilde{L}_{kp}^{R\dagger}-\tilde{L}_{kp}^R\tilde{L}_{jp}^{R\dagger})\sqrt{\rho_x^{\otimes p}}\|_1,2p\}$.

\section{Examples}\label{sec:example}
\subsection{Example 1}
%Consider a state $\rho_x=\frac{1}{2}I+\frac{1}{4}\sigma_3+x_1\sigma_1+x_2\sigma_2+x_3\sigma_3)$.
%{\color{blue}see appendix}

Consider a state $\rho_x=\frac{1}{2}(I+\delta\sigma_3+x_1\sigma_1+x_2\sigma_2+x_3\sigma_3)$, where the true values of the parameters, $x_1,x_2,x_3$ are all $0$ and $|\delta|<1$. The eigenvectors of $\rho_x$ are $|0\rangle$ and $|1\rangle$ with $\rho_x\ket{0}=\frac{1}{2}(1+\delta)\ket{0}$, $\rho_x\ket{1}=\frac{1}{2}(1-\delta)\ket{1}$.
The SLD %and RLD
operators corresponding to the parameters can be easily obtained as
\begin{equation}
  L_1 =\begin{pmatrix}
    0 & 1\\
    1 & 0
  \end{pmatrix},
  L_2 =\begin{pmatrix}
    0 & -i\\
    i & 0
  \end{pmatrix},
  L_3 =\begin{pmatrix}
    \frac{1}{1+\delta} & 0\\
    0 & \frac{-1}{1-\delta}
  \end{pmatrix},
\end{equation}
%\begin{equation}
%  L^R_1 =\begin{pmatrix}
%    0 & \frac{1}{1+\delta}\\
%    \frac{1}{1-\delta} & 0
%  \end{pmatrix},
%  L^R_2 =\begin{pmatrix}
%    0 & \frac{-i}{1+\delta}\\
%    \frac{i}{1-\delta} & 0
%  \end{pmatrix},
%  L^R_3 =\begin{pmatrix}
%    \frac{1}{1+\delta} & 0\\
%    0 & \frac{-1}{1-\delta}
%  \end{pmatrix}.
%\end{equation}
%And The QFI matrix $F_Q$ and the real part of the RLD matrix $F^{RLD}_{Re}$ can be computed as
from which we can get the QFIM
\begin{equation}
  F_Q=\begin{pmatrix}
    1 & 0 & 0\\
    0 & 1 & 0\\
    0 & 0 & \frac{1}{1-\delta^2}
  \end{pmatrix}.
 % F^{RLD}_{Re}=\frac{1}{1-\delta^2}\begin{pmatrix}
 %   1 & 0 & 0\\
 %   0 & 1 & 0\\
 %   0 & 0 & 1
 % \end{pmatrix}
\end{equation}
 The SLD under the reparametrization $\tilde{x}=F_Q^{\frac{1}{2}}x$ are given by%,  and RLD operators under the reparametrization is then given by
 \begin{equation}
  \tilde{L}_1 =\begin{pmatrix}
    0 & 1\\
    1 & 0
  \end{pmatrix},
  \tilde{L}_2 =\begin{pmatrix}
    0 & -i\\
    i & 0
  \end{pmatrix},
  \tilde{L}_3 =\begin{pmatrix}
    \sqrt{\frac{1-\delta}{1+\delta}} & 0\\
    0 & -\sqrt{\frac{1+\delta}{1-\delta}}
  \end{pmatrix}.
\end{equation}
%\begin{equation}
%  \tilde{L}^R_1 =\begin{pmatrix}
%    0 & \frac{1}{1+\delta}\\
%    \frac{1}{1-\delta} & 0
%  \end{pmatrix},
%  \tilde{L}^R_2 =\begin{pmatrix}
%    0 & \frac{-i}{1+\delta}\\
%    \frac{i}{1-\delta} & 0
%  \end{pmatrix},
%  \tilde{L}^R_3 =\begin{pmatrix}
%    \sqrt{\frac{1-\delta}{1+\delta}} & 0\\
%    0 & -\sqrt{\frac{1+\delta}{1-\delta}}
%  \end{pmatrix}.
%\end{equation}
From $(C_1)_{jk}=\frac{1}{2}\|\sqrt{\rho_x}[\tilde{L}_j,\tilde{L}_k]\sqrt{\rho_x}\|_1$ we have
\begin{equation}
  C_1=\begin{pmatrix}
    0 & 1 & 1\\
    1 & 0 & 1\\
    1 & 1 & 0
  \end{pmatrix},
%  C^{RLD}_1=\begin{pmatrix}
%    0 & \min\{\frac{1}{1-\delta^2},2\} & 1\\
%    \min\{\frac{1}{1-\delta^2},2\} & 0 & 1\\
%    1 & 1 & 0
%  \end{pmatrix},
\end{equation}
%where $C_1^{RLD}$ is obtained from $(C_1^{RLD})_{jk}=\min\{\frac{1}{2}\|\sqrt{\rho_x}(\tilde{L}_{j}^R\tilde{L}_{k}^{R\dagger}-\tilde{L}_{k}^R\tilde{L}_{j}^{R\dagger})\sqrt{\rho_x}\|_1,2\}$.
%The Frobenius norm of them can be easily computed as $\|C_1\|_F=\sqrt{6}$ and
%\begin{equation}
%  \|C_1^{RLD}\|_F=\sqrt{4+\min\{\frac{2}{(1-\delta^2)^2},8\}}=\min\{\sqrt{4+\frac{2}{(1-\delta^2)^2}},2\sqrt{3}\}，
%\end{equation}
which gives the tradeoff relation under the 1-local measurement as
\begin{equation}\label{eq:qubitC1}
  \frac{1}{\nu}Tr[F_Q^{-1}Cov(\hat{x})^{-1}]\le n-\frac{1}{4(n-1)}\|C_1\|_F^2=\frac{9}{4}.
\end{equation}

With
\begin{eqnarray}
\aligned
    (T_1)_{jk}&=\frac{1}{2}\{\frac{1+\delta}{2}|\langle 0|[\tilde{L}_j,\tilde{L}_k]|0\rangle|+\frac{1-\delta}{2}|\langle 1|[\tilde{L}_j,\tilde{L}_k]|1\rangle\},
        \endaligned
\end{eqnarray}
we can obtain the bound with $T_1$ as
%{\color{purple}
\begin{equation}
  \frac{1}{\nu}Tr[F_Q^{-1}Cov(\hat{x})^{-1}]\le n-\frac{1}{4(n-1)}\|T_1\|_F^2=\frac{11}{4}.
\end{equation}

%\begin{equation}
%  \begin{aligned}
%    \frac{1}{\nu}Tr[F_{Q}^{-1}Cov^{-1}(\hat{x})]& \le Tr[F_{Q}^{-1}F_{Re}^{RLD}]-\frac{1}{4(n-1)}\|C_1^{RLD}\|_F^2\\
%    &=\frac{3-\delta^2}{1-\delta^2}-\frac{1}{8}\min\{4+\frac{2}{(1-\delta^2)^2},12\}\\
%    &=\frac{5-\delta^2}{2(1-\delta^2)}-\frac{1}{4}\min\{\frac{1}{(1-\delta^2)^2},4\},
%  \end{aligned}
%\end{equation}
%where the bound given by RLD operators is the tightest when $\delta=0$, which gives the same result as the SLD bound.
If we choose a set of $\{\ket{u_q}\}$ as $\ket{u_0}=\begin{pmatrix} 1\\0 \end{pmatrix}$, $\ket{u_1}=\begin{pmatrix} 0\\1 \end{pmatrix}$, which satisfies $|u_0\rangle\langle u_0|+|u_1\rangle\langle u_1|=I$, from $(F_{u_q})_{jk}=\langle u_q |\sqrt{\rho_x}\tilde{L}_j\tilde{L}_k\sqrt{\rho_x}|u_q\rangle$ we can obtain
\begin{equation}
  \begin{aligned}
    F_{u_0}=\frac{1}{2}
    \begin{pmatrix}
      1+\delta & i(1+\delta) & 0\\
      -i(1+\delta) & 1+\delta & 0\\
      0 & 0 & 1-\delta
    \end{pmatrix},\\
    F_{u_1}=\frac{1}{2}
    \begin{pmatrix}
      1-\delta & -i(1-\delta) & 0\\
      i(1-\delta) & 1-\delta & 0\\
      0 & 0 & 1+\delta
    \end{pmatrix}.
  \end{aligned}
\end{equation}

% \begin{equation}
%   F_{u_0}^{RLD}=\frac{1}{2}
%   \begin{pmatrix}
%     1-\delta & i(1-\delta) & 0\\
%     -i(1-\delta) & 1-\delta & 0\\
%     0 & 0 & 1-\delta
%   \end{pmatrix},
%   F_{u_1}^{RLD}=\frac{1}{2}
%   \begin{pmatrix}
%     1+\delta & -i(1+\delta) & 0\\
%     i(1+\delta) & 1+\delta & 0\\
%     0 & 0 & 1+\delta
%   \end{pmatrix}.
% \end{equation}
We can choose $\mathbf{\bar{F}}=F_{u_0}+F_{u_1}^T$ whose imaginary part is  %and $\bar{F}^{RLD}=F_{u_0}^{RLD}+(F_{u_1}^{RLD})^T$,
\begin{equation}
  \bold{\bar{F}}_{Im}=\begin{pmatrix}
    0 & 1 & 0\\
    -1 & 0 & 0\\
    0 & 0 & 0
  \end{pmatrix},
\end{equation}
the tradeoff relation in Eq.(\ref{eq:mainImproved}) then gives
\begin{equation}
  \frac{1}{\nu}Tr[F_Q^{-1}Cov(\hat{x})^{-1}]\le n-\frac{(n-2)}{(n-1)^2}\|\bold{\bar{F}}_{Im}\|_F^2=\frac{5}{2}.
\end{equation}
%which is less tighter than Eq.(\ref{eq:qubitC1}).

For 2-local measurement, using $(C_2)_{jk}=\frac{1}{2}\|\sqrt{\rho_x^{\otimes 2}}[\tilde{L}_{j2},\tilde{L}_{k2}]\sqrt{\rho_x^{\otimes 2}}\|_1$ with $\tilde{L}_{j2}=\tilde{L}_{j}\otimes I+I\otimes\tilde{L}_{j}$, we can obtain
\begin{equation}
  \begin{aligned}
    C_2=&\begin{pmatrix}
      0 & 1+\delta^2 & \sqrt{1+\delta^2}\\
      1+\delta^2 & 0 & \sqrt{1+\delta^2}\\
      \sqrt{1+\delta^2} & \sqrt{1+\delta^2} & 0
    \end{pmatrix},%\\
%    C_2^{RLD}=&\begin{pmatrix}
%      0 & \min\{4,\frac{1-\delta^2+2|\delta|}{1-\delta^2}\} & \sqrt{1+4\delta^2}\\
%      \min\{4,\frac{1-\delta^2+2|\delta|}{1-\delta^2}\} & 0 & \sqrt{1+4\delta^2}\\
%      \sqrt{1+4\delta^2} & \sqrt{1+4\delta^2} & 0
%    \end{pmatrix},
  \end{aligned}
\end{equation}
which gives the tradeoff relation
\begin{eqnarray}\label{eq:qubitC2}
\aligned
  \frac{1}{\nu}Tr[F_Q^{-1}Cov(\hat{x})^{-1}]&\le n-\frac{1}{4(n-1)}\|\frac{C_2}{2}\|_F^2\\
  &=\frac{45}{16}-\frac{1}{4}\delta^2-\frac{1}{16}\delta^4.
\endaligned
\end{eqnarray}
%\begin{equation}
%  \begin{aligned}
%    \frac{1}{\nu}Tr[F_{Q}^{-1}Cov^{-1}(\hat{x})]&%\le Tr[F_{Q}^{-1}F_{Re}^{RLD}]-\frac{1}{4(n-1)}\|%\frac{C_2^{RLD}}{2}\|_F^2\\
%    &=\frac{7}{8}-\frac{\delta^2}{2}+\frac{2}{1-\%delta^2}-\frac{1}{16}\min\{16,\left(\frac{1-\delt%a^2+2|\delta|}{1-\delta^2}\right)^2\}
%  \end{aligned}
%\end{equation}
%In some regions, the bound given by RLD operators is tighter than the SLD bound.
%For example, when $\delta=0.15$, $\Gamma_2^S\approx 2.807$, $\Gamma_2^R\approx 2.803$, while for $\delta=0.3$, $\Gamma_2^S\approx 2.789$, $\Gamma_2^R\approx 2.856$.
From
\begin{equation}
  \begin{aligned}
    (T_2)_{jk}=&\frac{1}{2}\left(\frac{1+\delta}{2}\right)^2 |\langle 0|[\tilde{L}_j,\tilde{L}_k]|0\rangle + \langle 0|[\tilde{L}_j,\tilde{L}_k]|0\rangle|\\
    &+\frac{1+\delta}{2}\frac{1-\delta}{2}|\langle 0|[\tilde{L}_j,\tilde{L}_k]|0\rangle + \langle 1|[\tilde{L}_j,\tilde{L}_k]|1\rangle\}\\
    &+\frac{1}{2}\left(\frac{1-\delta}{2}\right)^2 |\langle 1|[\tilde{L}_j,\tilde{L}_k]|1\rangle + \langle 1|[\tilde{L}_j,\tilde{L}_k]|1\rangle| ,
  \end{aligned}
\end{equation}
we have
\begin{equation}
    T_2=\begin{pmatrix}
      0 & 1+\delta^2 & 0\\
      1+\delta^2 & 0 & 0\\
      0 & 0 & 0
    \end{pmatrix},
\end{equation}
which gives the tradeoff relation with $T_2$ as
\begin{equation}
  \frac{1}{\nu}Tr[F_Q^{-1}Cov(\hat{x})^{-1}]\le n-\frac{1}{4(n-1)}\|\frac{T_2}{2}\|_F^2=\frac{47}{16}-\frac{\delta^2}{8}-\frac{\delta^4}{16}.
\end{equation}

% \begin{equation}
%   \frac{1}{\nu}Tr[(F^{RLD}_{Re})^{-1}Cov(\hat{x})^{-1}]\le n-\frac{(n-2)}{(n-1)^2}\|\bar{F}_{Im}^{RLD}\|_F^2=\frac{5}{2},
% \end{equation}

If we choose a set of $\{\ket{u_q}\}$ in the two-qubit space as $\ket{u_0}=\ket{00}$, $\ket{u_1}=\ket{01}$, $\ket{u_2}=\ket{10}$, $\ket{u_3}=\ket{11}$, we can obtain %the matrix $F_{u_q}$ are given as
\begin{equation}
  \begin{aligned}
    F_{u_0}&=\frac{1}{2}
    \begin{pmatrix}
      (1+\delta)^2 & i(1+\delta)^2 & 0\\
      -i(1+\delta)^2 & (1+\delta)^2 & 0\\
      0 & 0 & 2(1-\delta^2)
    \end{pmatrix},\\
    F_{u_1}&=\frac{1}{2}
    \begin{pmatrix}
      1-\delta^2 & 0 & 0\\
      0 & 1-\delta^2 & 0\\
      0 & 0 & 2\delta^2
    \end{pmatrix},\\
    F_{u_2}&=\frac{1}{2}
    \begin{pmatrix}
      1-\delta^2 & 0 & 0\\
      0 & 1-\delta^2 & 0\\
      0 & 0 & 2\delta^2
    \end{pmatrix},\\
    F_{u_3}&=\frac{1}{2}
    \begin{pmatrix}
      (1-\delta)^2 & -i(1-\delta)^2 & 0\\
      i(1-\delta)^2 & (1-\delta)^2 & 0\\
      0 & 0 & 2(1-\delta^2)
    \end{pmatrix},
  \end{aligned}
\end{equation}
where the entries of $F_{u_q}$ are obtained as $(F_{u_q})_{jk}=\langle u_q |\sqrt{\rho_x^{\otimes 2}}\tilde{L}_{j2}\tilde{L}_{k2}\sqrt{\rho_x^{\otimes 2}}|u_q\rangle$.
% \begin{equation}
%   \begin{aligned}
%     F_{u_0}^{RLD}&=\frac{1}{2}
%     \begin{pmatrix}
%       1-\delta^2 & i(1-\delta^2) & 0\\
%       -i(1-\delta^2) & 1-\delta^2 & 0\\
%       0 & 0 & 2(1-\delta^2)
%     \end{pmatrix},
%     F_{u_1}^{RLD}=\frac{1}{2}
%     \begin{pmatrix}
%       1+\delta^2 & -2i\delta & 0\\
%       2i\delta & 1+\delta^2 & 0\\
%       0 & 0 & 2\delta^2
%     \end{pmatrix},\\
%     F_{u_2}^{RLD}&=\frac{1}{2}
%     \begin{pmatrix}
%       1+\delta^2 & -2i\delta & 0\\
%       2i\delta & 1+\delta^2 & 0\\
%       0 & 0 & 2\delta^2
%     \end{pmatrix},
%     F_{u_3}^{RLD}=\frac{1}{2}
%     \begin{pmatrix}
%       1-\delta^2 & -i(1-\delta^2) & 0\\
%       i(1-\delta^2) & 1-\delta^2 & 0\\
%       0 & 0 & 2(1-\delta^2)
%     \end{pmatrix}
%   \end{aligned}
% \end{equation}
%The optimal $\bar{F}_{Im}$ is then given by the imaginary part of
Let $\bold{\bar{F}}=F_{u_0}+F_{u_1}+F_{u_2}+F_{u_3}^T$, % and $\bar{F}^{RLD}=F_{u_0}^{RLD}+(F_{u_1}^{RLD})^T+(F_{u_2}^{RLD})^T+(F_{u_3}^{RLD})^T$,
which has the imaginary part as
\begin{equation}
  \begin{aligned}
    \bold{\bar{F}}_{Im2}=\begin{pmatrix}
      0 & 1+\delta^2 & 0\\
      -(1+\delta^2) & 0 & 0\\
      0 & 0 & 0
    \end{pmatrix}.
    % \bar{F}_{Im2}^{RLD}=\begin{pmatrix}
    %   0 & 1+2\delta-\delta^2 & 0\\
    %   -(1+2\delta-\delta^2) & 0 & 0\\
    %   0 & 0 & 0
    % \end{pmatrix},
  \end{aligned}
\end{equation}
%which gives the Frobenius norm as $\|\bar{F}_{Im2}\|_F=\sqrt{2}(2-\delta+\delta^2)$. %$\bar{F}_{Im2}^{RLD}=\sqrt{2}(1+2\delta-\delta^2)$.
This gives the tradeoff relation
\begin{equation}
\aligned
  \frac{1}{\nu}Tr[F_Q^{-1}Cov(\hat{x})^{-1}]&\le n-\frac{(n-2)}{(n-1)^2}\|\frac{\bold{\bar{F}}_{Im2}}{2}\|_F^2\\
  &=3-\frac{1}{8}(1+\delta^2)^2.
  \endaligned
\end{equation}
%which is less tighter than Eq.(\ref{eq:qubitC2}).

%{\color{purple}

%which is generally tighter than the bound given by $\Gamma_2^S$ and $\Gamma_2^R$.
% \begin{equation}
%   \frac{1}{\nu}Tr[(F^{RLD}_{Re})^{-1}Cov(\hat{x})^{-1}]\le n-\frac{(n-2)}{(n-1)^2}\|\frac{\bar{F}_{Im2}^{RLD}}{2}\|_F^2=\frac{23}{8}-\frac{1}{2}\delta(1-\frac{1}{2}\delta)(1+\delta-\frac{1}{2}\delta^2).
% \end{equation}

%The results above can be summarized in Table.\ref{tb:d=2full} for $\delta=0.15$ and $\delta=0.3$.

%\begin{table}[ht]\label{tb:d=2full}
%  \caption{Tradeoff relations for $p=1,2$}
%  \centering
%  \begin{tabular}{c@{\hskip 0.3in} c@{\hskip 0.3in} c@{\hskip 0.3in} c@{\hskip 0.3in} c@{\hskip 0.3in} c@{\hskip 0.3in} c@{\hskip 0.3in}}
%  \hline\hline
%   & $\Gamma_1^S$ & $\Gamma_1^R$ & $\bar{\Gamma}_1^S$
%   & $\Gamma_2^S$ & $\Gamma_2^R$ & $\bar{\Gamma}_2^S$  \\ [0.5ex]
%  \hline
%  $\delta=0.15$ & 2.25 & 2.28 & 2.5 & 2.81 & 2.80 & 2.56 \\
%  $\delta=0.3$  & 2.25 & 2.40 & 2.5 & 2.79 & 2.86 & 2.60 \\
%  \hline
%  \end{tabular}
%\end{table}
%When $p$ gets large, the computation gets hard. However,
When $\delta=0$, i.e., $\rho_x=\frac{1}{2}(I+x_1\sigma_1+x_2\sigma_2+x_3\sigma_3)$, the tradeoff relations can be analytically calculated under general $p$-local measurement.
%In this case the SLD bound the RLD bound equals to each other.
In this case the SLD operators under the reparametrization is given by $\tilde{L}_1=\sigma_1$, $\tilde{L}_2=\sigma_2$, $\tilde{L}_3=\sigma_3$, thus %We first compute $(C_p)_{12}$, which is given by
\begin{eqnarray}
\aligned
(C_p)_{12}&=  \frac{1}{2}\|\sqrt{\rho_x^{\otimes p}}[\tilde{L}_{1p},\tilde{L}_{2p}]\sqrt{\rho_x^{\otimes p}}\|_1\\
&=\frac{1}{2}\|\sqrt{\rho_x^{\otimes p}}[\sigma_{1p},\sigma_{2p}]\sqrt{\rho_x^{\otimes p}}\|_1\\
  &=\frac{1}{2^{p}}\|\sigma_{3p}\|_1,
  \endaligned
\end{eqnarray}
where $\sigma_{lp}=\sum_{r=1}^p\sigma_l^{(r)}$ for $l\in\{1,2,3\}$. As the eigenvalues of $\sigma_{lp}$ are  $-p+2s$ with multiplicity $\binom{p}{s}$, here $s=0,1,...,p$, thus %We can define the norm \mathcal{N}_p&=
\begin{equation}
  \begin{aligned}
    \|\sigma_{3p}\|_1&=\sum_{s=0}^p \binom{p}{s} |-p+2s|\\
    &=2\sum_{s=0}^{\left\lfloor{\frac{p}{2}}\right\rfloor} \binom{p}{s} (p-2s)=
    \begin{cases}
      2p\binom{p-1}{\frac{p-1}{2}}, &\text{if}\ p\ \text{is odd},\\
      p\binom{p}{\frac{p}{2}}, &\text{if}\ p\ \text{is even}.
    \end{cases}
  \end{aligned}
\end{equation}
Due to the symmetry, $(C_p)_{jk}$ takes the same value for all $j\neq k\in\{1,2,3\}$. %can then be written as
%\begin{equation}
%  \begin{aligned}
%    (C_p)_{jk}&=\frac{1}{2}\|\sqrt{\rho_x^{\otimes p}}[\tilde{L}_{jp},\tilde{L}_{kp}]\sqrt{\rho_x^{\otimes p}}\|_1=\frac{1}{2^p}\|\sigma_{lp}\|_1=\frac{\mathcal{N}_p}{2^p}.
%  \end{aligned}
%\end{equation}
The tradeoff relation under the $p$-local measurement is then given by
\begin{equation}
  \begin{aligned}
    \Gamma_p=\frac{1}{\nu}Tr[F_Q^{-1}Cov(\hat{x})^{-1}]&\le n-\frac{1}{4(n-1)}\|\frac{C_p}{p}\|_F^2\\
   % &= 3-\frac{1}{8}\|\frac{C_p}{p}\|_F^2\\
    &=3-\frac{3}{4}\left(\frac{\mathcal{N}_p}{p}\right)^2,
  \end{aligned}
\end{equation}
where $\mathcal{N}_p=\frac{1}{2^{p}}\|\sigma_{3p}\|_1$.

\begin{figure}
  \includegraphics[width=0.5\textwidth]{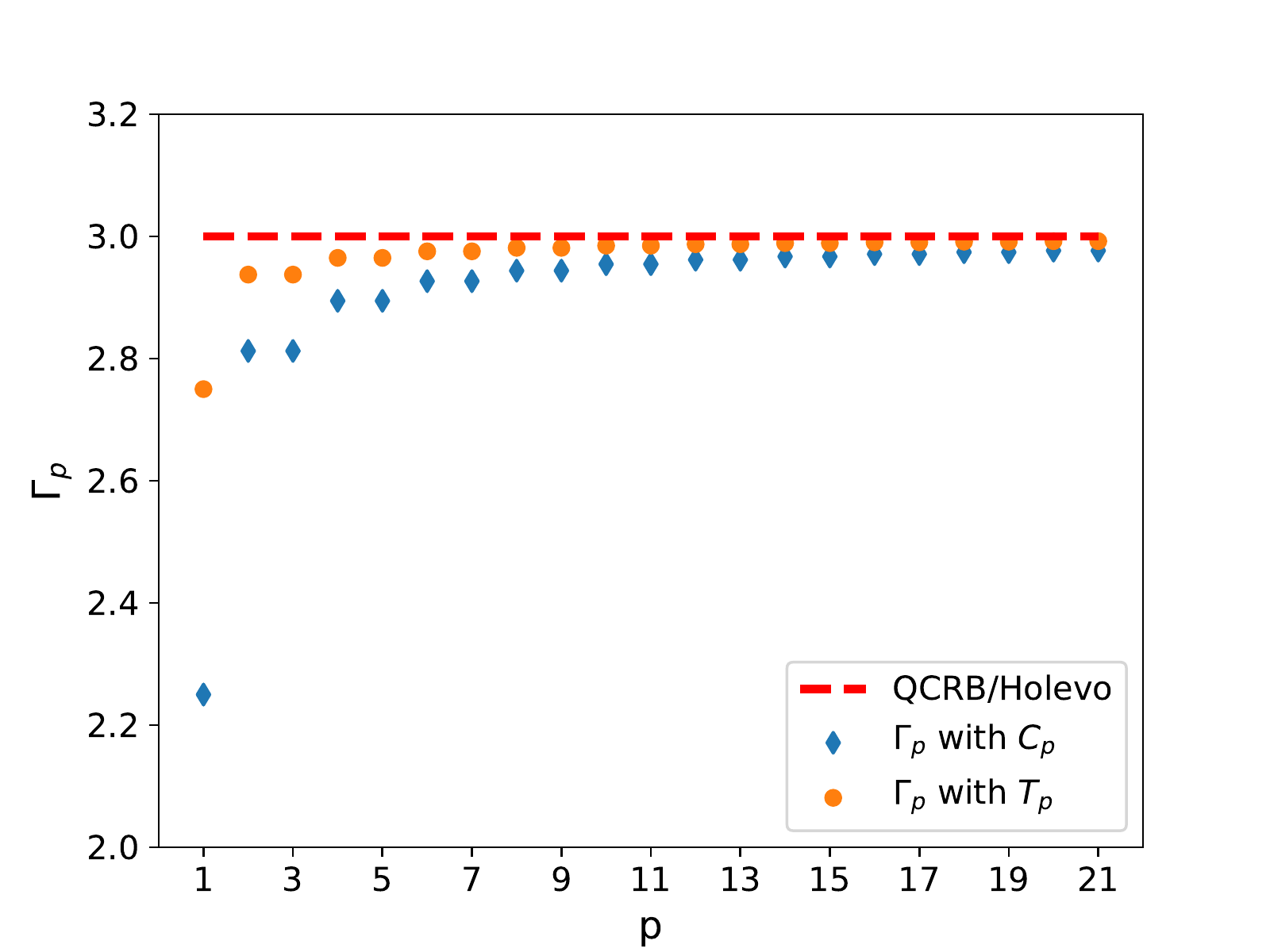}
  \caption{Upper bounds on $\Gamma_p$ obtained with $C_p$ and $T_p$, together with the QCRB/Holevo bound at the case $\delta=0$.}
  \label{fig:mainexample1}
\end{figure}

FOr the bound with $T_p$, we have %can be analytical obtained with general $\delta$. We have
\begin{equation}
    \begin{aligned}
      (T_p)_{12} &= \frac{1}{2}\sum_{s=0}^p \binom{p}{s} \left(\frac{1+\delta}{2}\right)^s\left(\frac{1-\delta}{2}\right)^{p-s}|2s-2(p-s)|\\
      &=\frac{1}{2^p}\sum_{s=0}^p \binom{p}{s} \left(1+\delta\right)^s\left(1-\delta\right)^{p-s}|2s-p|,\\
%      (T_p)_{13} &= 0\\
%      (T_p)_{23} &= 0.
    \end{aligned}
\end{equation}
and $(T_p)_{13} =(T_p)_{23}= 0$, thus
\begin{eqnarray}
\aligned
  \frac{1}{\nu}Tr[F_Q^{-1}Cov^{-1}(\hat{x})]&\le n-\frac{1}{4(n-1)}\|\frac{T_p}{p}\|_F^2\\
  &=3-\frac{1}{4p^2}(T_p)_{12}^2.
  \endaligned
\end{eqnarray}
%For each $p$, we can plot the corresponding precision bound in the same figure as
In Fig.\ref{fig:mainexample1} we plot the bounds as a function of $p$ in the case of $\delta=0$. Note that in this case the weak commutative condition holds, the Holevo bound equals to the QCRB, which is achievable when $p\rightarrow \infty$. For any finite $p$, however, the bounds are strictly less than 3, thus any collective measurement on finite copies can not saturate the Holevo bound which is only achievable with the collective measurement on genuinely infinite copies of states. It can also be seen that the difference between the bounds obtained from $C_p$ and $T_p$ is large for small $p$, but the difference decreases with $p$. %the Gill-Mazzar bound can only tell that the Holevo can not be achieved by 1-local measurement and Zhu-Hayashi bound only tells that the

\begin{figure}
  \includegraphics[width=0.5\textwidth]{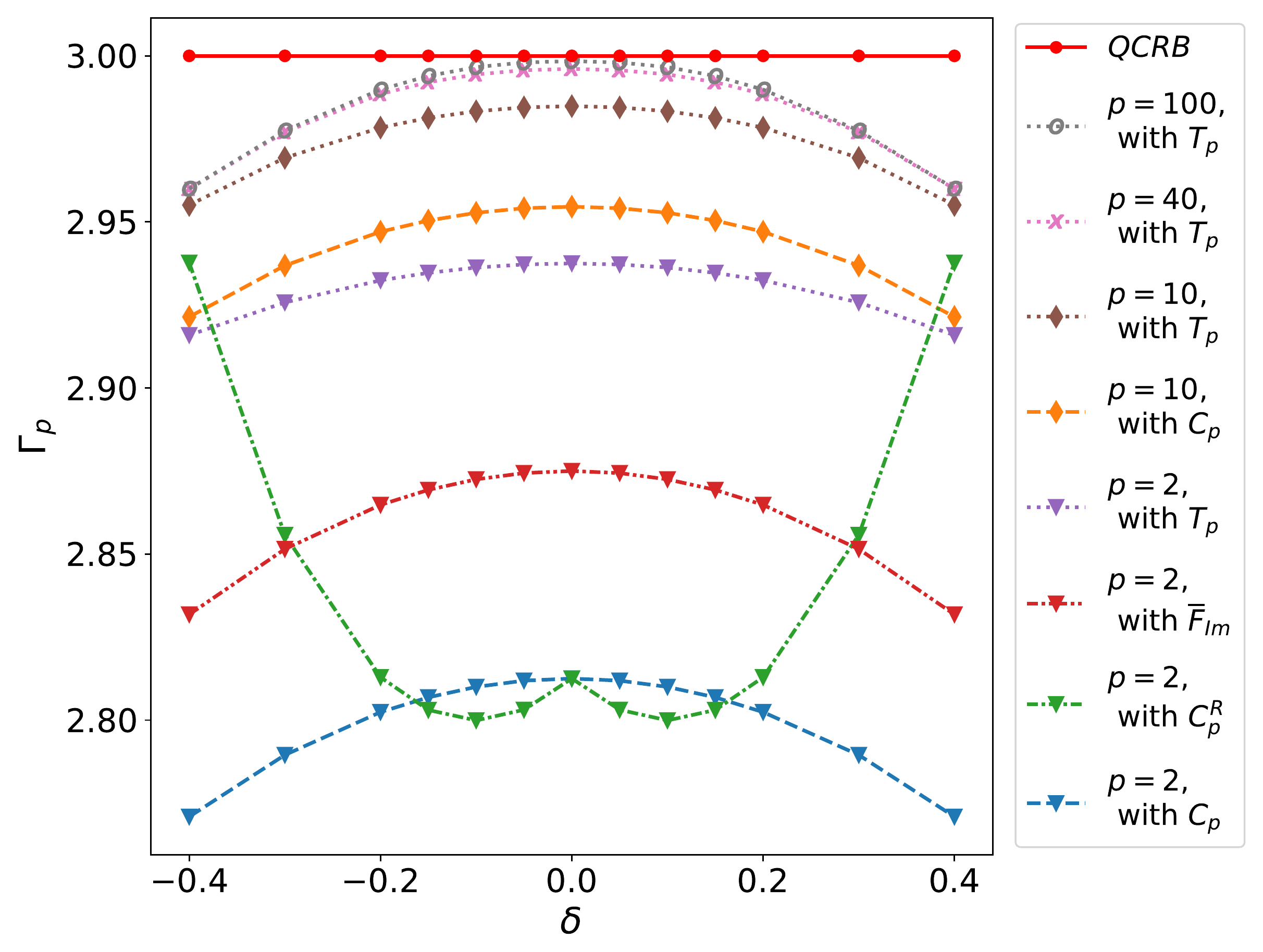}
  \caption{Comparison of different bounds of $\Gamma_p$ for the estimation of $\rho_x=\frac{1}{2}(I+\delta\sigma_3+x_1\sigma_1+x_2\sigma_2+x_3\sigma_3)$ at $x_1=x_2=x_3=0$.}
  \label{fig.mainexample1_delta}
\end{figure}

We also plot the bounds for the state $\rho_x=\frac{1}{2}(I+\delta\sigma_3+x_1\sigma_1+x_2\sigma_2+x_3\sigma_3)$ with general $\delta$ in Fig. \ref{fig.mainexample1_delta}. The complexity of calculating the bound with $C_p$, which we compute up to $p=10$, increases exponentially with $p$. As a comparison, the bound with $T_p$ is much easier to compute, which we compute up to $p=100$. Since the difference between these two bounds decreases with $p$, a good strategy is to use the bound with $C_p$ for small $p$ and use the bound with $T_p$ for large $p$. We also plot the bound with the RLD for $p=2$, it can be seen that the RLD bound can be either tighter or less tight than the bound with $C_p$. We can combine these bounds and choose the minimal of them to get a tighter bound.

\subsection{Example 2}
We consider another example with a three dimensional state, $\rho_x=\frac{1}{3}I+\sum_jx_jG_j$, where $G_j=\frac{1}{2}\Lambda_j$, here $\{\Lambda_j\}_{j=1}^8$ are the Gell-Mann matrices,
\begin{equation}
  % \begingroup
  % \renewcommand{\arraystretch}{0.8}
  % \setlength\arraycolsep{6pt}
  \begin{aligned}
    &\Lambda_1=\begin{pmatrix}
      0 & 1 & 0\\
      1 & 0 & 0\\
      0 & 0 & 0
    \end{pmatrix},
    \Lambda_2=\begin{pmatrix}
      0 & -i & 0\\
      i & 0 & 0\\
      0 & 0 & 0
    \end{pmatrix},
    \Lambda_3=\begin{pmatrix}
      1 & 0 & 0\\
      0 & -1 & 0\\
      0 & 0 & 0
    \end{pmatrix},\\
    &\Lambda_4=\begin{pmatrix}
      0 & 0 & 1\\
      0 & 0 & 0\\
      1 & 0 & 0
    \end{pmatrix},
    \Lambda_5=\begin{pmatrix}
      0 & 0 & -i\\
      0 & 0 & 0\\
      i & 0 & 0
    \end{pmatrix},
    \Lambda_6=\begin{pmatrix}
      0 & 0 & 0\\
      0 & 0 & 1\\
      0 & 1 & 0
    \end{pmatrix},\\
    &\Lambda_7=\begin{pmatrix}
      0 & 0 & 0\\
      0 & 0 & -i\\
      0 & i & 0
    \end{pmatrix},
    \Lambda_8=\frac{1}{\sqrt{3}}\begin{pmatrix}
      1 & 0 & 0\\
      0 & 1 & 0\\
      0 & 0 & -2
    \end{pmatrix},
  \end{aligned}
  %\endgroup
\end{equation}
which form a basis for 3 by 3 Hermitian matrices.
When the true values of the parameters are all 0, the SLDs can be obtained as $L_j=3G_j$, %(note that in this case the RLDs are the same as the SLDs, thus the tradeoff relations from the SLDs and RLDs are the same).
and $F_Q=\frac{3}{2}I$. The SLDs after the reparametrization which makes $\tilde{F}_Q=I$ are given by $\tilde{L}_j=\sqrt{\frac{2}{3}}L_j=\sqrt{6}G_j$.
Since
\begin{equation}
  (C_1)_{jk}=\frac{1}{2}\|\sqrt{\rho_x}[\tilde{L}_j,\tilde{L}_k]\sqrt{\rho_x}\|_1=\|[G_j,G_k]\|_1,
\end{equation}
we have
\begin{equation}
  C_1=
  \begin{pmatrix}
    0 & 1 & 1 & \frac{1}{2} & \frac{1}{2} & \frac{1}{2} & \frac{1}{2} & 0\\
    1 & 0 & 1 & \frac{1}{2} & \frac{1}{2} & \frac{1}{2} & \frac{1}{2} & 0\\
    1 & 1 & 0 & \frac{1}{2} & \frac{1}{2} & \frac{1}{2} & \frac{1}{2} & 0\\
    \frac{1}{2} & \frac{1}{2} & \frac{1}{2} & 0 & 1 & \frac{1}{2} & \frac{1}{2} & \frac{\sqrt{3}}{2}\\
    \frac{1}{2} & \frac{1}{2} & \frac{1}{2} & 1 & 0 & \frac{1}{2} & \frac{1}{2} & \frac{\sqrt{3}}{2}\\
    \frac{1}{2} & \frac{1}{2} & \frac{1}{2} & \frac{1}{2} & \frac{1}{2} & 0 & 1 & \frac{\sqrt{3}}{2}\\
    \frac{1}{2} & \frac{1}{2} & \frac{1}{2} & \frac{1}{2} & \frac{1}{2} & 1 & 0 & \frac{\sqrt{3}}{2}\\
    0 & 0 & 0 & \frac{\sqrt{3}}{2} & \frac{\sqrt{3}}{2} & \frac{\sqrt{3}}{2} & \frac{\sqrt{3}}{2} & 0
  \end{pmatrix}.
\end{equation}
% For $p\ge 2$, it can be computed that $C_2=\frac{4}{3}C_1$ and $C_3=\frac{5}{3}C_1$.
This gives the tradeoff relation under the 1-local measurement as
\begin{equation}
  \begin{aligned}
    \frac{1}{\nu}Tr[F_Q^{-1}Cov^{-1}(\hat{x})]&\le n-\frac{1}{4(n-1)}\|C_1\|_F^2\\
    &%= 8-\frac{1}{28}\times 2(5\times 1 +4\times\frac{3}{4}+16 \times \frac{1}{4})
    =\frac{50}{7}\approx 7.14.
  \end{aligned}
\end{equation}

For $p$-local measurement, we can similarly obtain
\begin{equation}
  (C_p)_{jk}=\frac{1}{2}\|\sqrt{\rho_x^{\otimes p}}[\tilde{L}_{jp},\tilde{L}_{kp}]\sqrt{\rho_x^{\otimes p}}\|_1=\frac{1}{3^{p-1}}\|[G_{jp},G_{kp}]\|_1,
\end{equation}
where $[G_{jp},G_{kp}]=\sum_{r=1}^p[G_j^{(r)},G_k^{(r)}]$. Since the eigenvalues of $[G_{j},G_{k}]$ are $\{-\lambda,0,\lambda\}$, where $\lambda=\frac{1}{2}(C_1)_{jk}$, %$ or $\frac{1}{4}$ or $\frac{\sqrt{3}}{4}$ for different $j,k$.
the eigenvalues of $[G_{jp},G_{kp}]$ are given by $\lambda s$ with multiplicity $\binom{p}{s}_2$, for $s=-p,-p+1,\cdots,p$, here
%\begin{equation}
  $\binom{p}{s}_2=\sum_{i=0}^p (-1)^i\binom{p}{i}\binom{2p-2i}{p-s-i}$
%\end{equation}
is the trinomial coefficient, which can be obtained as the $(j+p)$-th coefficient of the polynomial $(1+x+x^2)^p$
(see appendix for details). We thus have
\begin{eqnarray}
\aligned
  \|[G_{jp},G_{kp}]\|_1&=\sum_{s=-p}^p |\lambda s|\binom{p}{s}_2=2\lambda\sum_{s=0}^p s\binom{p}{s}_2\\
  &=(C_1)_{jk}\sum_{s=0}^p s\binom{p}{s}_2,
  \endaligned
\end{eqnarray}
where we have used the fact that $\binom{p}{s}_2=\binom{p}{-s}_2$.
Denote $\mathcal{N}_p=\sum_{s=0}^p s\binom{p}{s}_2$, we then have
\begin{equation}
  (C_p)_{jk}=\frac{1}{3^{p-1}}\|[G_{jp},G_{kp}]\|_1=(C_1)_{jk}\frac{\mathcal{N}_p}{3^{p-1}},
\end{equation}
which gives the Frobenius norm of $C_p$ as
\begin{equation}
  \begin{aligned}
    \|C_p\|_F&=\sqrt{\sum_{jk}(C_p)_{jk}^2}=\sqrt{\sum_{jk}\left((C_1)_{jk}\frac{1}{3^{p-1}}\mathcal{N}_p\right)^2}\\
    &=\frac{1}{3^{p-1}}\mathcal{N}_p\sqrt{\sum_{jk}\left((C_1)_{jk}\right)^2}=\frac{1}{3^{p-1}}\mathcal{N}_p\|C_1\|_F.
  \end{aligned}
\end{equation}
The tradeoff relation under the $p$-local measurement is then given by
\begin{equation}
  \begin{aligned}
    \frac{1}{\nu}Tr[F_Q^{-1}Cov^{-1}(\hat{x})]&\le n-\frac{1}{4(n-1)}\|\frac{C_p}{p}\|_F^2\\
    &= n-\frac{1}{4(n-1)}\|C_1\|_F^2\left(\frac{1}{p3^{p-1}}\mathcal{N}_p\right)^2\\
    &= 8-\frac{6}{7}\left(\frac{1}{p3^{p-1}}\mathcal{N}_p\right)^2.
  \end{aligned}
\end{equation}
%For $p=2$, this gives $\frac{1}{\nu}Tr[F_Q^{-1}Cov(\hat{x})^{-1}]\leq \frac{160}{21}\approx 7.62$ and for $p=3$, this gives $\frac{1}{\nu}Tr[F_Q^{-1}Cov(\hat{x})^{-1}]\leq \frac{1462}{189}\approx 7.74$, and
Here $\frac{1}{p3^{p-1}}\mathcal{N}_p$ monotonically decreases with $p$ and it is only equal to $0$ when $p\rightarrow \infty$. The Holevo bound, which equals to the QCRB in this case since the weak commutative condition holds, can thus only be achieved with collective measurement on genuinely infinite number of quantum states in this case. %i.e.,  $\frac{1}{\nu}Tr[F_Q^{-1}Cov^{-1}(\hat{x})]\leq 8$ when $p\rightarrow \infty$.

If there are only three parameters, for example, $\{x_1,x_2,x_5\}$,
the associated matrices are given by the $3\times 3$ submatrices of the original ones. Under the 1-local measurement we have
\begin{equation}
  \begin{aligned}
    C_1=
    \begin{pmatrix}
      0 & 1 & \frac{1}{2}\\
      1 & 0 & \frac{1}{2}\\
      \frac{1}{2} & \frac{1}{2} & 0
    \end{pmatrix},
%    \bar{F}_{Im}=
%    \begin{pmatrix}
%      0 & 1 & 0\\
%      -1 & 0 & 0\\
%      0 & 0  & 0
%    \end{pmatrix},
  \end{aligned}
\end{equation}
%which gives $\|C_1\|_F=\sqrt{3}$, $\|\bar{F}_{Im}\|_F=\sqrt{2}$.
which gives the tradeoff relation
\begin{eqnarray}
\aligned
  \frac{1}{\nu}Tr[F_Q^{-1}Cov(\hat{x})^{-1}]&\le n-\frac{1}{4(n-1)}\|C_1\|_F^2\\
  &=3-\frac{3}{8}=2.625.
  \endaligned
\end{eqnarray}
%\begin{equation}
%  \frac{1}{\nu}Tr[F_Q^{-1}Cov(\hat{x})^{-1}]\le n-\frac{(n-2)}{(n-1)^2}\|\bar{F}_{Im}\|_F^2=3-\frac{1}{2}=\frac{5}{2}=2.5.
%\end{equation}

\begin{figure}
  \includegraphics[width=0.5\textwidth]{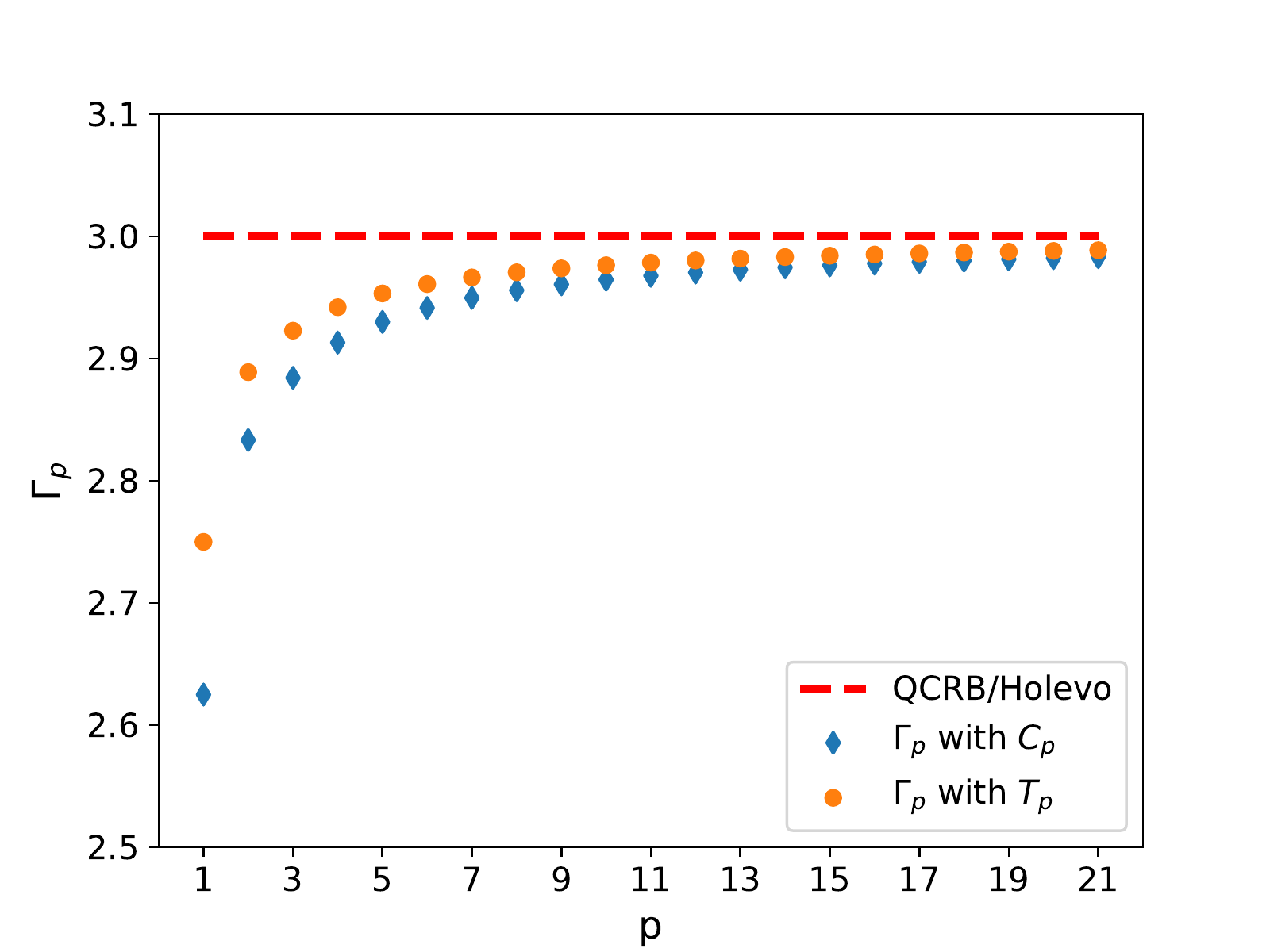}
  \caption{Precision bounds $\Gamma_p$ for $p$-local measurements and the Holevo bound when $n=3$.}
 \label{fig:example2}
\end{figure}

Under the $p$-local measurement, we have
\begin{equation}
  \begin{aligned}
    \frac{1}{\nu}Tr[F_Q^{-1}Cov(\hat{x})^{-1}]&\le n-\frac{1}{4(n-1)}\|\frac{C_p}{p}\|_F^2\\
    &= n-\frac{1}{4(n-1)}\|C_1\|_F^2\left(\frac{1}{p3^{p-1}}\mathcal{N}_p\right)^2\\
    &= 3-\frac{3}{8}\left(\frac{1}{p3^{p-1}}\mathcal{N}_p\right)^2.
  \end{aligned}
\end{equation}
For $p=2$, this gives
\begin{eqnarray}
\aligned
  \frac{1}{\nu}Tr[F_Q^{-1}Cov(\hat{x})^{-1}]&\le n-\frac{1}{4(n-1)}\|\frac{C_2}{2}\|_F^2\\
  &=3-\frac{3}{8}\times \frac{1}{4}\times\frac{16}{9}\\
  &=\frac{17}{6}\approx 2.83.
  \endaligned
\end{eqnarray}
%which is tighter than the Zhu-Hayashi bound in this case.
%{\color{purple}
The bound with $T_p$ can be similarly calculated as
$(T_p)_{jk}=\frac{1}{2}\sum_{s=0}^p\sum_{r=0}^{p-s}\binom{p}{s}\binom{p-s}{r}\left(\frac{1+3\delta}{3}\right)^s\left(\frac{1-3\delta}{3}\right)^r\left(\frac{1}{3}\right)^{p-s-r}|s\times\bra{0}[\tilde{L}_j,\tilde{L}_k]\ket{0} + r\times\bra{1}[\tilde{L}_j,\tilde{L}_k]\ket{1}+(p-s-r)\times\bra{2}[\tilde{L}_j,\tilde{L}_k]\ket{2}|$.
For $\delta=0$, the equation can be simplified as
\begin{equation}
    \begin{aligned}
      (T_p)_{12}=\frac{1}{2}\left(\frac{1}{3}\right)^p\sum_{s=0}^p\sum_{r=0}^{p-s}\binom{p}{s}\binom{p-s}{r}|3s-3r|
    \end{aligned}
\end{equation}
and $(T_p)_{13}=0$, $(T_p)_{23}=0$. This then gives
%The bound in Eq.(\ref{eq:mainTp}) then gives
\begin{eqnarray}
\aligned
  \frac{1}{\nu}Tr[F_Q^{-1}Cov^{-1}(\hat{x})]&\le n-\frac{1}{4(n-1)}\|\frac{T_p}{p}\|_F^2\\
  &=3-\frac{1}{4p^2}(T_p)_{12}^2.
  \endaligned
\end{eqnarray}
We plot the bounds for $n=3$ as a typical case in Fig\ref{fig:example2}. It can be seen that the Holevo bound, which equals to the QCRB as the weak commutative condition holds, is only achievable when $p\rightarrow \infty$. For any finite $p$, the bounds are strictly less than $n$.

%}
%and for $p=3$,
%\begin{eqnarray}
%\aligned
%  \frac{1}{\nu}Tr[F_Q^{-1}Cov(\hat{x})^{-1}]&\le n-\frac{1}{4(n-1)}\|\frac{C_3}{3}\|_F^2\\
%  &=3-\frac{3}{8}\times \frac{1}{9}\times\frac{25}{9}\\
%  &=\frac{623}{216}\approx 2.88.
%\endaligned
%\end{eqnarray}

If there are only two parameters,
the associated matrices are then given by the $2\times 2$ submatrices of the original ones. For example, suppose the two parameters are  $\{x_1,x_2\}$, we then have
\begin{equation}
  \begin{aligned}
    C_1=
    \begin{pmatrix}
      0 & 1\\
      1 & 0
    \end{pmatrix}
  \end{aligned}
\end{equation}
%which further gives $\|C_1\|_F=\sqrt{2}$.
the tradeoff relation under the 1-local measurement is then given by
\begin{equation}
  \frac{1}{\nu}Tr[F_Q^{-1}Cov(\hat{x})^{-1}]\le n-\frac{1}{4(n-1)}\|C_1\|_F^2=\frac{3}{2},
\end{equation}
in this case it is tighter than the Gill-Massar bound .

Under general $p$-local measurement, we have
\begin{equation}
  \begin{aligned}
    \frac{1}{\nu}Tr[F_Q^{-1}Cov(\hat{x})^{-1}]&\le n-\frac{1}{4(n-1)}\|\frac{C_p}{p}\|_F^2\\
    &= n-\frac{1}{4(n-1)}\|C_1\|_F^2\left(\frac{1}{p3^{p-1}}\mathcal{N}_p\right)^2\\
    &= 2-\frac{1}{2}\left(\frac{1}{p3^{p-1}}\mathcal{N}_p\right)^2.
  \end{aligned}
\end{equation}
For $p=2$, we have
\begin{equation}
  \frac{1}{\nu}Tr[F_Q^{-1}Cov(\hat{x})^{-1}]\le %n-\frac{1}{4(n-1)}\|\frac{C_2}{2}\|_F^2=2-\frac{1}{2}\times \frac{1}{4}\times\frac{16}{9}=
  \frac{16}{9}\approx 1.78,
\end{equation}
and for $p=3$,
\begin{equation}
  \frac{1}{\nu}Tr[F_Q^{-1}Cov(\hat{x})^{-1}]\le %n-\frac{1}{4(n-1)}\|\frac{C_3}{3}\|_F^2=2-\frac{1}{2}\times \frac{1}{9}\times\frac{25}{9}=
  \frac{299}{162}\approx 1.85.
\end{equation}

%For each $p$, we can plot the corresponding precision bound in the same figure as Fig.\ref{fig:example2} and it can be seen that with $p\rightarrow
%\infty$, the precision bound approaches the Holevo bound.

\begin{figure}
  \includegraphics[width=0.5\textwidth]{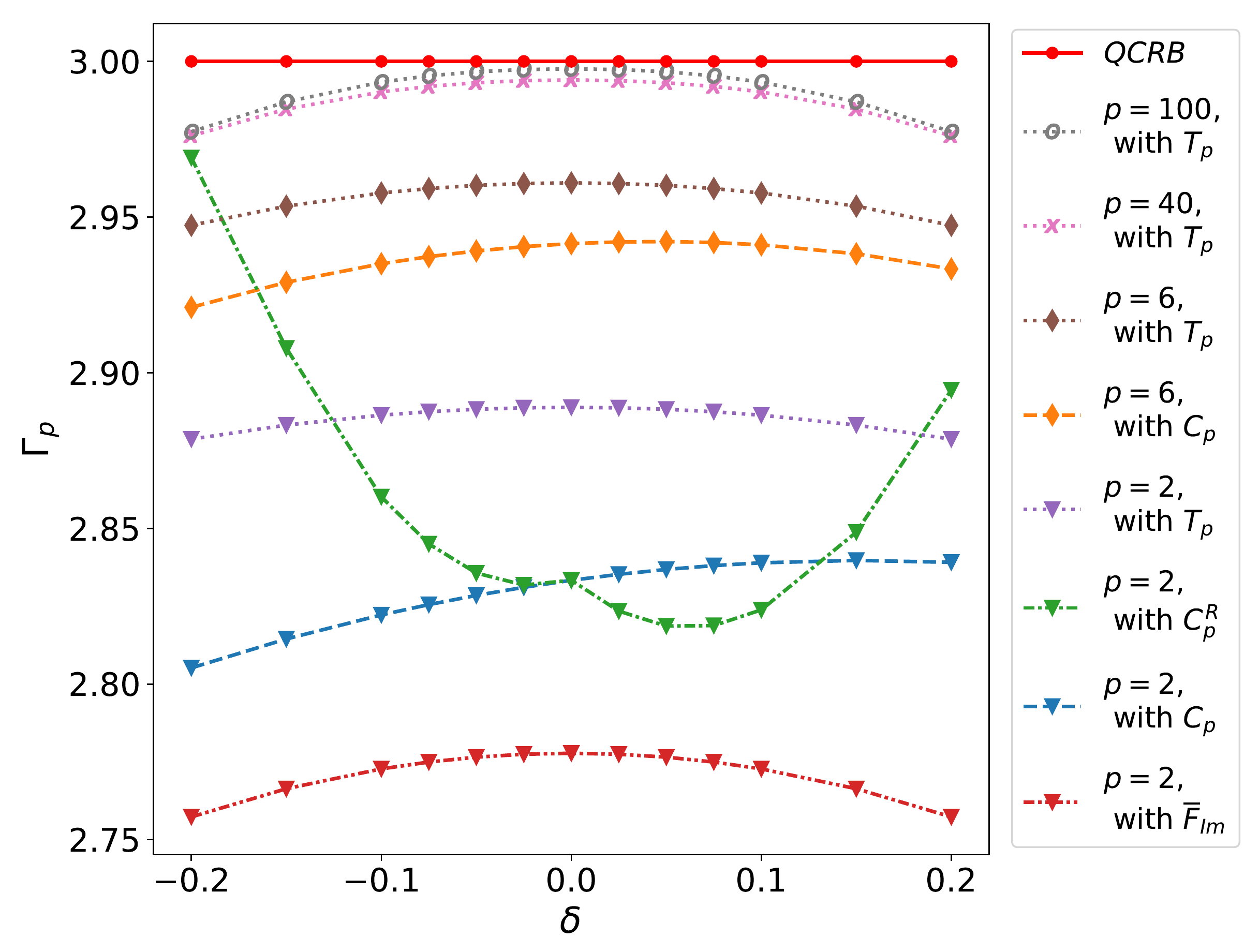}
  \caption{Comparison of different bounds of $\Gamma_p$ for the estimation of $\rho_x=\frac{1}{3}I+\delta G_3 + x_1G_1+x_2G_2+x_5G_5$ at $x_1=x_2=x_5=0$.}
  \label{fig:example2_delta}
\end{figure}

%{
%\color{purple}
%Similarly as example 1,
Similar as the previous example, we also consider the estimation of the state $\rho_x=\frac{1}{3}I+\delta G_3 + x_1G_1+x_2G_2+x_5G_5$ with general $\delta$ and plot the precision bounds in Fig. \ref{fig:example2_delta}, where we plotted the bounds with $C_p$ up to $p=6$ and the bounds with $T_p$ up to $p=100$. We also plotted the bounds with RLDs and $\mathbf{\bar{F}}_{Im}$ for $p=2$(see appendix for detailed calculations), as it can be seen the bound given by $\frac{1}{\nu}Tr[F_Q^{-1}Cov(\hat{x})^{-1}]\le n-\frac{(n-2)}{(n-1)^2}\|\frac{\bar{F}_{Im2}}{2}\|_F^2$ is tighter than the bounds given by $C_2$ and $T_2$ in this case. %which have similar behaviours as the previous example.

%$\rho_x=\frac{1}{3}(I+x_1\Lambda_1+x_2\Lambda_2+x_3\Lambda_5)$.
%{\color{blue}see appendix}
%\subsection{Example 3}

\section{Summary}\label{sec:summary}
%The various upper bounds on $\frac{1}{\nu}Tr[F_Q^{-1}Cov^{-1}(\hat{x})]$ can be used to obtain lower bounds for $\nu Tr[F_QCov(\hat{x})]$ through the Cauchy-Schwarz inequality in Eq.(\ref{eq:CS}) as
%\begin{eqnarray}
%\aligned
%\nu Tr[F_QCov(\hat{x})]\geq \frac{n^2}{\frac{1}{\nu}Tr[F_Q^{-1}Cov^{-1}(\hat{x})]}.
%\endaligned
%\end{eqnarray}

%Previous studies on multi-parameter quantum estimation have been mostly focused on the Holevo bound\cite{Hole82book}.
%While the Holevo bound characterizes the precision limit with collective measurements on infinite number of quantum states, the precision limits under general $p$-local measurements remained largely unexplored.
The presented framework provided a versatile tool to obtain bounds on the precision limit in multi-parameter quantum estimation under general $p$-local measurements, which significantly increased our knowledge on the incompatibility in multi-parameter quantum estimation. The relation between the partial commutative condition and the weak commutative condition is also clarified. Future studies includes improving the bounds by exploring different choices of $\{|u_q\rangle\}$ and operators in $\mathbf{\bar{S}}$, clarifying whether the partial commutative condition is sufficient for the saturation of the QCRB, and identifying the ultimate precision under general $p$-local measurements. The approach can also be used to strengthen the uncertainty relations for multiple observables, which is another interesting directions to pursue.

%\bibliographystyle{apsrev4-1}
%\bibliographystyle{unsrt}
%\bibliography{reference}
\begin{acknowledgements}
This work is partially supported by the Research Grants Council of Hong Kong with the Grant No. 14307420.
\end{acknowledgements}

%\end{thebibliography}

\begin{widetext}
\appendix

\section{Tradeoff Relations}\label{sec:tradeoffappendix}
We derive the tradeoff relation from
\begin{equation}
    S=\left(\begin{array}{cc}
      A & B  \\
      B^\dagger & F \\
          \end{array}\right)\geq 0,
\end{equation}
where $A,B,F$ are $n\times n$ matrices with $Cov(\hat{x})\geq A$, $B=I+iB_{Im}$ and $F=F_Q+iF_{Im}$. %and , i.e., the real part of B
We note that the derivation below works regardless whether $S$ is obtained from pure states or mixed states.
%with the entries given by
%\begin{eqnarray}
%\aligned
%(A)_{kj}&=Tr(\rho_x\hat{X}_k\hat{X}_j),\\
%(B)_{kj}&=Tr(\rho_x\hat{X}_kL_j),\\
%(F)_{kj}&=Tr(\rho_xL_kL_j),
%\endaligned
%\end{eqnarray}
%here $\{X_j\}$ are Hermitian operators satisfy the locally unbiased condition
%\begin{eqnarray}
%\langle\Psi_x|\hat{X}_j|\Psi_x\rangle=0,
%Tr(\rho_x\hat{X}_j)=0, \qquad j=1,\cdots, n%\langle\Psi|\hat{X}_2|\Psi\rangle=0,
%\end{eqnarray}
%and
%\begin{eqnarray}\label{eq:local}
%\aligned
%Tr(\partial_{x_j}\rho_x\hat{X}_k)=\delta^j_k,
%\endaligned
%\end{eqnarray}
%$\{L_j\}$ are the SLDs. %for $x_j$, here $j\in\{1,\cdots, n\}$,

 %more specifically it works for all $S=\left(\begin{array}{cc}
%      A & B  \\
%      B^\dagger & F \\
%          \end{array}\right)\geq 0$ with $Cov(\hat{x})\geq A$ and $B$ has the form $B=I+iB_{Im}$.

Since $Cov(\hat{x})\geq A$, we have
\begin{equation}
    \left(\begin{array}{cc}
      Cov(\hat{x}) & B  \\
      B^\dagger & F \\
          \end{array}\right)=\left(\begin{array}{cc}
      Cov(\hat{x})-A & 0  \\
      0 & 0 \\
          \end{array}\right)+\left(\begin{array}{cc}
      A & B  \\
      B^\dagger & F \\
          \end{array}\right)\geq 0.
\end{equation}
This implies that $F-B^\dagger Cov^{-1}(\hat{x})B\geq 0$. Since $F=F_Q+iF_{Im}$, $B=I+iB_{Im}$, we thus have
\begin{eqnarray}
\label{eq:schur}
F_Q+iF_{Im}-[Cov^{-1}(\hat{x})+B_{Im}^TCov^{-1}(\hat{x})B_{Im}+i(B_{Im}^TCov^{-1}(\hat{x})-Cov^{-1}(\hat{x})B_{Im})]\geq 0,
\end{eqnarray}
which implies the real part is positive semidefinite, i.e.,
\begin{equation}
F_Q-Cov^{-1}(\hat{x})-B_{Im}^TCov^{-1}(\hat{x})B_{Im}\geq 0.
\end{equation}
This can be written as
$F_Q-Cov^{-1}(\hat{x})\geq B_{Im}^TCov^{-1}(\hat{x})B_{Im}\geq 0$, which is stronger than the quantum Cramer-Rao bound $F_Q-Cov^{-1}(\hat{x})\geq 0$, typically written as $Cov(\hat{x})\geq F_Q^{-1}$. To saturate the bound, i.e., $Cov(\hat{x})=F_Q^{-1}$, we need to have $B_{Im}^TCov(\hat{x})^{-1}B_{Im}=0$. When the covariance matrix is full rank, which is always the case when $F_Q$ is invertable, this requires $B_{Im}=0$, Eq.(\ref{eq:schur}) then becomes
\begin{eqnarray}
F_Q+iF_{Im}-Cov^{-1}(\hat{x})\geq 0.
\end{eqnarray}
The saturation of the quantum Cramer-Rao bound then requires
$iF_{Im}\geq 0$. Since $F_{Im}$ is anti-symmetric and its eigenvalues are in the form of $\pm i\beta$ with $\beta\in \mathbb{R}$, $iF_{Im}\geq 0$ is only possible when all the eigenvalues are zero, i.e., $F_{Im}=0$. This is exactly the weak commutative condition for the saturation of the quantum Cramer-Rao bound.

When $F_{Im}\neq0$, the QCRB is not saturable. %We have $F_Q-Cov(\hat{x})^{-1}\geq B_{Im}^TCov(\hat{x})^{-1}B_{Im}\neq 0$. %Suppose $Cov(\hat{x})$ is achieved under a POVM and the corresponding classical Fisher information matrix is $F_C$, i.e.,
Denote $F_C=Cov^{-1}(\hat{x})$, we write Eq.(\ref{eq:schur}) as
\begin{eqnarray}
%\label{eq:schur}
F_Q-F_C-B_{Im}^TF_CB_{Im}+i(F_{Im}+B_{Im}^TF_C-F_CB_{Im})]\geq 0.
\end{eqnarray}
By multiplying $F_Q^{-\frac{1}{2}}$ from both the left and the right, we get
 \begin{eqnarray}
%\label{eq:schur}
I-F_Q^{-\frac{1}{2}}F_CF_Q^{-\frac{1}{2}}-F_Q^{-\frac{1}{2}}B_{Im}^TF_CB_{Im}F_Q^{-\frac{1}{2}}+i(F_Q^{-\frac{1}{2}}F_{Im}F_Q^{-\frac{1}{2}}+F_Q^{-\frac{1}{2}}B_{Im}^TF_CF_Q^{-\frac{1}{2}}-F_Q^{-\frac{1}{2}}F_CB_{Im}F_Q^{-\frac{1}{2}})]\geq 0.
\end{eqnarray}
Denote $\tilde{F}_C=F_Q^{-\frac{1}{2}}F_CF_Q^{-\frac{1}{2}}$, $\tilde{B}_{Im}=F_Q^{\frac{1}{2}}B_{Im}F_Q^{-\frac{1}{2}}$, $\tilde{F}_{Im}=F_Q^{-\frac{1}{2}}F_{Im}F_Q^{-\frac{1}{2}}$, we can write the inequality as
 \begin{eqnarray}
%\label{eq:schur}
I-\tilde{F}_C-\tilde{B}_{Im}^T\tilde{F}_C\tilde{B}_{Im}+i(\tilde{F}_{Im}+\tilde{B}_{Im}^T\tilde{F}_C-\tilde{F}_C\tilde{B}_{Im}]\geq 0.
\end{eqnarray}
Since $F_C\leq F_Q$, we have $\tilde{F}_C\leq I$, thus $\tilde{F}_C\geq \tilde{F}_C^2$ and  $\tilde{B}_{Im}^T\tilde{F}_C\tilde{B}_{Im}\geq \tilde{B}_{Im}^T\tilde{F}_C^2\tilde{B}_{Im}$. We then have
 \begin{eqnarray}
%\label{eq:schur}
I-\tilde{F}_C-\tilde{B}_{Im}^T\tilde{F}_C^2\tilde{B}_{Im}+i(\tilde{F}_{Im}+\tilde{B}_{Im}^T\tilde{F}_C-\tilde{F}_C\tilde{B}_{Im}]\geq 0.
\end{eqnarray}
Now denote $\tilde{F}_C\tilde{B}_{Im}$ as $D$,
 \begin{eqnarray}\label{eq:boundsupp}
%\label{eq:schur}
I-\tilde{F}_C-D^TD+i(\tilde{F}_{Im}+D^T-D)\geq 0.
\end{eqnarray}
Since any two by two principle submatrix of a positive semidefinite matrix is also positive semidefinite, 
 \begin{eqnarray}
 \aligned
%\label{eq:schur}
%\left(\begin{array}{cc}
%      1 & 0  \\
%      0 & 1 \\
%          \end{array}\right)-\left(\begin{array}{cc}
 %     (\tilde{F}_C)_{jj} & (\tilde{F}_C)_{jk}  \\
 %     (\tilde{F}_C)_{kj} & (\tilde{F}_C)_{kk} \\
 %         \end{array}\right)-
 \left(\begin{array}{cc}
      1-(\tilde{F}_C)_{jj}-(D^TD)_{jj} & -(\tilde{F}_C)_{jk}-(D^TD)_{jk} \\
      -(\tilde{F}_C)_{kj}-(D^TD)_{kj} & 1-(\tilde{F}_C)_{kk}-(D^TD)_{kk} \\
          \end{array}\right)&+i\left(\begin{array}{cc}
      0 & (\tilde{F}_{Im})_{jk}+D_{kj}-D_{jk}  \\
      -(\tilde{F}_{Im})_{jk}-D_{kj}+D_{jk} & 0 \\
          \end{array}\right)%\\
         % &\geq 0.
          \endaligned
\end{eqnarray}
is then positive semidefinite. Note that $\tilde{F}_C$ and $D^TD$ are symmetric and the determination of a positive semidefinite matrix is nonnegative, we thus have
\begin{eqnarray}
[1-(\tilde{F}_C)_{jj}-(D^TD)_{jj}][1-(\tilde{F}_C)_{kk}-(D^TD)_{kk}]\geq [(\tilde{F}_C)_{jk}+(D^TD)_{jk}]^2+[(\tilde{F}_{Im})_{jk}+D_{kj}-D_{jk}]^2,
\end{eqnarray}
from which we can get
\begin{eqnarray}
\aligned
&[1-(\tilde{F}_C)_{jj}-(D^TD)_{jj}]+[1-(\tilde{F}_C)_{kk}-(D^TD)_{kk}]\\
&\geq 2\sqrt{[1-(\tilde{F}_C)_{jj}-(D^TD)_{jj}][1-(\tilde{F}_C)_{kk}-(D^TD)_{kk}]}\\
&\geq 2\sqrt{[(\tilde{F}_C)_{jk}+(D^TD)_{jk}]^2+[(\tilde{F}_{Im})_{jk}+D_{kj}-D_{jk}]^2}\\
&\geq 2|(\tilde{F}_{Im})_{jk}+D_{kj}-D_{jk}|,
\endaligned
\end{eqnarray}
i.e.,
\begin{eqnarray}\label{eq:jk}
1-(\tilde{F}_C)_{jj}+1-(\tilde{F}_C)_{kk}
\geq 2|(\tilde{F}_{Im})_{jk}+D_{kj}-D_{jk}|+(D^TD)_{jj}+(D^TD)_{kk}.
\end{eqnarray}
As $(D^TD)_{jj}=\sum_p D_{pj}^2\geq D_{kj}^2$ and $(D^TD)_{kk}=\sum_p D_{pk}^2\geq D_{jk}^2$, we have
\begin{equation}
(D^TD)_{jj}+(D^TD)_{kk}=\sum_p (D_{pj}^2+D_{pk}^2)\geq D_{kj}^2+D_{jk}^2=\frac{1}{2}(D_{kj}-D_{jk})^2+\frac{1}{2}(D_{kj}+D_{jk})^2,
\end{equation}
and from $I+i\tilde{F}_{Im}=F_Q^{-\frac{1}{2}}FF_Q^{-\frac{1}{2}}\geq 0$, we have $|(\tilde{F}_{Im})_{jk}|\leq 1$. Thus
\begin{eqnarray}\label{eq:tradeoffsupp}
\aligned
1-(\tilde{F}_C)_{jj}+1-(\tilde{F}_C)_{kk}
&\geq 2|(\tilde{F}_{Im})_{jk}+D_{kj}-D_{jk}|+(D^TD)_{jj}+(D^TD)_{kk}\\
&\geq 2|(\tilde{F}_{Im})_{jk}+D_{kj}-D_{jk}|+\frac{1}{2}(D_{kj}-D_{jk})^2\\
&\geq \frac{1}{2}|(\tilde{F}_{Im})_{jk}|^2,
\endaligned
\end{eqnarray}
where the last inequality we used the fact that $2|y+x|+\frac{1}{2}x^2\geq \frac{1}{2}y^2$ when $|y|\leq 1$, since
\begin{eqnarray}
\aligned
2|y+x|+\frac{1}{2}x^2&=2|y+x|+\frac{1}{2}(y+x-y)^2\\
&=2|y+x|+\frac{1}{2}(y+x)^2-y(x+y)+\frac{1}{2}y^2\\
&\geq 2|y+x|-|y(x+y)|+\frac{1}{2}y^2\\
&=(2-|y|)|x+y|+\frac{1}{2}y^2\\
&\geq \frac{1}{2}y^2.
\endaligned
\end{eqnarray}
This provides a tradeoff relation between $(\tilde{F}_C)_{jj}$ and $(\tilde{F}_C)_{kk}$. When $F_{Im}=0$, the quantum Cram\'er-Rao bound is saturable, $F_C$ can reach $F_Q$, in this case $\tilde{F}_C=I$, $(\tilde{F}_C)_{jj}$ and $(\tilde{F}_C)_{kk}$ can reach the maximal value simultaneously, which is 1. When $(\tilde{F}_{Im})_{jk}\neq 0$, $(\tilde{F}_C)_{jj}$ and $(\tilde{F}_C)_{kk}$ can not simultaneously reach 1, Eq.(\ref{eq:tradeoffsupp}) puts a tradeoff between them.

By summing Eq.(\ref{eq:tradeoffsupp}) over different choice of $j,k$ directly, we can get
\begin{equation}
2(n-1)\sum_j [1-(\tilde{F}_C)_{jj}]\geq \frac{1}{2}\sum_{j,k,j\neq k}|(\tilde{F}_{Im})_{jk}|^2=\frac{1}{2}\|\tilde{F}_{Im}\|_F^2,    \end{equation}
which gives
\begin{equation}
    Tr(\tilde{F}_C)\leq n-\frac{1}{4(n-1)}\|\tilde{F}_{Im}\|_F^2.
\end{equation}
where $\|\tilde{F}_{Im}\|_F^2=Tr(\tilde{F}_{Im}^T\tilde{F}_{Im})$. This can be rewritten as
%\begin{equation}
%    Tr(F_Q^{-1}F_C)\leq n-\frac{1}{4(n-1)}Tr(F_Q^{-1}F_{Im}^TF_Q^{-1}F_{Im}),
%    \end{equation}
%i.e.,
\begin{eqnarray}%\label{eq:tradeoffpure}
\aligned
Tr[F_Q^{-1}Cov^{-1}(\hat{x})]&\leq n-\frac{1}{4(n-1)}\|F_Q^{-\frac{1}{2}}F_{Im}F_Q^{-\frac{1}{2}}\|_F^2\\ &=n-\frac{1}{4(n-1)}Tr(F_Q^{-1}F_{Im}^TF_Q^{-1}F_{Im}),
\endaligned
\end{eqnarray}
%\begin{eqnarray}
%    Tr[F_Q^{-1}Cov^{-1}(\hat{x})]\leq n-\frac{1}{4(n-1)}Tr(F_Q^{-1}F_{Im}^TF_Q^{-1}F_{Im}).
 %   \end{eqnarray}
The same relation can be obtained by including the number of copies of the state, $\nu$, explicitly, essentially just replace $F_Q$ and $F_{Im}$ with $\nu F_Q$ and $\nu F_{Im}$. The tradeoff relation with $\nu$ copies of the state is then
\begin{equation}
    \frac{1}{\nu}Tr[F_Q^{-1}Cov^{-1}(\hat{x})]\leq n-\frac{1}{4(n-1)}\|F_Q^{-\frac{1}{2}}F_{Im}F_Q^{-\frac{1}{2}}\|_F^2.%Tr(F_Q^{-1}F_{Im}^TF_Q^{-1}F_{Im}).
    \end{equation}

 %We note that the Gill-Massar inequality holds for the separable measurement, while the tradeoff relation here holds for any measurement.

When the number of the parameters, $n\geq 3$, the tradeoff can be tightened by keeping all terms in $(D^TD)_{jj}$ and $(D^TD)_{kk}$ in Eq.(\ref{eq:jk}) as
\begin{eqnarray}
\sum_{j, k,j\neq k} [1-(\tilde{F}_C)_{jj}+1-(\tilde{F}_C)_{kk}]
\geq \sum_{j,k,j\neq k} [2|(\tilde{F}_{Im})_{jk}+D_{kj}-D_{jk}|+(D^TD)_{jj}+(D^TD)_{kk}],
\end{eqnarray}
here
\begin{equation}
(D^TD)_{jj}+(D^TD)_{kk}=\sum_p (D_{pj}^2+D_{pk}^2),
\end{equation}
which not only includes the correlations between the $j,k$-th entry, but also with other entries.
By summing over all choice of $j,k$, we have
\begin{eqnarray}
\aligned
2(n-1)\sum_j [1-(\tilde{F}_C)_{jj}]
&\geq \sum_{j,k,j\neq k} 2|(\tilde{F}_{Im})_{jk}+D_{kj}-D_{jk}|+2(n-1)\sum_j(D^TD)_{jj}\\
&=\sum_{j,k,j\neq k} 2|(\tilde{F}_{Im})_{jk}+D_{kj}-D_{jk}|+2(n-1)\sum_{j,k} D_{jk}^2\\
&\geq\sum_{j,k, j\neq k}\{2|(\tilde{F}_{Im})_{jk}+D_{kj}-D_{jk}|+(n-1)(D_{jk}^2+D_{kj}^2)\}\\
&=\sum_{j,k,j\neq k}\{2|(\tilde{F}_{Im})_{jk}+D_{kj}-D_{jk}|+\frac{n-1}{2}(D_{kj}-D_{jk})^2+\frac{n-1}{2}(D_{kj}+D_{jk})^2\}\\
&\geq \frac{2(n-2)}{n-1} \sum_{j,k,j\neq k}|(\tilde{F}_{Im})_{jk}|^2\\
&=\frac{2(n-2)}{n-1} \|\tilde{F}_{Im}\|_F^2,
\endaligned
\end{eqnarray}
where the last inequality we used the fact that
\begin{eqnarray}
\aligned
2|y+x|+\frac{n-1}{2}x^2&=2|y+x|+\frac{n-1}{2}(y+x-y)^2\\
&=2|y+x|+\frac{n-1}{2}(y+x)^2-(n-1)y(x+y)+\frac{n-1}{2}y^2\\
&\geq \frac{n-1}{2}(y+x)^2+ 2|y||y+x|-(n-1)|y(x+y)|+\frac{n-1}{2}y^2\\
&= \frac{n-1}{2}(y+x)^2- (n-3)|y||x+y|+\frac{n-1}{2}y^2\\
&= \frac{n-1}{2}(|y+x|-\frac{n-3}{n-1}|y|)^2+\frac{2(n-2)}{n-1} y^2\\
&\geq \frac{2(n-2)}{n-1} y^2.
\endaligned
\end{eqnarray}
This then gives a tradeoff relation on $\tilde{F}_C$ as
\begin{equation}\label{eq:suppF}
    Tr(\tilde{F}_C)\leq n-\frac{n-2}{(n-1)^2}\|\tilde{F}_{Im}\|_F^2.
\end{equation}
With $\nu$ copies of the state, this can be equivalently written %in a similar form to the Gill-Massar inequality as
\begin{equation}
    \frac{1}{\nu}Tr[F_Q^{-1}Cov^{-1}(\hat{x})]\leq n-\frac{n-2}{(n-1)^2}\|F_Q^{-\frac{1}{2}}F_{Im}F_Q^{-\frac{1}{2}}\|_F^2.%Tr(F_Q^{-1}F_{Im}^TF_Q^{-1}F_{Im}).
\end{equation}

The bound can be further improved. From Eq.(\ref{eq:boundsupp}),
\begin{eqnarray}
%\label{eq:schur}
I-\tilde{F}_C-D^TD+i(\tilde{F}_{Im}+D^T-D)\geq 0,
\end{eqnarray}
we have
\begin{equation}
    I-\tilde{F}_C-D^TD\geq -i(\tilde{F}_{Im}+D^T-D),
\end{equation}
from which we can obtain
\begin{equation}\label{eq:suppineq}
    Tr(I-\tilde{F}_C-D^TD)\geq \|\tilde{F}_{Im}+D^T-D\|_1.
\end{equation}
Note that $\tilde{F}_{Im}+D^T-D$ is skew symmetric with purely imaginary eigenvalues, and the singular values are just the amplitude of the eigenvalues as $\{|\lambda_1|,\cdots, |\lambda_n|\}$. Since
\begin{eqnarray}
-i(\tilde{F}_{Im}+D^T-D)\leq I-\tilde{F}_C-D^TD\leq I,
\end{eqnarray}
we have $|\lambda_j|\leq 1$, thus $|\lambda_j|\geq |\lambda_j|^2$. As
\begin{eqnarray}
\aligned
\|\tilde{F}_{Im}+D^T-D\|_1=\sum_{j=1}^n|\lambda_j|
&\geq \sum_{j=1}^n|\lambda_j|^2\\&=Tr[(\tilde{F}_{Im}+D^T-D)^T(\tilde{F}_{Im}+D^T-D)]\\
&=\sum_{jk}[(\tilde{F}_{Im}+D^T-D)_{jk}]^2,
\endaligned
\end{eqnarray}
from Eq.(\ref{eq:suppineq}) we then have
\begin{eqnarray}
\aligned
    Tr(I-\tilde{F}_C)&\geq Tr(D^TD)+\|\tilde{F}_{Im}+D^T-D\|_1\\
    &\geq \sum_{k}(D^TD)_{kk}+\sum_{jk}[(\tilde{F}_{Im}+D^T-D)_{jk}]^2\\
    &=\sum_{jk} D_{jk}^2+[(\tilde{F}_{Im})_{jk}+D_{kj}-D_{jk}]^2\\
    &=\sum_{jk} \frac{1}{2}(D_{jk}^2+D_{kj}^2)+[(\tilde{F}_{Im})_{jk}+D_{kj}-D_{jk}]^2\\
    &\geq \sum_{jk} \frac{1}{4}(D_{jk}-D_{kj})^2+[(\tilde{F}_{Im})_{jk}+D_{kj}-D_{jk}]^2\\
    &\geq \sum_{jk}\frac{1}{5} (\tilde{F}_{Im})_{jk}^2\\
    &=\frac{1}{5}\|\tilde{F}_{Im}\|_F^2,
\endaligned
\end{eqnarray}
where the last inequality we used the fact that
\begin{eqnarray}
\aligned
\frac{1}{4}x^2+(y+x)^2&=\frac{5}{4}x^2+2xy+y^2\\
&=\frac{5}{4}(x+\frac{4}{5}y)^2+\frac{1}{5}y^2\\
&\geq \frac{1}{5}y^2.
\endaligned
\end{eqnarray}
We thus have
\begin{equation}
    Tr(\tilde{F}_C)\leq n-\frac{1}{5}\|\tilde{F}_{Im}\|_F^2.
\end{equation}
For $\nu$ copies of the state, this gives the tradeoff relation
\begin{equation}
    \frac{1}{\nu}Tr[F_Q^{-1}Cov^{-1}(\hat{x})]\leq n-\frac{1}{5}\|F_Q^{-\frac{1}{2}}F_{Im}F_Q^{-\frac{1}{2}}\|_F^2,
\end{equation}
which is tighter than Eq.(\ref{eq:suppF}) when $n\geq 5$.

\section{Connection between the partial commutative condition and the weak commutative condition}\label{sec:appconnection}
Here we show %that the partial commutative condition reduces to the weak commutative condition when $p\rightarrow \infty$, specifically we show that
\begin{equation}
\lim_{p\rightarrow \infty}\frac{\|\sqrt{\rho_x^{\otimes p}}[L_{jp}, L_{kp}]\sqrt{\rho_x^{\otimes p}}\|_1}{p}=|Tr(\rho_x[L_j,L_k])|.
\end{equation}
%The condition $\frac{C_p}{p}=0$ then reduces to the weak commutative condition, $Tr(\rho_x[L_j,L_k])=0$, $\forall j, k$, when $p\rightarrow \infty$.  This then also suggests that the partial commutative condition is likely also sufficient for the saturation of QCRB under $p$-local measurements.

%For pure states, there is no difference between the partial commutative condition and the weak commutative condition. %This is consistent with the fact that for pure states the optimal measurement can be taken as 1-local measurement, thus if the QCRB can be saturated by any measurement, it can also be saturated by 1-local measurement, which is a subset of $p$-local measurement.
%For mixed states,
We write the state in the eigenvalue decomposition as $\rho_x=\sum_{q=1}^m\lambda_q|\Psi_q\rangle\langle\Psi_q|$ with $\lambda_q>0$ and $\sum_{q=1}^m\lambda_q=1$. Then $\sqrt{\rho_x}=\sum_q\sqrt{\lambda_q}|\Psi_q\rangle\langle\Psi_q|$,
\begin{eqnarray}
\aligned
\|\sqrt{\rho_x^{\otimes p}}[L_{jp},L_{kp}]\sqrt{\rho_x^{\otimes p}}\|_1=&\|\sqrt{\rho_x^{\otimes p}}\sum_{r=1}^p[L_{j}^{(r)},L_{k}^{(r)}]\sqrt{\rho_x^{\otimes p}}\|_1 \\
=&\|\sum_{r=1}^p\rho_x^{\otimes (r-1)}\otimes (\sqrt{\rho_x}[L_j,L_k]\sqrt{\rho_x})\otimes \rho_x^{\otimes (p-r)}\|_1,
\endaligned
\end{eqnarray}
here $L_{j}^{(r)}=I^{\otimes(r-1)}\otimes L_j\otimes I^{\otimes (p-r)}$. The support of $\sum_{r=1}^p\rho_x^{\otimes (r-1)}\otimes (\sqrt{\rho_x}[L_j,L_k]\sqrt{\rho_x})\otimes \rho_x^{\otimes (p-r)}$ is in the subspace spanned by $\{|\Psi_{v_1}\Psi_{v_2}\cdots \Psi_{v_p}\rangle\}$, with $|\Psi_{v_1}\rangle, \cdots, |\Psi_{v_p}\rangle\in\{|\Psi_1\rangle, \cdots,|\Psi_m\rangle\}$, here $\{|\Psi_1\rangle, \cdots,|\Psi_m\rangle\}$ are the eigenvectors of $\rho_x$ with nonzero eigenvalues. We can focus on the support space and calculate the entries of $\sum_{r=1}^p\rho_x^{\otimes (r-1)}\otimes \sqrt{\rho_x}[L_j,L_k]\sqrt{\rho_x}\otimes \rho_x^{\otimes (p-r)}$ in the basis of $|\Psi_{v_1}\Psi_{v_2}\cdots \Psi_{v_p}\rangle$ with $v_1,\cdots, v_p\in\{1,\cdots, m\}$ and show that when $p\rightarrow \infty$, $\frac{\|\sqrt{\rho_x^{\otimes p}}[L_{jp},L_{kp}]|\sqrt{\rho_x^{\otimes p}}\|_1}{p}\rightarrow |Tr(\rho_x[L_j,L_k])|$.

The entries of $\sqrt{\rho_x^{\otimes p}}[L_{jp},L_{kp}]|\sqrt{\rho_x^{\otimes p}}$ is given by
%\begin{widetext}
\begin{eqnarray}
\aligned
%&\langle\Psi_{j_p} \cdots \Psi_{j_1}| (C_p)_{jk} |\Psi_{k_1}\rangle \cdots \Psi_{j_p}\rangle\\
&\langle\Psi_{\tilde{v}_1} \cdots \Psi_{\tilde{v}_p}| \sum_{r=1}^p\rho_x^{\otimes (r-1)}\otimes \sqrt{\rho_x}[L_j,L_k]\sqrt{\rho_x}\otimes \rho_x^{\otimes (p-r)} |\Psi_{v_1} \cdots \Psi_{v_p}\rangle\\
%&=\sum_{r=1}^p \otimes_{q=1}^{r-1}\langle\Psi_{\tilde{v}_q}|\rho_x|\Psi_{v_q}\rangle\otimes \langle\Psi_{\tilde{v}_r}|\sqrt{\rho_x}[L_j,L_k]\sqrt{\rho_x}|\Psi_{v_r}\rangle\otimes_{q=r+1}^{p}\langle\Psi_{\tilde{v}_q}|\rho_x|\Psi_{v_q}\rangle  \\
&=\sum_{r=1}^p  [\langle\Psi_{\tilde{v}_r}|\sqrt{\rho_x}[L_j,L_k]\sqrt{\rho_x}|\Psi_{v_r}\rangle \prod_{q\neq r} (\delta_{\tilde{v}_q}^{v_q}\lambda_{v_q})].  %\prod_{q=r+1}^{p}(\delta_{\tilde{v}_q}^{v_q}\lambda_{v_q})]
\endaligned
\end{eqnarray}
%\end{widetext}
It is easy to see that when the indexes $\{v_1,v_2\cdots, v_p\}$ and $\{\tilde{v}_1,\cdots, \tilde{v}_p\}$ differ at two or more entries, the corresponding matrix entry equals to 0. When the two indexes differ at only one entry, for example, $v_r\neq \tilde{v}_r$ but $v_q=\tilde{v}_q$ for all $q\neq r$, the corresponding matrix entry equals to
\begin{eqnarray}
\aligned
%&\sum_{s=1}^p  [\langle\Psi_{\tilde{v}_s}|\sqrt{\rho_x}[L_j,L_k]\sqrt{\rho_x}|\Psi_{v_s}\rangle \prod_{q\neq s} (\delta_{\tilde{v}_q}^{v_q}\lambda_{v_q})]\\
\langle\Psi_{\tilde{v}_r}|\sqrt{\rho_x}[L_j,L_k]\sqrt{\rho_x}|\Psi_{v_r}\rangle \prod_{q\neq r} \lambda_{\tilde{v}_q}=\frac{\langle\Psi_{\tilde{v}_r}|\sqrt{\rho_x}[L_j,L_k]\sqrt{\rho_x}|\Psi_{v_r}\rangle}{\lambda_{\tilde{v}_r}}\prod_{q=1}^p \lambda_{\tilde{v}_q}.%\\
%&\leq l_{\max}\prod_{q=1}^p \lambda_{\tilde{v}_q},
\endaligned
\end{eqnarray}
%here $l_{\max}=\max_{\tilde{v}_r\neq v_r} |\frac{\langle\Psi_{\tilde{v}_r}|\sqrt{\rho_x}[L_j,L_k]\sqrt{\rho_x}|\Psi_{v_r}\rangle}{\lambda_{\tilde{v}_r}}|$.
%$|\Psi_{j_k}\rangle\in \{|\Psi_1,\cdots,|\Psi_m\}$ are the eigenvectors of $\rho_x$.
%it also converges to 0 when $p\rightarrow \infty$ since $\prod_{q=1}^p \lambda_{v_q}\rightarrow 0$. Thus when $p\rightarrow \infty$, $\sum_{r=1}^p\rho_x^{\otimes (r-1)}\otimes \sqrt{\rho_x}[L_j,L_k]\sqrt{\rho_x}\otimes \rho_x^{\otimes (p-r)}$ converges to a diagonal matrix and the diagonal entries are given by
When the indexes $\{v_1,v_2\cdots, v_p\}$ and $\{\tilde{v}_1,\cdots, \tilde{v}_p\}$ are the same, we get the diagonal entries of $\sqrt{\rho_x^{\otimes p}}[L_{jp},L_{kp}]|\sqrt{\rho_x^{\otimes p}}$ as
\begin{eqnarray}
\aligned
%&\langle\Psi_{j_p} \cdots \Psi_{j_1}| (C_p)_{jk} |\Psi_{k_1}\rangle \cdots \Psi_{j_p}\rangle\\
%&\langle\Psi_{v_p} \cdots \Psi_{v_1}| \sum_{r=1}^p\rho_x^{\otimes (r-1)}\otimes \sqrt{\rho_x}[L_j,L_k]\sqrt{\rho_x}\otimes \rho_x^{\otimes (p-r)} |\Psi_{v_1} \cdots \Psi_{v_p}\rangle\\
\sum_{r=1}^p  (\langle\Psi_{v_r}|\sqrt{\rho_x}[L_j,L_k]\sqrt{\rho_x}|\Psi_{v_r}\rangle \prod_{q\neq r} \lambda_{v_q})
&=\sum_{r=1}^p (\langle\Psi_{v_r}|[L_j,L_k]|\Psi_{v_r}\rangle \prod_{q=1}^p\lambda_{v_q})\\
&=(\prod_{q=1}^p\lambda_{v_q}) \sum_{r=1}^p \langle\Psi_{v_r}|[L_j,L_k]|\Psi_{v_r}\rangle.
\endaligned
\end{eqnarray}
%for diagonal matrices,
Next we write $\sqrt{\rho_x^{\otimes p}}[L_{jp},L_{kp}]|\sqrt{\rho_x^{\otimes p}}=D_p^{(jk)}+O_p^{(jk)}$ with $D_p^{(jk)}$ as the diagonal part of the matrix and $O_p^{(jk)}$ as the off-diagonal part of the matrix. %in the basis of $|\Psi_{v_1}\Psi_{v_2}\cdots \Psi_{v_p}\rangle$ with $v_1,\cdots, v_p\in\{1,\cdots, m\}$,
We then use the inequality
\begin{eqnarray}
\|D_p^{(jk)}\|_1\leq \|D_p^{(jk)}+O_p^{(jk)}\|_1\leq \|D_p^{(jk)}\|_1+\|O_p^{(jk)}\|_1
\end{eqnarray}
to bound $\|\sqrt{\rho_x^{\otimes p}}[L_{jp},L_{kp}]|\sqrt{\rho_x^{\otimes p}}\|_1$, here  the first inequality comes from the fact that for any matrix, $M$, $\|M\|_1\geq \sum_q |M_{qq}|$, and for diagonal matrix $\|D_p^{(jk)}\|_1=\sum_q|(D_p^{(jk)})_{qq}|$, the second inequality is from the triangle inequality of the trace norm.

The singular values of the diagonal matrix, $D_p^{(jk)}$,  are just the absolute value of the diagonal entries, which are  $\{(\prod_{r=1}^p\lambda_{v_r}) |\sum_{r=1}^p \langle\Psi_{v_r}|[L_j,L_k]|\Psi_{v_r}\rangle|\}$. These entries can be interpreted as the absolute value of the summation of $p$ randomly chosen $\langle\Psi_{v_r}|[L_j,L_k]|\Psi_{v_r}\rangle$ multiplied with the corresponding probabilities, where each term $\langle\Psi_{v_r}|[L_j,L_k]|\Psi_{v_r}\rangle$ is selected with probability $\lambda_{v_r}$. For a given diagonal entry with a particular choice of $p$ terms, $|\sum_{r=1}^p \langle\Psi_{v_r}|[L_j,L_k]|\Psi_{v_r}\rangle|$, the associated probability is $\prod_{r=1}^p\lambda_{v_r}$. $\|D_p^{(jk)}\|_1$, which equals to the summation of the absolute value of all diagonal entries, then corresponds to the expected value of $|\sum_{r=1}^p \langle\Psi_{v_r}|[L_j,L_k]|\Psi_{v_r}\rangle|$ with each $|\Psi_{v_r}\rangle$ selected with probability $\lambda_{v_r}$. When $p\rightarrow \infty$, by the law of large numbers, $\frac{\sum_{r=1}^p \langle\Psi_{v_r}|[L_j,L_k]|\Psi_{v_r}\rangle}{p}$ converges to the expected value of $\langle\Psi_{v_r}|[L_j,L_k]|\Psi_{v_r}\rangle$ with probability one, i.e., with probability one
\begin{eqnarray}
\aligned
\frac{\sum_{r=1}^p \langle\Psi_{v_r}|[L_j,L_k]|\Psi_{v_r}\rangle}{p}&\rightarrow E[\langle\Psi_{v_r}|[L_j,L_k]|\Psi_{v_r}\rangle]\\
&=\sum_{q=1}^m \lambda_q\langle\Psi_{q}|[L_j,L_k]|\Psi_{q}\rangle\\
&=\sum_{q=1}^m \lambda_qTr(|\Psi_{q}\rangle\langle\Psi_{q}|[L_j,L_k])\\
&=Tr(\rho_x[L_j,L_k]).
\endaligned
\end{eqnarray}
Thus when $p\rightarrow \infty$,
\begin{eqnarray}
\aligned
%(C_{p})_{jk}&=\frac{1}{2}\|\sqrt{\rho_x^{\otimes p}}[L_{jp},L_{kp}]|\sqrt{\rho_x^{\otimes p}}\|_1\\
\|D_p^{(jk)}\|_1&=  E[|\sum_{r=1}^p \langle\Psi_{v_r}|[L_j,L_k]|\Psi_{v_r}\rangle|]\\
&\rightarrow 1\times |E[\sum_{r=1}^p \langle\Psi_{v_r}|[L_j,L_k]|\Psi_{v_r}\rangle]|\\
%&=|\sum_{r}p\lambda_r \langle\Psi_{r}|[L_j,L_k]|\Psi_{r}\rangle|\\
&=p|Tr(\rho_x[L_j,L_k])|.
\endaligned
\end{eqnarray}

For the off-diagonal part, note that for any matrix, we have $\|M\|_1\leq \sum_{j}\sqrt{\sum_k|M_{jk}|^2}$, and \begin{eqnarray}
\aligned
&|\frac{\langle\Psi_{\tilde{v}_r}|\sqrt{\rho_x}[L_j,L_k]\sqrt{\rho_x}|\Psi_{v_r}\rangle}{\lambda_{\tilde{v}_r}}\prod_{q=1}^p \lambda_{\tilde{v}_q}|
\leq& l_{\max}\prod_{q=1}^p \lambda_{\tilde{v}_q},
\endaligned
\end{eqnarray}
here $l_{\max}=\max_{\tilde{v}_r\neq v_r} \{|\frac{\langle\Psi_{\tilde{v}_r}|\sqrt{\rho_x}[L_j,L_k]\sqrt{\rho_x}|\Psi_{v_r}\rangle}{\lambda_{\tilde{v}_r}}|\}$, we then have
\begin{eqnarray}
\aligned
\|O_p^{(jk)}\|_1%&\leq \sum_{\tilde{v}_1,\cdots,\tilde{v}_p}\sqrt{\sum_{r=1}^p \sum_{v_r\neq \tilde{v}_r} |\langle\Psi_{\tilde{v}_r}|\sqrt{\rho_x}[L_j,L_k]\sqrt{\rho_x}|\Psi_{v_r}\rangle \prod_{q\neq r} (\lambda_{\tilde{v}_q})|^2}\\
&\leq \sum_{\tilde{v}_1,\cdots,\tilde{v}_p}\sqrt{\sum_{r=1}^p \sum_{v_r\neq \tilde{v}_r} l_{\max}^2\prod_{q=1}^p \lambda^2_{\tilde{v}_q}}\\
&=\sum_{\tilde{v}_1,\cdots,\tilde{v}_p} \sqrt{p(m-1)l_{\max}^2\prod_{q=1}^p \lambda^2_{\tilde{v}_q}}\\
&=\sqrt{(m-1)p}l_{\max}\sum_{\tilde{v}_1,\cdots,\tilde{v}_p}\prod_{q=1}^p \lambda_{\tilde{v}_q}\\
&=\sqrt{(m-1)p}l_{\max}\prod_{q=1}^p(\sum_{\tilde{v}_q=1}^m\lambda_{\tilde{v}_q})\\
&=\sqrt{(m-1)p}l_{\max}.
\endaligned
\end{eqnarray}
Thus when $p\rightarrow \infty$,  %$\|D\|_1\leq \|D+O\|_1\leq \|D\|_1+\|O\|_1$ reduces to
%Thus for $(C_p)_{jk}=\frac{1}{2}\|\sqrt{\rho_x^{\otimes p}}[L_{jp},L_{kp}]|\sqrt{\rho_x^{\otimes p}}\|_1=\frac{1}{2}\|D+O\|_1$,
%\begin{eqnarray}
%\aligned
%\frac{1}{2}\|D\|_1\leq (C_p)_{jk}\leq \frac{1}{2} (\|D\|_1+\|O\|_1),
%\endaligned
%\end{eqnarray}
%when $p\rightarrow \infty$, $\|D\|_1\rightarrow p|Tr(\rho_x[L_j,L_k])|$, we then have
\begin{eqnarray}
\aligned
\frac{\|D_p^{(jk)}+O_p^{(jk)}\|_1}{p}&\geq \frac{\|D_p^{(jk)}\|_1}{p}=|Tr(\rho_x[L_j,L_k])|,\\
\frac{\|D_p^{(jk)}+O_p^{(jk)}\|_1}{p}&\leq \frac{\|D_p^{(jk)}\|_1+\|O_p^{(jk)}\|_1}{p}
\leq  |Tr(\rho_x[L_j,L_k])|+\frac{\sqrt{(m-1)}l_{\max}}{\sqrt{p}}, \endaligned
\end{eqnarray}
i.e.,
\begin{eqnarray}
\aligned
%\frac{\|\sqrt{\rho_x^{\otimes p}}[L_{jp},L_{kp}]|\sqrt{\rho_x^{\otimes p}}\|_1}{p}\geq
%|Tr(\rho_x[L_j,L_k])|,\\
|Tr(\rho_x[L_j,L_k])|\leq \frac{\|\sqrt{\rho_x^{\otimes p}}[L_{jp},L_{kp}]|\sqrt{\rho_x^{\otimes p}}\|_1}{p}\leq |Tr(\rho_x[L_j,L_k])|+\frac{\sqrt{(m-1)}l_{\max}}{\sqrt{p}}.
\endaligned
\end{eqnarray}
From which it is easy to see that $\lim_{p\rightarrow \infty}\frac{\|\sqrt{\rho_x^{\otimes p}}[L_{jp},L_{kp}]|\sqrt{\rho_x^{\otimes p}}\|_1}{p}= |Tr(\rho_x[L_j,L_k])|$. The condition, $\frac{\|\sqrt{\rho_x^{\otimes p}}[L_{jp},L_{kp}]|\sqrt{\rho_x^{\otimes p}}\|_1}{p}=0$, then reduces to the weak commutative condition, $Tr(\rho_x[L_j,L_k])=0$, when $p\rightarrow \infty$. %This solves an open question in \cite{Yang2019}.

%\section{Alternative tradeoff relations}
%The tradeoff relation in Eq.(\ref{eq:tradeoffp}) requires the calculation of $C_p$, which involves operators whose dimension increases exponentially with $p$.
It can also be seen that $\frac{\|D_p^{(jk)}\|_1}{p}$ provides a lower bound on $\frac{\|\sqrt{\rho_x^{\otimes p}}[L_{jp},L_{kp}]|\sqrt{\rho_x^{\otimes p}}\|_1}{p}$ and the difference between them is in the order of $O(\frac{1}{\sqrt{p}})$. We can thus use $\|D_p^{(jk)}\|_1$ to provide an alternative tradeoff relation, which is less tight but easier to compute. Under p-local measurements %with collective measurements on at most $p$ copies of the state, under the parametrization that $F_Q=I$
the tradeoff relation can be written as
\begin{equation}
    \frac{1}{\nu}Tr[F_Q^{-1}Cov^{-1}(\hat{x})]\leq n-\frac{1}{4(n-1)} \|\frac{T_p}{p}\|_F^2,
\end{equation}
where
\begin{eqnarray}
\aligned
    (T_p)_{jk}&=\frac{1}{2}\|D_p^{(jk)}\|_1\\
    &=\frac{1}{2}\sum_{v_1,\cdots, v_p%\in\{1,\cdots, m\}
    }(\prod_{r=1}^p\lambda_{v_r}) |\sum_{r=1}^p \langle\Psi_{v_r}|[\tilde{L}_j,\tilde{L}_k]|\Psi_{v_r}\rangle|,
    \endaligned
\end{eqnarray}
where $\tilde{L}_{j}=\sum_{\mu}(F_Q^{-{\frac{1}{2}}})_{j\mu}L_{\mu}$ and $\tilde{L}_{k}=\sum_{\mu}(F_Q^{-{\frac{1}{2}}})_{k\mu}L_{\mu}$. Compared to $C_p$, $T_p$ is expressed only with operators on a single copy of the state. We note that $T_p$ can be equivalently obtained by choosing the set of $\{|u_j\rangle\}$ in Eq.(\ref{eq:Covu}) as the eigenvectors of $\rho_x$, instead of the eigenvectors of $\sqrt{\rho_x}[\tilde{L}_j,\tilde{L}_k]\sqrt{\rho_x}$.

\section{Bound on the trace norm}\label{sec:appbound}
For completeness, here we include a proof for the inequality $\sum_j|M_{jj}|\leq \|M\|_1\leq \sum_j\sqrt{\sum_k |M_{jk}|^2}$, which is used in the derivation that $\frac{C_p}{p}=0$ reduces to the weak commutative condition when $p\rightarrow \infty$.

We first show $\|M\|_1\leq \sum_j\sqrt{\sum_k |M_{jk}|^2}$.

From the singular value decomposition, $M=U\Lambda V$, we have
\begin{eqnarray}
\aligned
\|M\|_1&=Tr(\Lambda)=Tr(U^\dagger M V^\dagger)=Tr(V^\dagger U^\dagger M)=Tr(WM),
\endaligned
\end{eqnarray}
where $W=V^\dagger U^\dagger$ is a unitary matrix. Note that
\begin{eqnarray}
\aligned
(WM)_{jj}&=\sqrt{|(WM)_{jj}|^2}\leq \sqrt{\sum_k |(WM)_{jk}|^2}\\
&=\|(WM)_{j}\|_2\\
&=\|WM_{j}\|_2\\
&=\|M_{j}\|_2\\
&=\sqrt{\sum_k M^2_{jk}},
\endaligned
\end{eqnarray}
where we used $()_{j}$ to denote the $j$-th column of a matrix(thus $(WM)_j$ is the $j$-th column of $WM$ which equals to $WM_j$, $W$ multiples the $j$-th column of $M$), and $\|v\|_2=\sqrt{\sum_k |v_k|^2}$ as the $l_2$ norm for a vector.
It is then straightforward to see
\begin{eqnarray}
\aligned
\|M\|_1&=Tr(WM)=\sum_{j}(WM)_{jj}\leq \sum_j\sqrt{\sum_k M^2_{jk}},
\endaligned
\end{eqnarray}

Next we show $\sum_j|M_{jj}|\leq \|M\|_1$. From the singular value decomposition, $M=U\Lambda V$, we have
\begin{eqnarray}
  M_{jj}=\sum_k U_{jk}\Lambda_{kk}V_{kj},
\end{eqnarray}
thus
\begin{eqnarray}
\aligned
\sum_j|M_{jj}|&=\sum_j |\sum_k U_{jk}\Lambda_{kk}V_{kj}|\\
&\leq \sum_j \sum_k |U_{jk}\Lambda_{kk}V_{kj}|\\
&=\sum_k\sum_j \Lambda_{kk}|U_{jk}V_{kj}|\\
&\leq \sum_k \Lambda_{kk} \sqrt{(\sum_j |U_{jk}|^2)(\sum_j |V_{kj}|^2)}\\
&=\sum_k \Lambda_{kk}\\
&=\|M\|_1.
\endaligned
\end{eqnarray}

\section{Proof of $Cov_u\geq A_u$}\label{sec:cov>A}
For a mixed state, $\rho_x$, with $x=(x_1,\cdots, x_n)$, given any POVM, $\{M_\alpha\}$, and any $|u\rangle$, we define $Cov_u$ as a $n\times n$ matrix with the $jk$-th entry given by
\begin{equation}
    (Cov_u)_{jk}=\sum_{\alpha}(\hat{x}_j(\alpha)-x_j)(\hat{x}_k(\alpha)-x_k)\langle u|\sqrt{\rho_x}M_{\alpha} \sqrt{\rho_x}|u\rangle,
\end{equation}
and $A_u$ as a $n\times n$ matrix with the $jk$-th entry given by
\begin{eqnarray}\label{eq:Au}
\aligned
&(A_u)_{jk}=\langle u|\sqrt{\rho_x}X_j^\dagger X_k\sqrt{\rho_x}|u\rangle\\
&=\frac{1}{2}\langle u|\sqrt{\rho_x}\{X_j,X_k\}\sqrt{\rho_x}|u\rangle+i\frac{1}{2i}\langle u|\sqrt{\rho_x}[X_j,X_k]\sqrt{\rho_x}|u\rangle,
\endaligned
\end{eqnarray}
here $X_j=\sum_{\alpha}[\hat{x}_j(\alpha)-x_j]M_{\alpha}$ is locally unbiased.

%We first note that for any set of $\{|u_q\rangle\}$ that satisfies $\sum_q |u_q\rangle\langle u_q|=I$, we have $Cov(\hat{x})=\sum_q Cov_{u_q}$. %in particular this holds when $\{|u_q\rangle\}$ form a complete basis. %, $\{|u_1\rangle, \cdots, |u_d\rangle\}$,
%This can be verified by comparing $\sum_q (Cov_{u_q})_{jk}$ and $Cov(\hat{x})_{jk}$ as
%\begin{eqnarray}\label{eq:Covu}
%\aligned
%&\sum_q (Cov_{u_q})_{jk}\\
%&=\sum_q\sum_{\alpha}(\hat{x}_j(\alpha)-x_j)(\hat{x}_k(\alpha)-x_k)\langle u_q|\sqrt{\rho_x}M_{\alpha} \sqrt{\rho_x}|u_q\rangle\\
%&=\sum_{\alpha}(\hat{x}_j(\alpha)-x_j)(\hat{x}_k(\alpha)-x_k)Tr(\rho_xM_{\alpha})\\
%&=Cov(\hat{x})_{jk}.
%\endaligned
%\end{eqnarray}

We then have $Cov_u\geq A_u$ since
for any vector $b=(b_1, \cdots, b_n)^T$,
%\begin{widetext}
\begin{eqnarray}
\aligned
&b^\dagger Cov_u b-b^\dagger A_u b\\
=&\langle u|\sum_{j,k} b_j^*b_k\{\sum_{\alpha}(\hat{x}_j(\alpha)-x_j)(\hat{x}_k(\alpha)-x_k)\sqrt{\rho_x}M_{\alpha} \sqrt{\rho_x}- \sum_{\beta}(\hat{x}_j(\beta)-x_j)\sqrt{\rho_x}M_{\beta}\sum_{\gamma}(\hat{x}_k(\gamma)-x_k)M_{\gamma}]\sqrt{\rho_x}\}|u\rangle \\
=&\langle u|\sum_{j,k} \{\sum_{\alpha}(\hat{x}_j(\alpha)-x_j)b_j^*(\hat{x}_k(\alpha)-x_k)b_k\sqrt{\rho_x}M_{\alpha} \sqrt{\rho_x}\\&- \sum_{\beta}(\hat{x}_j(\beta)-x_j)b_j^*\sqrt{\rho_x}M_{\beta}(\sum_{\alpha}M_{\alpha})\sum_{\gamma}(\hat{x}_k(\gamma)-x_k)b_kM_{\gamma}]\sqrt{\rho_x}\}|u\rangle \\
=&\langle u|\sum_{\alpha}\{[\sum_j(\hat{x}_j(\alpha)-x_j)b_j^*\sqrt{\rho_x}-\sum_j\sum_{\beta}(\hat{x}_j(\beta)-x_j)b_j^*\sqrt{\rho_x}M_{\beta}]M_{\alpha}[\sum_k (x_k(\alpha)-x_k)b_k\sqrt{\rho_x}\\& -\sum_k \sum_{\gamma}(\hat{x}_k(\gamma)-x_k)b_kM_{\gamma}\sqrt{\rho_x}]\}|u\rangle\\
=&\langle u|\sum_{\alpha} M^\dagger(b) M_{\alpha} M(b)]|u\rangle \geq 0,
\endaligned
\end{eqnarray}
%\end{widetext}
here $M(b)=\sum_k (x_k(\alpha)-x_k)b_k\sqrt{\rho_x}-\sum_k \sum_{\gamma}(\hat{x}_k(\gamma)-x_k)b_kM_{\gamma}\sqrt{\rho_x}$. %This then implies $Cov_u\geq A_u$.

\section{Tradeoff relations with RLDs}\label{sec:appRLD}
%We prove the tradeoff relation from
Let
\begin{equation}
    S_u=\left(\begin{array}{cc}
      A_u & B_u  \\
      B^\dagger_u & F_u \\
          \end{array}\right)\geq 0,
\end{equation}
with $(A_u)_{jk}=\langle u|\sqrt{\rho_x} X_jX_k\sqrt{\rho_x}|u\rangle$, $(B_u)_{jk}=\langle u|\sqrt{\rho_x} X_jL_k^{R\dagger}\sqrt{\rho_x}|u\rangle$, $(F_u)_{jk}=\langle u|\sqrt{\rho_x} L_j^{R}L_k^{R\dagger}\sqrt{\rho_x}|u\rangle$, where $L_j^R$ is the RLD corresponding to the parameter $x_j$.

If we choose a complete basis, $\{|u_1\rangle,\cdots, |u_d\rangle\}$, and let $S=\sum_j S_{u_j}=\left(\begin{array}{cc}
A & B\\
B^\dagger & F^{RLD}
\end{array}\right)\geq 0$,
where $(A)_{jk}=Tr(\rho_xX_jX_k)$, $(B)_{jk}=Tr(\rho_xX_jL_k^{R\dagger})=I$, $(F^{RLD})_{jk}=Tr(\rho_xL_j^RL_k^{R\dagger})$, we obtain the RLD bound
\begin{equation}
    Cov(\hat{x})\geq A\geq (F^{RLD})^{-1}.
\end{equation}
This can be equivalently written as
\begin{equation}
    Cov^{-1}(\hat{x})\leq F^{RLD}=F^{RLD}_{Re}+iF^{RLD}_{Im},
\end{equation}
with $F^{RLD}_{Re}$ and $F^{RLD}_{Im}$ as the real and imaginary part of $F^{RLD}$ respectively, $F^{RLD}_{Re}=\frac{1}{2}[F^{RLD}+(F^{RLD})^T]$ is real symmetric and  $F^{RLD}_{Im}=\frac{1}{2}[F^{RLD}-(F^{RLD})^T]$ is real skew-symmetric. By taking the transpose, we also have(note $Cov(\hat{x})$ is symmetric)
\begin{equation}
    Cov^{-1}(\hat{x})\leq (F^{RLD})^T=F^{RLD}_{Re}-iF^{RLD}_{Im}.
\end{equation}
From which we get
\begin{eqnarray}
\aligned
F_Q^{-\frac{1}{2}}Cov^{-1}(\hat{x})F_Q^{-\frac{1}{2}}\leq F_Q^{-\frac{1}{2}}F^{RLD}_{Re}F_Q^{-\frac{1}{2}}\pm iF_Q^{-\frac{1}{2}}F^{RLD}_{Im}F_Q^{-\frac{1}{2}}.
\endaligned
\end{eqnarray}
Then for any vector, $|w\rangle$, we have
\begin{equation}
    \langle w|F_Q^{-\frac{1}{2}}Cov^{-1}(\hat{x})F_Q^{-\frac{1}{2}}|w\rangle\leq \langle w|F_Q^{-\frac{1}{2}}F^{RLD}_{Re}F_Q^{-\frac{1}{2}}|w\rangle- |\langle w|F_Q^{-\frac{1}{2}}F^{RLD}_{Im}F_Q^{-\frac{1}{2}}|w\rangle|.
\end{equation}
By choosing $|w\rangle$ as all the eigenvectors of $F_Q^{-\frac{1}{2}}F^{RLD}_{Im}F_Q^{-\frac{1}{2}}$ and making a summation, we obtain the tradeoff relation from the standard RLD as
\begin{equation}\label{eq:tradeoffRLD1supp}
    Tr[F_Q^{-1}Cov^{-1}(\hat{x})]\leq Tr[F_Q^{-1}F^{RLD}_{Re}]-\|F_Q^{-\frac{1}{2}}F^{RLD}_{Im}F_Q^{-\frac{1}{2}}\|_1.
\end{equation}
When there are $\nu$ copies of the state, this gives
\begin{equation}\label{eq:tradeoffRLDnusupp}
    \frac{1}{\nu}Tr[F_Q^{-1}Cov^{-1}(\hat{x})]\leq Tr[F_Q^{-1}F^{RLD}_{Re}]-\|F_Q^{-\frac{1}{2}}F^{RLD}_{Im}F_Q^{-\frac{1}{2}}\|_1.
\end{equation}

The bound can be improved by taking transposes on any $S_u$. We choose a complete basis, $\{|u_1\rangle, \cdots, |u_d\rangle\}$, as the orthonormal eigenvectors of  $\sqrt{\rho_x}(L_j^RL_k^{R\dagger}-L_k^RL_j^{R\dagger})\sqrt{\rho_x}$. As mentioned in the main text, for any $|u_q\rangle$,
$\frac{1}{2i}\langle u_q|\sqrt{\rho_x}(L_j^RL_k^{R\dagger}-L_k^RL_j^{R\dagger})\sqrt{\rho_x}|u_q\rangle$, which is the imaginary part of $(F_{u_q})_{jk}$, is a real number, which we denote as $t^q_{jk}$. We then define %$\bar{S}_{u_q}:=S_{u_q}$ when $\frac{1}{2i}\langle u_q|\sqrt{\rho_x}[L_j,L_k]\sqrt{\rho_x}|u_q\rangle\geq 0$, $\bar{S}_{u_q}:=S_{u_q}^T$ when $\frac{1}{2i}\langle u_q|\sqrt{\rho_x}[L_j,L_k]\sqrt{\rho_x}|u_q\rangle<0$. %$\bar{S}_{u_j}$ thus has the same real part as $S_{u_j}$ and
\begin{eqnarray}
\bar{S}_{u_q}:=\{\begin{array}{cc}
S_{u_q},  \text{when } t^q_{jk}\geq 0,\\%\frac{1}{2i}\langle u_q|\sqrt{\rho_x}(L_j^RL_k^{R\dagger}-L_k^RL_j^{R\dagger})\sqrt{\rho_x}|u_q\rangle\geq 0\\
S^T_{u_q}, \text{when } t^q_{jk}<0.%\frac{1}{2i}\langle u_q|\sqrt{\rho_x}(L_j^RL_k^{R\dagger}-L_k^RL_j^{R\dagger})\sqrt{\rho_x}|u_q\rangle< 0
\end{array}
\end{eqnarray}
By summing $\bar{S}_{u_q}$ we get %$\bar{S}=\sum_j \bar{S}_{u_j}$.
\begin{eqnarray}\label{eq:suqRLDsupp}
\bar{S}=\sum_q \bar{S}_{u_q}=\left(\begin{array}{cc}
\bar{A} & \bar{B}\\
\bar{B}^\dagger & \bar{F}^{RLD}
\end{array}\right),
\end{eqnarray}
here $\bar{B}=I+i\bar{B}_{Im}$, $\bar{F}^{RLD}=\sum_q\bar{F}_{u_q}$ with $\bar{F}_{u_q}$ equals to either $F_{u_q}$ or $F^T_{u_q}$ so that the imaginary part of $(\bar{F}_{u_q})_{jk}$ is always positive. %which equals to one-half of the absolute value of the eigenvalue that corresponds to the eigenvector $|u_q\rangle$ of $\sqrt{\rho_x}[L_j,L_k]\sqrt{\rho_x}$.
The imaginary part of the $jk$-th entry of $\bar{F}^{RLD}$ is then given by %which is the sum of the $jk$-th entry of $\bar{F}_{u_q}$, is just
\begin{equation}\label{eq:Fjksupp}
(\bar{F}_{Im})_{jk}=\frac{1}{2}\|\sqrt{\rho_x}(L_j^RL_k^{R\dagger}-L_k^RL_j^{R\dagger})\sqrt{\rho_x}\|_1,
\end{equation}
and the real part of $\bar{F}^{RLD}$ remains the same as $F^{RLD}_{Re}$.

%where $\| \|_1$ is the trace norm which equals to the sum of singular values and for the skew-Hermitian matrix just equals to the sum of the absolute value of the eigenvalues.
%look at the $jk$-th entry of $F_u=F_$
%In this case  $Cov(\hat{x})\geq\bar{A}\geq(\bar{F}^{RLD})^{-1}$, i.e., $Cov^{-1}(\hat{x})\leq \bar{F}^{RLD}=F^{RLD}_{Re}+i\bar{F}_{Im}^{RLD}$. Since any $2\times 2$ submatrix of $F^{RLD}_{Re}+i\bar{F}_{Im}^{RLD}-Cov^{-1}(\hat{x})$ is also positive semidefinite, i.e.,

By the Schur's complement we then have $\bar{F}^{RLD}-\bar{B}^\dagger Cov^{-1}(\hat{x})\bar{B}\geq 0$, which can be equivalently written as
\begin{eqnarray}
%\label{eq:schur}
\bar{F}_{Re}^{RLD}+i\bar{F}_{Im}^{RLD}-[Cov^{-1}(\hat{x})+\bar{B}_{Im}^TCov^{-1}(\hat{x})\bar{B}_{Im}+i(\bar{B}_{Im}^TCov^{-1}(\hat{x})-Cov^{-1}(\hat{x})\bar{B}_{Im})]\geq 0.
\end{eqnarray}
%By multiplying $F_Q^{-\frac{1}{2}}$ from both left and right, it gives
%\begin{eqnarray}
%\label{eq:schur}
%\tilde{F}_{Re}^{RLD}+i\tilde{F}_{Im}^{RLD}-[\tilde{Cov}^{-1}(\hat{x})+\tilde{B}_{Im}^T\tilde{Cov}^{-1}(\hat{x})\tilde{B}_{Im}+i(\tilde{B}_{Im}^T\tilde{Cov}^{-1}(\hat{x})-\tilde{Cov}^{-1}(\hat{x})\tilde{B}_{Im})]\geq 0,
%\end{eqnarray}
%where $\tilde{F}_{Re}^{RLD}=F_Q^{-\frac{1}{2}}\bar{F}_{Re}^{RLD}F_Q^{-\frac{1}{2}}$, $\tilde{F}_{Im}^{RLD}=F_Q^{-\frac{1}{2}}\bar{F}_{Im}^{RLD}F_Q^{-\frac{1}{2}}$, $\tilde{Cov}^{-1}(\hat{x})=F_Q^{-\frac{1}{2}}Cov^{-1}(\hat{x})F_Q^{-\frac{1}{2}}$, $\tilde{B}_{Im}=F_Q^{\frac{1}{2}}\bar{B}_{Im} F_Q^{-\frac{1}{2}}$.
We first assume $F_Q=I$, in this case $Cov^{-1}(\hat{x})\leq F_Q=I$. Then by following the same procedure as previous, we denote $Cov^{-1}(\hat{x})\bar{B}_{Im}$ as $D$ and get
 \begin{eqnarray}
%\label{eq:schur}
\bar{F}_{Re}^{RLD}-Cov^{-1}(\hat{x})-D^TD+i(\bar{F}_{Im}^{RLD}+D^T-D]\geq 0.
\end{eqnarray}
By taking a $2\times 2$ principle submatrix %of a positive semidefinite matrix is also positive semidefinite,
we have
 \begin{eqnarray}
%\label{eq:schur}
%\left(\begin{array}{cc}
%      1 & 0  \\
%      0 & 1 \\
%          \end{array}\right)-\left(\begin{array}{cc}
 %     (\tilde{F}_C)_{jj} & (\tilde{F}_C)_{jk}  \\
 %     (\tilde{F}_C)_{kj} & (\tilde{F}_C)_{kk} \\
 %         \end{array}\right)-
 \aligned
 &\left(\begin{array}{cc}
      (\bar{F}_{Re}^{RLD})_{jj}-Cov^{-1}(\hat{x})_{jj}-(D^TD)_{jj} & -Cov^{-1}(\hat{x})_{jk}-(D^TD)_{jk} \\
      -Cov^{-1}(\hat{x})_{kj}-(D^TD)_{kj} & (\bar{F}_{Re}^{RLD})_{kk}-Cov^{-1}(\hat{x})_{kk}-(D^TD)_{kk} \\
          \end{array}\right)\\
          +&i\left(\begin{array}{cc}
      0 & (\bar{F}_{Im}^{RLD})_{jk}+D_{kj}-D_{jk}  \\
      -(\bar{F}_{Im}^{RLD})_{jk}-D_{kj}+D_{jk} & 0 \\
          \end{array}\right)\geq 0.
          \endaligned
\end{eqnarray}
%Note that $\tilde{F}_C$ and $D^TD$ are symmetric and the determination of a positive semidefinite matrix is nonnegative, we thus have
From the positiveness of the determinant, we have
\begin{eqnarray}
\aligned
&[(\bar{F}_{Re}^{RLD})_{jj}-Cov^{-1}(\hat{x})_{jj}-(D^TD)_{jj}][(\bar{F}_{Re}^{RLD})_{kk}-Cov^{-1}(\hat{x})_{kk}-(D^TD)_{kk}]\\
\geq & [Cov^{-1}(\hat{x})_{jk}+(D^TD)_{jk}]^2+[(\bar{F}_{Im}^{RLD})_{jk}+D_{kj}-D_{jk}]^2,
\endaligned
\end{eqnarray}
from which we can get
\begin{eqnarray}
\aligned
&[(\bar{F}_{Re}^{RLD})_{jj}-Cov^{-1}(\hat{x})_{jj}-(D^TD)_{jj}]+[(\bar{F}_{Re}^{RLD})_{kk}-Cov^{-1}(\hat{x})_{kk}-(D^TD)_{kk}]\\
&\geq 2\sqrt{[(\bar{F}_{Re}^{RLD})_{jj}-Cov^{-1}(\hat{x})_{jj}-(D^TD)_{jj}][(\bar{F}_{Re}^{RLD})_{kk}-Cov^{-1}(\hat{x})_{kk}-(D^TD)_{kk}]}\\
&\geq 2\sqrt{[Cov^{-1}(\hat{x})_{jk}+(D^TD)_{jk}]^2+[(\bar{F}_{Im}^{RLD})_{jk}+D_{kj}-D_{jk}]^2}\\
&\geq 2|(\bar{F}_{Im}^{RLD})_{jk}+D_{kj}-D_{jk}|,
\endaligned
\end{eqnarray}
i.e.,
\begin{eqnarray}%\label{eq:jk}
(\bar{F}_{Re}^{RLD})_{jj}-Cov^{-1}(\hat{x})_{jj}+(\bar{F}_{Re}^{RLD})_{kk}-Cov^{-1}(\hat{x})_{kk}
\geq 2|(\bar{F}_{Im}^{RLD})_{jk}+D_{kj}-D_{jk}|+(D^TD)_{jj}+(D^TD)_{kk}.
\end{eqnarray}
Again as $(D^TD)_{jj}=\sum_p D_{pj}^2\geq D_{kj}^2$ and $(D^TD)_{kk}=\sum_p D_{pk}^2\geq D_{jk}^2$, we have
\begin{equation}
(D^TD)_{jj}+(D^TD)_{kk}=\sum_p (D_{pj}^2+D_{pk}^2)\geq D_{kj}^2+D_{jk}^2=\frac{1}{2}(D_{kj}-D_{jk})^2+\frac{1}{2}(D_{kj}+D_{jk})^2.
\end{equation}
%and from $I+i\tilde{F}_{Im}=F_Q^{-\frac{1}{2}}FF_Q^{-\frac{1}{2}}\geq 0$, we have $|(\tilde{F}_{Im})_{jk}|\leq 1$.
Thus
\begin{eqnarray}\label{eq:tradeoffRLDsupp}
\aligned
&(\bar{F}_{Re}^{RLD})_{jj}-Cov^{-1}(\hat{x})_{jj}+(\bar{F}_{Re}^{RLD})_{kk}-Cov^{-1}(\hat{x})_{kk}
\geq &2|(\bar{F}_{Im}^{RLD})_{jk}+D_{kj}-D_{jk}|+(D^TD)_{jj}+(D^TD)_{kk}.
\\
\geq &2|(\bar{F}_{Im}^{RLD})_{jk}+D_{kj}-D_{jk}|+\frac{1}{2}(D_{kj}-D_{jk})^2\\
\geq & \min \{\frac{1}{2}|(\bar{F}_{Im})_{jk}|^2,2\},
\endaligned
\end{eqnarray}
where the last inequality we used the fact that when $|y|\leq 2$, $2|y+x|+\frac{1}{2}x^2\geq \frac{1}{2}y^2$ since
\begin{eqnarray}
\aligned
2|y+x|+\frac{1}{2}x^2&=2|y+x|+\frac{1}{2}(y+x-y)^2\\
&=2|y+x|+\frac{1}{2}(y+x)^2-y(x+y)+\frac{1}{2}y^2\\
&\geq 2|y+x|-|y(x+y)|+\frac{1}{2}y^2\\
&=(2-|y|)|x+y|+\frac{1}{2}y^2\\
&\geq \frac{1}{2}y^2,
\endaligned
\end{eqnarray}
while when $|y|\geq 2$, $2|y+x|+\frac{1}{2}x^2\geq 2$ since
\begin{eqnarray}
\aligned
2|y+x|+\frac{1}{2}x^2&\geq 2(|y|-|x|)+\frac{1}{2}x^2\\
&=\frac{1}{2}(|x|-2)^2-2+2|y|\\
&\geq 2|y|-2\\
&\geq 2.
\endaligned
\end{eqnarray}
From which we can get
\begin{eqnarray}
\aligned
(\bar{F}^{RLD}_{Re})_{jj}-Cov^{-1}(\hat{x})_{jj}+(\bar{F}^{RLD}_{Re})_{kk}-Cov^{-1}(\hat{x})_{kk}
\geq \min \{\frac{1}{2}|\bar{F}^{RLD}_{Im})_{jk}|^2, 2\}.
\endaligned
\end{eqnarray}
By repeating the procedure for different choices of $j,k$ and make a summation, we then get the tradeoff relation, under the parametrization that $F_Q=I$, as
%\begin{equation}
%    Tr[Cov^{-1}(\hat{x})]\leq Tr[\bar{F}^{RLD}_{Re}]-\frac{1}{2(n-1)}\sum_{j,k}|(C^{RLD}_1)_{jk}|.
%\end{equation}
%with $(C^{RLD}_1)_{jk}=\min\{\frac{1}{8}\|\sqrt{\rho_x}(L_j^RL_k^{R\dagger}-L_k^RL_j^{R\dagger})\sqrt{\rho_x}\|_1^2,2\}$, which can also be equivalently written as
%{\color{blue}
%or equivalently as,
\begin{equation}
    Tr[Cov^{-1}(\hat{x})]\leq Tr[\bar{F}^{RLD}_{Re}]-\frac{1}{4(n-1)}\|C^{RLD}_1\|_F^2,
\end{equation}
with $(C^{RLD}_1)_{jk}=\min\{\frac{1}{2}\|\sqrt{\rho_x}(L_j^RL_k^{R\dagger}-L_k^RL_j^{R\dagger})\sqrt{\rho_x}\|_1,2\}$.

If we repeat the 1-local measurement on $\nu$ copies of the state, the tradeoff relation under the 1-local measurement, with the parametrization that $F_Q=I$, is then
\begin{equation}\label{eq:supptradeoffRLD}
    \frac{1}{\nu}Tr[Cov^{-1}(\hat{x})]\leq Tr[F^{RLD}_{Re}]-\frac{1}{4(n-1)}\|C^{RLD}_1\|_F^2
%    \frac{1}{2(n-1)}\sum_{j,k}|(C^{RLD}_1)_{jk}|.
\end{equation}
When $F_Q\neq I$ initially, we can first make a reparametrization with $\tilde{x}=F_Q^{-\frac{1}{2}}x$.  The tradeoff relation in Eq.(\ref{eq:supptradeoffRLD}) can then be expressed in the original parametrization as
\begin{eqnarray}\label{eq:tradeoffRLDoriginsupp}
\aligned
    &\frac{1}{\nu}Tr[F_Q^{-1}Cov(\hat{x})^{-1}]\\
    \leq &Tr[F_Q^{-1}F^{RLD}_{Re}]-\frac{1}{4(n-1)}\|C^{RLD}_1\|_F^2
    %\frac{1}{2(n-1)} \sum_{j,k}|(C^{RLD}_{1})_{jk}|,
    \endaligned
\end{eqnarray}
with the entries of $C^{RLD}_1$ given by
\begin{eqnarray}
\aligned
(C^{RLD}_1)_{jk}&=\min \{\frac{1}{2}\|\sqrt{\rho_x}(\tilde{L}_j^R\tilde{L}_k^{R\dagger}-\tilde{L}_k^R\tilde{L}_j^{R\dagger})\sqrt{\rho_x}\|_1, 2\}
%&=\frac{1}{2}\|\sqrt{\rho_x}[\sum_q (F_Q^{-\frac{1}{2}})_{jq}L^{RLD}_q,\sum_q (F_Q^{-\frac{1}{2}})_{kq}L^{RLD}_q]\sqrt{\rho_x}\|_1.
\endaligned
\end{eqnarray}
where $\tilde{L}^R_j=\sum_q (F_Q^{-\frac{1}{2}})_{jq}L^R_q$ and $\tilde{L}^R_k=\sum_q (F_Q^{-\frac{1}{2}})_{kq}L^R_q$.

%{\color{blue}
For p-local measurements, we can similarly get
\begin{equation}
  \frac{1}{\nu}Tr[F_{Q}^{-1}Cov^{-1}(\hat{x})]\le Tr[F_{Q}^{-1}F_{Re}^{RLD}]-\frac{1}{4(n-1)}\|\frac{C_p^{RLD}}{p}\|_F^2,
\end{equation}
where $(C_p^{RLD})_{jk}=\min\{\frac{1}{2}\|\sqrt{\rho_x^{\otimes p}}(\tilde{L}_{jp}^R\tilde{L}_{kp}^{R\dagger}-\tilde{L}_{kp}^R\tilde{L}_{jp}^{R\dagger})\sqrt{\rho_x^{\otimes p}}\|_1,2p\}$.

\section{Example 2}

Here we provided more detailed calculations for example 2.

For mixed states $\rho_x=\frac{1}{3}I+\sum_jx_jG_j$ with $G_j=\frac{1}{2}\Lambda_j$, where $\{\Lambda_j\}_{j=1}^8$ are the Gell-Mann matrices,
\begin{equation}
  % \begingroup
  % \renewcommand{\arraystretch}{0.8}
  % \setlength\arraycolsep{6pt}
  \begin{aligned}
    &\Lambda_1=\begin{pmatrix}
      0 & 1 & 0\\
      1 & 0 & 0\\
      0 & 0 & 0
    \end{pmatrix},
    \Lambda_2=\begin{pmatrix}
      0 & -i & 0\\
      i & 0 & 0\\
      0 & 0 & 0
    \end{pmatrix},
    \Lambda_3=\begin{pmatrix}
      1 & 0 & 0\\
      0 & -1 & 0\\
      0 & 0 & 0
    \end{pmatrix},\\
    &\Lambda_4=\begin{pmatrix}
      0 & 0 & 1\\
      0 & 0 & 0\\
      1 & 0 & 0
    \end{pmatrix},
    \Lambda_5=\begin{pmatrix}
      0 & 0 & -i\\
      0 & 0 & 0\\
      i & 0 & 0
    \end{pmatrix},\\
    &\Lambda_6=\begin{pmatrix}
      0 & 0 & 0\\
      0 & 0 & 1\\
      0 & 1 & 0
    \end{pmatrix},
    \Lambda_7=\begin{pmatrix}
      0 & 0 & 0\\
      0 & 0 & -i\\
      0 & i & 0
    \end{pmatrix},
    \Lambda_8=\frac{1}{\sqrt{3}}\begin{pmatrix}
      1 & 0 & 0\\
      0 & 1 & 0\\
      0 & 0 & -2
    \end{pmatrix}.
  \end{aligned}
  %\endgroup
\end{equation}
If the parameters $x_j$ are all close to 0, the SLDs and RLDs are all given by $L_j=3G_j$.
Thus the tradeoff relations from the SLDs and RLDs will be the same.
The QFI matrix is given as $F_Q=F^{RLD}=\frac{3}{2}I$, thus $\tilde{L}_j=\sqrt{\frac{2}{3}}L_j=\sqrt{6}G_j$.
The entries of $C_1$ is given by
\begin{equation}
  (C_1)_{jk}=\frac{1}{2}\|\sqrt{\rho_x}[\tilde{L}_j,\tilde{L}_k]\sqrt{\rho_x}\|_1=\|[G_j,G_k]\|_1,
\end{equation}
from which the matrix form of $C_1$ can be computed as
\begin{equation}
  C_1=
  \begin{pmatrix}
    0 & 1 & 1 & \frac{1}{2} & \frac{1}{2} & \frac{1}{2} & \frac{1}{2} & 0\\
    1 & 0 & 1 & \frac{1}{2} & \frac{1}{2} & \frac{1}{2} & \frac{1}{2} & 0\\
    1 & 1 & 0 & \frac{1}{2} & \frac{1}{2} & \frac{1}{2} & \frac{1}{2} & 0\\
    \frac{1}{2} & \frac{1}{2} & \frac{1}{2} & 0 & 1 & \frac{1}{2} & \frac{1}{2} & \frac{\sqrt{3}}{2}\\
    \frac{1}{2} & \frac{1}{2} & \frac{1}{2} & 1 & 0 & \frac{1}{2} & \frac{1}{2} & \frac{\sqrt{3}}{2}\\
    \frac{1}{2} & \frac{1}{2} & \frac{1}{2} & \frac{1}{2} & \frac{1}{2} & 0 & 1 & \frac{\sqrt{3}}{2}\\
    \frac{1}{2} & \frac{1}{2} & \frac{1}{2} & \frac{1}{2} & \frac{1}{2} & 1 & 0 & \frac{\sqrt{3}}{2}\\
    0 & 0 & 0 & \frac{\sqrt{3}}{2} & \frac{\sqrt{3}}{2} & \frac{\sqrt{3}}{2} & \frac{\sqrt{3}}{2} & 0
  \end{pmatrix},
\end{equation}
% For $p\ge 2$, it can be computed that $C_2=\frac{4}{3}C_1$ and $C_3=\frac{5}{3}C_1$.
Thus we have
\begin{equation}
  \begin{aligned}
    \frac{1}{\nu}Tr[F_Q^{-1}Cov(\hat{x})^{-1}]&\le n-\frac{1}{4(n-1)}\|C_1\|_F^2\\
    &= 8-\frac{1}{28}\times 2(5\times 1 +4\times\frac{3}{4}+16 \times \frac{1}{4})=\frac{50}{7}\approx 7.14.
  \end{aligned}
\end{equation}

For $p$-local measurements on $\rho_x$, the entries of $C_p$ is given by
\begin{equation}
  (C_p)_{jk}=\frac{1}{2}\|\sqrt{\rho_x^{\otimes p}}[\tilde{L}_{jp},\tilde{L}_{kp}]\sqrt{\rho_x^{\otimes p}}\|_1=\frac{1}{3^{p-1}}\|[G_{jp},G_{kp}]\|_1,
\end{equation}
For all $j,k$, the eigenvalues of $[G_{j},G_{k}]$ are $\{-\lambda,0,\lambda\}$, where $\lambda=\frac{1}{2}$ or $\frac{1}{4}$ or $\frac{\sqrt{3}}{4}$.
%{\color{purple}
Suppose that the eigenvectors corresponding to eigenvalues $\{-\lambda,0,\lambda\}$ can be written as $\{\ket{\Phi_l}\}_{l\in\{-\lambda,0,\lambda\}}$.
The eigenvectors of $[G_{jp},G_{kp}]=\sum_{r=1}^p[G_j^{(r)},G_k^{(r)}]$ are then given by $\otimes_{r=1}^p\ket{\Phi_{l_r}}$ with the corresponding eigenvalues $\sum_{r=1}^p l_r$, here $l_r\in\{-\lambda,0,\lambda\}$.
, %whose possible values can be generated by adding $-\lambda$, $0$, or $\lambda$ to all of the possible values of $\sum_{r=1}^{p-1} l_r$.
The recursive relation to obtain the eigenvalues is depicted in Fig. \ref{fig.trinomial}(a), where in Fig. \ref{fig.trinomial}(b) a few possible values of $\sum_{r=1}^p l_r$ have been listed(note that the $(p+1)$-th row in Fig. \ref{fig.trinomial}(b) corresponds to all possible values of $\sum_{r=1}^p l_r$). The multiplicity of each eigenvalue can be obtained as Fig. \ref{fig.trinomial}(c), which is just the trinomial triangle that corresponds to the coefficients of $(1+x+x^2)^p$.
\begin{figure}
  \includegraphics[width=0.7\textwidth]{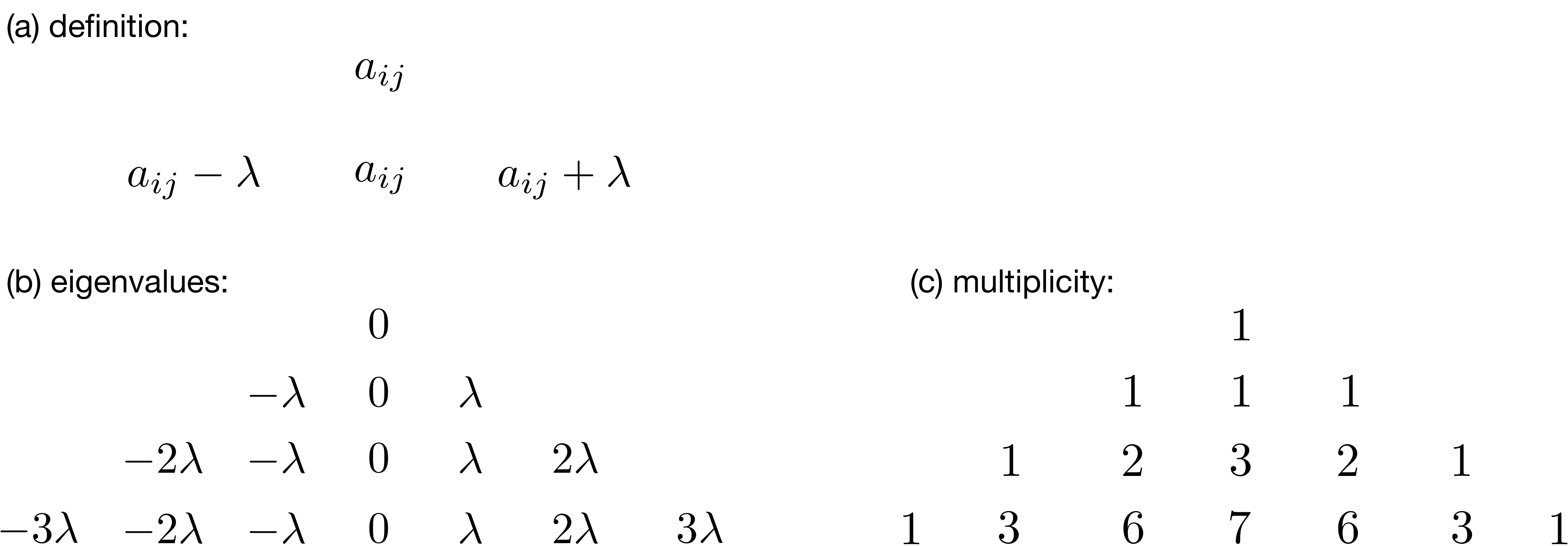}
  \caption{Eigenvalues and multiplicities of $\sum_{r=1}^p[G_j^{(r)},G_k^{(r)}]$.}
  \label{fig.trinomial}
\end{figure}
%}
Hence the eigenvalues of $[G_{jp},G_{kp}]=\sum_{r=1}^p[G_j^{(r)},G_k^{(r)}]$ are $\lambda s$ with multiplicity $\binom{p}{s}_2$ for $s=-p,-p+1,\cdots,p$, here
%\begin{equation}
  $\binom{p}{s}_2=\sum_{i=0}^p (-1)^i\binom{p}{i}\binom{2p-2i}{p-s-i}$
%\end{equation}
is the trinomial coefficient.% which is also the $(j+p)$-th coefficient of the polynomial $(1+x+x^2)^p$.

Denote $\mathcal{N}_p=\sum_{s=0}^p s\binom{p}{s}_2$, we then have
\begin{equation}
  (C_p)_{jk}=\frac{1}{3^{p-1}}\|[G_{jp},G_{kp}]\|_1=(C_1)_{jk}\frac{\mathcal{N}_p}{3^{p-1}}.
\end{equation}
The Frobenius norm of $C_p$ is then given by
\begin{equation}
  \begin{aligned}
    \|C_p\|_F&=\sqrt{\sum_{jk}(C_p)_{jk}^2}=\sqrt{\sum_{jk}\left((C_1)_{jk}\frac{1}{3^{p-1}}\mathcal{N}_p\right)^2}\\
    &=\frac{1}{3^{p-1}}\mathcal{N}_p\sqrt{\sum_{jk}\left((C_1)_{jk}\right)^2}=\frac{1}{3^{p-1}}\mathcal{N}_p\|C_1\|_F.
  \end{aligned}
\end{equation}

%We define the norm
%\begin{equation}
%  \mathcal{N}_p(\lambda)=\|[G_{jp},G_{kp}]\|_1=\sum_{j=-p}^p |\lambda j|\binom{p}{j}_2=2\lambda\sum_{j=0}^p j\binom{p}{j}_2,
%\end{equation}
%where we have used the fact that $\binom{p}{j}_2=\binom{p}{-j}_2$.
%It is then easy to see that
%\begin{equation}
%  (C_p)_{jk}=\frac{1}{3^{p-1}}\mathcal{N}_p(\lambda)=\frac{1}{3^{p-1}}\mathcal{N}_p(\frac{1}{2})(C_1)_{jk}.
%\end{equation}
%The Frobenius norm of $C_p$ can be computed as
%\begin{equation}
%  \begin{aligned}
%    \|C_p\|_F&=\sqrt{\sum_{jk}(C_p)_{jk}^2}=\sqrt{\sum_{jk}\left(\frac{1}{3^{p-1}}\mathcal{N}_p(\frac{1}{2})(C_1)_{jk}\right)^2}\\
%    &=\frac{1}{3^{p-1}}\mathcal{N}_p(\frac{1}{2})\sqrt{\sum_{jk}\left((C_1)_{jk}\right)^2}=\frac{1}{3^{p-1}}\mathcal{N}_p(\frac{1}{2})\|C_1\|_F
%  \end{aligned}
%\end{equation}
Using the tradeoff relations for $p$-local measurements, we have
\begin{equation}
  \begin{aligned}
    \frac{1}{\nu}Tr[F_Q^{-1}Cov(\hat{x})^{-1}]&\le n-\frac{1}{4(n-1)}\|\frac{C_p}{p}\|_F^2\\
    &= n-\frac{1}{4(n-1)}\|C_1\|_F^2\left(\frac{1}{p3^{p-1}}\mathcal{N}_p\right)^2\\
    &= 8-\frac{6}{7}\left(\frac{1}{p3^{p-1}}\mathcal{N}_p\right)^2
  \end{aligned}
\end{equation}
Specifically, for 2-local measurements,
\begin{equation}
  \frac{1}{\nu}Tr[F_Q^{-1}Cov(\hat{x})^{-1}]\le n-\frac{1}{4(n-1)}\|\frac{C_2}{2}\|_F^2=8-\frac{6}{7}\times \frac{1}{4}\times\frac{16}{9}=\frac{160}{21}\approx 7.62
\end{equation}
and for 3-local measurements,
\begin{equation}
  \frac{1}{\nu}Tr[F_Q^{-1}Cov(\hat{x})^{-1}]\le n-\frac{1}{4(n-1)}\|\frac{C_3}{3}\|_F^2=8-\frac{6}{7}\times \frac{1}{9}\times\frac{25}{9}=\frac{1462}{189}\approx 7.74
\end{equation}

If we choose the basis $\{\ket{u}_q\}$ as computational basis $\ket{u}_0=\ket{0}$, $\ket{u}_1=\ket{1}$, $\ket{u}_2=\ket{2}$, the matrices $F_{u_q}$ are given as
\begin{equation}
  F_{u_0}=\frac{1}{2}
  \begin{pmatrix}
    1 & i & 0 & 0 & 0 & 0 & 0 & 0\\
    -i & 1 & 0 & 0 & 0 & 0 & 0 & 0\\
    0 & 0 & 1 & 0 & 0 & 0 & 0 & \frac{\sqrt{3}}{3}\\
    0 & 0 & 0 & 1 & i & 0 & 0 & 0\\
    0 & 0 & 0 & -i & 1 & 0 & 0 & 0\\
    0 & 0 & 0 & 0 & 0 & 0 & 0 & 0\\
    0 & 0 & 0 & 0 & 0 & 0 & 0 & 0\\
    0 & 0 & \frac{\sqrt{3}}{3} & 0 & 0 & 0 & 0 & \frac{1}{3}
  \end{pmatrix},
  F_{u_1}=\frac{1}{2}
  \begin{pmatrix}
    1 & -i & 0 & 0 & 0 & 0 & 0 & 0\\
    i & 1 & 0 & 0 & 0 & 0 & 0 & 0\\
    0 & 0 & 1 & 0 & 0 & 0 & 0 & -\frac{\sqrt{3}}{3}\\
    0 & 0 & 0 & 0 & 0 & 0 & 0 & 0\\
    0 & 0 & 0 & 0 & 0 & 0 & 0 & 0\\
    0 & 0 & 0 & 0 & 0 & 1 & i & 0\\
    0 & 0 & 0 & 0 & 0 & -i & 1 & 0\\
    0 & 0 & -\frac{\sqrt{3}}{3} & 0 & 0 & 0 & 0 & \frac{1}{3}
  \end{pmatrix}
\end{equation}
\begin{equation}
  F_{u_2}=\frac{1}{2}
  \begin{pmatrix}
    0 & 0 & 0 & 0 & 0 & 0 & 0 & 0\\
    0 & 0 & 0 & 0 & 0 & 0 & 0 & 0\\
    0 & 0 & 0 & 0 & 0 & 0 & 0 & 0\\
    0 & 0 & 0 & 1 & -i & 0 & 0 & 0\\
    0 & 0 & 0 & i & 1 & 0 & 0 & 0\\
    0 & 0 & 0 & 0 & 0 & 1 & -i & 0\\
    0 & 0 & 0 & 0 & 0 & i & 1 & 0\\
    0 & 0 & 0 & 0 & 0 & 0 & 0 & \frac{4}{3}
  \end{pmatrix},
  \bar{F}_{Im}=
  \begin{pmatrix}
    0 & 1 & 0 & 0 & 0 & 0 & 0 & 0\\
    -1 & 0 & 0 & 0 & 0 & 0 & 0 & 0\\
    0 & 0 & 0 & 0 & 0 & 0 & 0 & 0\\
    0 & 0 & 0 & 0 & 1 & 0 & 0 & 0\\
    0 & 0 & 0 & -1 & 0 & 0 & 0 & 0\\
    0 & 0 & 0 & 0 & 0 & 0 & 0 & 0\\
    0 & 0 & 0 & 0 & 0 & 0 & 0 & 0\\
    0 & 0 & 0 & 0 & 0 & 0 & 0 & 0
  \end{pmatrix}
\end{equation}
Let $\bar{F}=F_{u_0}+F_{u_1}^T+F_{u_2}^T$, this gives a bound as
\begin{equation}
  \frac{1}{\nu}Tr[F_Q^{-1}Cov(\hat{x})^{-1}]\le n-\frac{(n-2)}{(n-1)^2}\|\bar{F}_{Im}\|_F^2=8-\frac{6}{49}\times 4\approx 7.51,
\end{equation}

If we only estimate $\{x_1,x_2,x_4,x_5\}$,
the associated matrices are given by the $4\times 4$ submatrices of the original ones,
\begin{equation}
  \begin{aligned}
    C_1=
    \begin{pmatrix}
      0 & 1 & \frac{1}{2} & \frac{1}{2}\\
      1 & 0 & \frac{1}{2} & \frac{1}{2}\\
      \frac{1}{2} & \frac{1}{2} & 0 & 1\\
      \frac{1}{2} & \frac{1}{2} & 1 & 0
    \end{pmatrix},
    \bar{F}_{Im}=
    \begin{pmatrix}
      0 & 1 & 0 & 0\\
      -1 & 0 & 0 & 0\\
      0 & 0 & 0 & 1 \\
      0 & 0 & -1 & 0
    \end{pmatrix},
  \end{aligned}
\end{equation}
which further gives $\|C_1\|_F=\sqrt{6}$, $\|\bar{F}_{Im}\|_F=2$.
Then we have
\begin{equation}
  \frac{1}{\nu}Tr[F_Q^{-1}Cov(\hat{x})^{-1}]\le n-\frac{1}{4(n-1)}\|C_1\|_F^2=4-\frac{1}{2}=\frac{7}{2}=3.5,
\end{equation}
\begin{equation}
  \frac{1}{\nu}Tr[F_Q^{-1}Cov(\hat{x})^{-1}]\le n-\frac{(n-2)}{(n-1)^2}\|\bar{F}_{Im}\|_F^2=4-\frac{8}{9}=\frac{28}{9}\approx 3.11.
\end{equation}

For $p$-local measurements, by following the same derivation as the previous case, we have
\begin{equation}
  \begin{aligned}
    \frac{1}{\nu}Tr[F_Q^{-1}Cov(\hat{x})^{-1}]&\le n-\frac{1}{4(n-1)}\|\frac{C_p}{p}\|_F^2\\
    &= n-\frac{1}{4(n-1)}\|C_1\|_F^2\left(\frac{1}{p3^{p-1}}\mathcal{N}_p\right)^2\\
    &= 4-\frac{1}{2}\left(\frac{1}{p3^{p-1}}\mathcal{N}_p\right)^2
  \end{aligned}
\end{equation}
Specifically, for p=2 we have %2-local measurements,
\begin{equation}
  \frac{1}{\nu}Tr[F_Q^{-1}Cov(\hat{x})^{-1}]\le n-\frac{1}{4(n-1)}\|\frac{C_2}{2}\|_F^2=4-\frac{1}{2}\times \frac{1}{4}\times\frac{16}{9}=\frac{34}{9}\approx 3.78,
\end{equation}
and for p=3,
\begin{equation}
  \frac{1}{\nu}Tr[F_Q^{-1}Cov(\hat{x})^{-1}]\le n-\frac{1}{4(n-1)}\|\frac{C_3}{3}\|_F^2=4-\frac{1}{2}\times \frac{1}{9}\times\frac{25}{9}=\frac{623}{162}\approx 3.85.
\end{equation}

If we choose the basis $\{\ket{u_q}\}$ as the computational basis $\ket{u_0}=\ket{00}$, $\ket{u_1}=\ket{01}$, $\ket{u_2}=\ket{02}$, $\ket{u_3}=\ket{10}$,..., $\ket{u_8}=\ket{22}$, the imaginary part of the matrices $F_{u_q}$ are given as
\begin{equation}
  \begin{aligned}
    F_{u_0Im}=\begin{pmatrix}
      0 & \frac{1}{3} & 0 & 0\\
      -\frac{1}{3} & 0 & 0 & 0\\
      0 & 0 & 0 & \frac{1}{3}\\
      0 & 0 & -\frac{1}{3} & 0
    \end{pmatrix},
    F_{u_4Im}=\begin{pmatrix}
      0 & -\frac{1}{3} & 0 & 0\\
      \frac{1}{3} & 0 & 0 & 0\\
      0 & 0 & 0 & 0\\
      0 & 0 & 0 & 0
    \end{pmatrix},
    F_{u_8Im}=\begin{pmatrix}
      0 & 0 & 0 & 0\\
      0 & 0 & 0 & 0\\
      0 & 0 & 0 & -\frac{1}{3}\\
      0 & 0 & \frac{1}{3} & 0
    \end{pmatrix},\\
    F_{u_1Im}=F_{u_3Im}=\frac{1}{2}\begin{pmatrix}
      0 & 0 & 0 & 0\\
      0 & 0 & 0 & 0\\
      0 & 0 & 0 & \frac{1}{3}\\
      0 & 0 & -\frac{1}{3} & 0
    \end{pmatrix},
    F_{u_2Im}=F_{u_6Im}=\frac{1}{2}\begin{pmatrix}
      0 & \frac{1}{3} & 0 & 0\\
      -\frac{1}{3} & 0 & 0 & 0\\
      0 & 0 & 0 & 0\\
      0 & 0 & 0 & 0
    \end{pmatrix},\\
    F_{u_5Im}=F_{u_7Im}=\frac{1}{2}\begin{pmatrix}
      0 & -\frac{1}{3} & 0 & 0\\
      \frac{1}{3} & 0 & 0 & 0\\
      0 & 0 & 0 & -\frac{1}{3}\\
      0 & 0 & \frac{1}{3} & 0
    \end{pmatrix}.
  \end{aligned}
\end{equation}
The optimal $\bar{F}_{Im2}$ is then given by $\bar{F}_{Im2}=F_{u_0Im}+F_{u_4Im}^T+F_{u_8Im}^T+(F_{u_1Im}+F_{u_3Im})+(F_{u_2Im}+F_{u_6Im})+(F_{u_5Im}+F_{u_7Im})^T$, i.e.,
\begin{equation}
  \bar{F}_{Im2}=\begin{pmatrix}
    0 & \frac{4}{3} & 0 & 0\\
    -\frac{4}{3} & 0 & 0 & 0\\
    0 & 0 & 0 & \frac{4}{3}\\
    0 & 0 & -\frac{4}{3} & 0
  \end{pmatrix},
\end{equation}
which gives a tighter bound as
\begin{equation}
  \frac{1}{\nu}Tr[F_Q^{-1}Cov(\hat{x})^{-1}]\le n-\frac{(n-2)}{(n-1)^2}\|\frac{\bar{F}_{Im2}}{2}\|_F^2=4-\frac{2}{9}\times\frac{1}{4}\times 4\times \frac{16}{9}=4-\frac{32}{81}\approx 3.60.
\end{equation}

If we only estimate $\{x_1,x_2,x_5\}$,
the associated matrices are given by the $3\times 3$ submatrices of the original ones,
\begin{equation}
  \begin{aligned}
    C_1=
    \begin{pmatrix}
      0 & 1 & \frac{1}{2}\\
      1 & 0 & \frac{1}{2}\\
      \frac{1}{2} & \frac{1}{2} & 0
    \end{pmatrix},
    \bar{F}_{Im}=
    \begin{pmatrix}
      0 & 1 & 0\\
      -1 & 0 & 0\\
      0 & 0  & 0
    \end{pmatrix},
  \end{aligned}
\end{equation}
which further gives $\|C_1\|_F=\sqrt{3}$, $\|\bar{F}_{Im}\|_F=\sqrt{2}$.
Then we have the tradeoff relations
\begin{equation}
  \frac{1}{\nu}Tr[F_Q^{-1}Cov(\hat{x})^{-1}]\le n-\frac{1}{4(n-1)}\|C_1\|_F^2=3-\frac{3}{8}=\frac{21}{8}=2.625,
\end{equation}
\begin{equation}
  \frac{1}{\nu}Tr[F_Q^{-1}Cov(\hat{x})^{-1}]\le n-\frac{(n-2)}{(n-1)^2}\|\bar{F}_{Im}\|_F^2=3-\frac{1}{2}=\frac{5}{2}=2.5.
\end{equation}

For $p$-local measurements, we have
\begin{equation}
  \begin{aligned}
    \frac{1}{\nu}Tr[F_Q^{-1}Cov(\hat{x})^{-1}]&\le n-\frac{1}{4(n-1)}\|\frac{C_p}{p}\|_F^2\\
    &= n-\frac{1}{4(n-1)}\|C_1\|_F^2\left(\frac{1}{p3^{p-1}}\mathcal{N}_p\right)^2\\
    &= 3-\frac{3}{8}\left(\frac{1}{p3^{p-1}}\mathcal{N}_p\right)^2
  \end{aligned}
\end{equation}
Specifically, for p=2,
\begin{equation}
  \frac{1}{\nu}Tr[F_Q^{-1}Cov(\hat{x})^{-1}]\le n-\frac{1}{4(n-1)}\|\frac{C_2}{2}\|_F^2=3-\frac{3}{8}\times \frac{1}{4}\times\frac{16}{9}=\frac{17}{6}\approx 2.83,
\end{equation}
and for p=3,
\begin{equation}
  \frac{1}{\nu}Tr[F_Q^{-1}Cov(\hat{x})^{-1}]\le n-\frac{1}{4(n-1)}\|\frac{C_3}{3}\|_F^2=3-\frac{3}{8}\times \frac{1}{9}\times\frac{25}{9}=\frac{623}{216}\approx 2.88.
\end{equation}

For 2-local measurements, if we choose the basis $\{\ket{u_q}\}$ as the computational basis $\ket{u_0}=\ket{00}$, $\ket{u_1}=\ket{01}$, $\ket{u_2}=\ket{02}$, $\ket{u_3}=\ket{10}$,..., $\ket{u_8}=\ket{22}$, the imaginary part of the matrices $F_{u_q}$ are given as
\begin{equation}
  \begin{aligned}
    F_{u_0Im}=\begin{pmatrix}
      0 & \frac{1}{3} & 0\\
      -\frac{1}{3} & 0 & 0\\
      0 & 0 & 0
    \end{pmatrix},
    F_{u_4Im}=\begin{pmatrix}
      0 & -\frac{1}{3} & 0\\
      \frac{1}{3} & 0 & 0\\
      0 & 0 & 0
    \end{pmatrix},
    F_{u_1Im}=F_{u_3Im}=F_{u_8Im}=\mathbf{0},\\
    F_{u_2Im}=F_{u_6Im}=\frac{1}{2}\begin{pmatrix}
      0 & \frac{1}{3} & 0\\
      -\frac{1}{3} & 0 & 0\\
      0 & 0 & 0
    \end{pmatrix},
    F_{u_5Im}=F_{u_7Im}=\frac{1}{2}\begin{pmatrix}
      0 & -\frac{1}{3} & 0\\
      \frac{1}{3} & 0 & 0\\
      0 & 0 & 0
    \end{pmatrix}.
  \end{aligned}
\end{equation}
The optimal $\bar{F}_{Im2}$ is then given by $\bar{F}_{Im2}=F_{u_0Im}+F_{u_4Im}^T+(F_{u_2Im}+F_{u_6Im})+(F_{u_5Im}+F_{u_7Im})^T$, i.e.,
\begin{equation}
  \bar{F}_{Im2}=\begin{pmatrix}
    0 & \frac{4}{3} & 0\\
    -\frac{4}{3} & 0 & 0\\
    0 & 0 & 0
  \end{pmatrix},
\end{equation}
which gives
\begin{equation}
  \frac{1}{\nu}Tr[F_Q^{-1}Cov(\hat{x})^{-1}]\le n-\frac{(n-2)}{(n-1)^2}\|\frac{\bar{F}_{Im2}}{2}\|_F^2=3-\frac{1}{4}\times\frac{1}{4}\times 2\times \frac{16}{9}=3-\frac{2}{9}\approx 2.78.
\end{equation}
This is tighter than the bound given by $C_2$.
%which
%}

If we only estimate $\{x_1,x_2\}$,
the associated matrices are given by the $2\times 2$ submatrices of the original ones,
\begin{equation}
  \begin{aligned}
    C_1=
    \begin{pmatrix}
      0 & 1\\
      1 & 0
    \end{pmatrix}
  \end{aligned}
\end{equation}
which further gives $\|C_1\|_F=\sqrt{2}$.
Then we have the tradeoff relation
\begin{equation}
  \frac{1}{\nu}Tr[F_Q^{-1}Cov(\hat{x})^{-1}]\le n-\frac{1}{4(n-1)}\|C_1\|_F^2=2-\frac{1}{2}=\frac{3}{2}.
\end{equation}

For $p$-local measurements, we have
\begin{equation}
  \begin{aligned}
    \frac{1}{\nu}Tr[F_Q^{-1}Cov(\hat{x})^{-1}]&\le n-\frac{1}{4(n-1)}\|\frac{C_p}{p}\|_F^2\\
    &= n-\frac{1}{4(n-1)}\|C_1\|_F^2\left(\frac{1}{p3^{p-1}}\mathcal{N}_p\right)^2\\
    &= 2-\frac{1}{2}\left(\frac{1}{p3^{p-1}}\mathcal{N}_p\right)^2
  \end{aligned}
\end{equation}
Specifically, for p=2,
\begin{equation}
  \frac{1}{\nu}Tr[F_Q^{-1}Cov(\hat{x})^{-1}]\le n-\frac{1}{4(n-1)}\|\frac{C_2}{2}\|_F^2=2-\frac{1}{2}\times \frac{1}{4}\times\frac{16}{9}=\frac{16}{9}\approx 1.78,
\end{equation}
and for p=3,
\begin{equation}
  \frac{1}{\nu}Tr[F_Q^{-1}Cov(\hat{x})^{-1}]\le n-\frac{1}{4(n-1)}\|\frac{C_3}{3}\|_F^2=2-\frac{1}{2}\times \frac{1}{9}\times\frac{25}{9}=\frac{299}{162}\approx 1.85.
\end{equation}

We plot the bound with different $p$ in Fig.\ref{fig:suppexample2}. It can be seen that the Holevo bound, which equals to the QCRB since the weak commutative condition holds in this case, is only achievable when $p\rightarrow
\infty$.
\begin{figure}
  \includegraphics[width=0.8\textwidth]{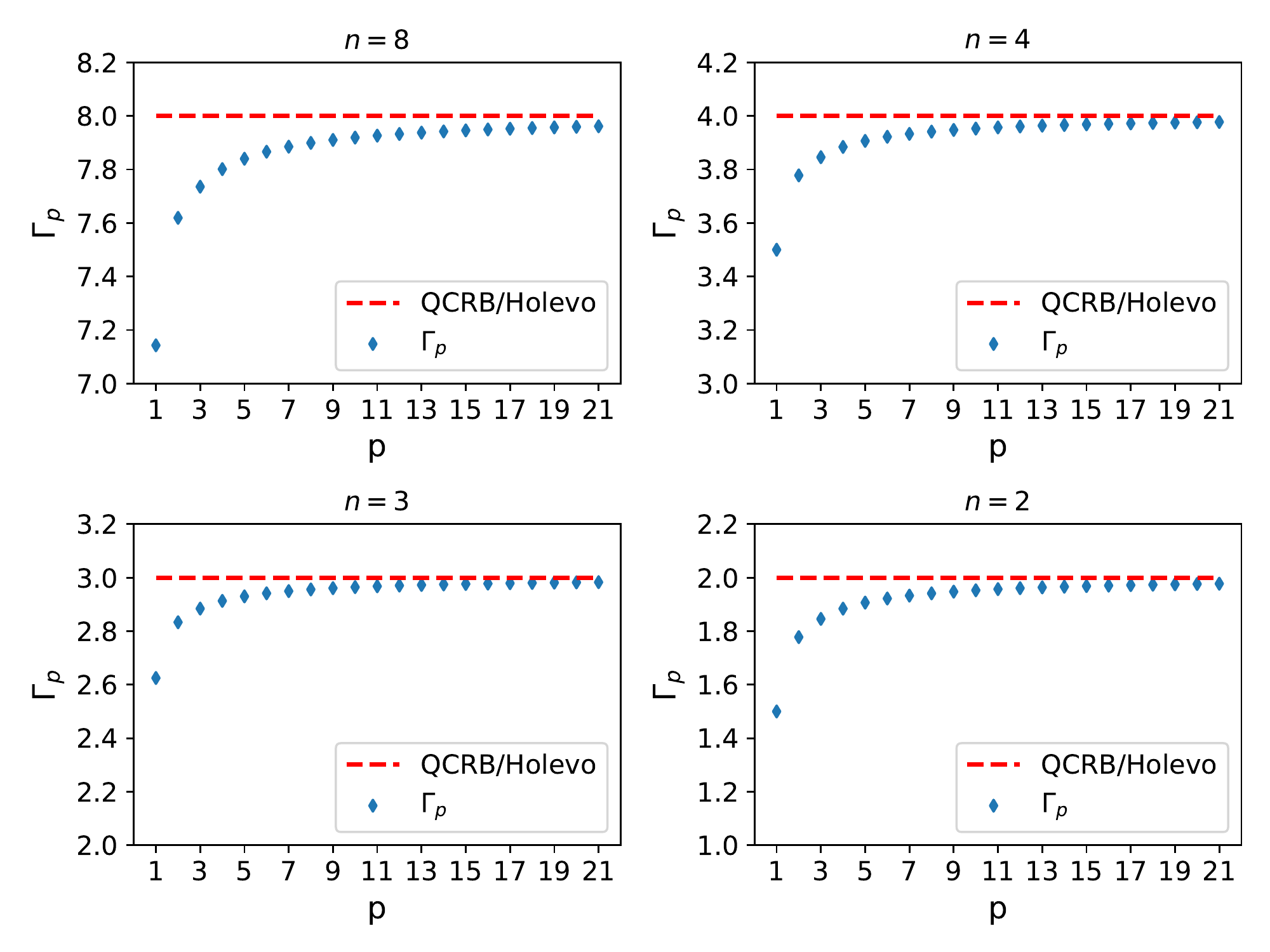}
  \caption{Upper bound on $\Gamma_p$ and the QCRB/Holevo bound with the number of parameters equal to $8, 4, 3, 2$ respectively.}
  \label{fig:suppexample2}
\end{figure}

\end{widetext}

\end{document}